\documentclass[prd,reprint,superscriptaddress,altaffillsymbol,amsmath,nofootinbib,longbibliography]{revtex4-1}
\pdfoutput=1
\usepackage{lipsum}
\usepackage{ifthen}
\usepackage{epsfig}
\usepackage{booktabs} 
\usepackage{natbib}
\usepackage{wasysym}
\usepackage{slashed} 
\usepackage{bbm}
\usepackage{mathrsfs}
\usepackage{amssymb}
\usepackage{dsfont}
\usepackage[export]{adjustbox}

\usepackage{comment}
\usepackage{glossaries}

\usepackage{xcolor}

\definecolor{OurBlue}{RGB}{13,115,178}
\definecolor{OurRed}{RGB}{213,91,1}     
\definecolor{OurOrange}{RGB}{222,143,6}    
\definecolor{OurGreen}{RGB}{19,158,114}   
\definecolor{OurPurple}{RGB}{204,120,188}  
\definecolor{OurBrown}{RGB}{202,145,96}   
\definecolor{OurLightBlue}{RGB}{86,180,233}   
\definecolor{OurPink}{RGB}{251,175,228}  
\definecolor{OurGray}{RGB}{148,148,148}  
\definecolor{OurYellow}{RGB}{236,225,50} 

\usepackage[
colorlinks=true, 
linkcolor=OurBlue,
citecolor=OurRed,
urlcolor=OurBlue,
]{hyperref}

\allowdisplaybreaks

\newcommand{\erf}{\mathop{\mathrm{erf}}}
\newcommand{\beqra}{\begin{eqnarray}}
\newcommand{\eeqra}{\end{eqnarray}}
\newcommand{\beq}{\begin{equation}}
\newcommand{\eeq}{\end{equation}}
\newcommand{\dd}{\mathrm{d}}

\newcommand{\ddd}{\mathrm{d}^3\,}

\newcommand{\boldell}{\boldsymbol{\ell}}

\renewcommand{\epsilon}{\varepsilon}

\renewcommand{\vec}[1]{\mathbf{#1}}

\newcommand{\vPerpEl}{\mathbf{v}_{\rm el}^\perp}

\newcount\colveccount

\newcommand*\columnvector[1]{
        \global\colveccount#1
        \begin{pmatrix}
        \colvecnext
}
\def\colvecnext#1{
        #1
        \global\advance\colveccount-1
        \ifnum\colveccount>0
                \\[2mm]
                \expandafter\colvecnext
        \else
                \end{pmatrix}
        \fi
}

\begin{document}

\title{Direct searches for general dark matter-electron interactions with graphene detectors:\\
Part II. Sensitivity studies}

\author{Riccardo Catena}
\email{catena@chalmers.se}
\affiliation{Chalmers University of Technology, Department of Physics, SE-412 96 G\"oteborg, Sweden}

\author{Timon Emken}
\email{timon.emken@fysik.su.se}
\affiliation{The Oskar Klein Centre, Department of Physics, Stockholm University, AlbaNova, SE-10691 Stockholm, Sweden}

\author{Marek Matas}
\email{marek.matas@mat.ethz.ch}
\affiliation{Department of Materials, ETH Z\"urich, CH-8093 Z\"urich, Switzerland}

\author{Nicola A. Spaldin}
\email{nicola.spaldin@mat.ethz.ch}
\affiliation{Department of Materials, ETH Z\"urich, CH-8093 Z\"urich, Switzerland}

\author{Einar Urdshals}
\email{urdshals@chalmers.se}
\affiliation{Chalmers University of Technology, Department of Physics, SE-412 96 G\"oteborg, Sweden}

\begin{abstract}
We use a formalism that describes electron ejections from graphene-like targets by dark matter (DM) scattering for general forms of scalar and spin 1/2 DM-electron interactions in combination with state-of-the-art density functional calculations to produce predictions and reach estimates for various possible carbon-based detector designs. Our results indicate the importance of a proper description of the target electronic structure. In addition, we find a strong dependence of the predicted observed signal for different DM candidate masses and interaction types on the detailed geometry and design of the detector. Combined with directional background vetoing, these dependencies will enable the identification of DM particle properties once a signal has been established.
\end{abstract}
\maketitle

\section{Introduction}
\label{sec:introduction}

In the quest for dark matter (DM), the lack of detection of Weakly Interacting Massive Particles (WIMPs) has motivated the exploration of alternative {\it theoretical}, and {\it experimental} frameworks.~On the theoretical side, the focus is gradually shifting towards DM candidates that are lighter than a nucleon~\cite{Battaglieri:2017aum}, and thereby too light to be observed in conventional nuclear recoil experiments~\cite{Drukier:1983gj,Goodman:1984dc}.~On the experimental side, the emphasis is being placed on the search for sub-GeV DM via electronic transitions induced by the scattering of Milky Way DM particles by the electrons bound to a target material~\cite{Essig:2011nj,Essig:2015cda}.

The most common target materials used in low-background experiments are liquid argon~\cite{Agnes:2018oej} or xenon~\cite{Essig:2012yx,Essig:2017kqs,Aprile:2019xxb}, and semiconductor crystals~\cite{Essig:2011nj,Graham:2012su,Lee:2015qva,Essig:2015cda,Crisler:2018gci,Agnese:2018col,Abramoff:2019dfb,Aguilar-Arevalo:2019wdi,Amaral:2020ryn,Andersson:2020uwc}.
In addition, more exotic materials such as Dirac materials~\cite{Hochberg:2017wce,Geilhufe:2019ndy}, graphene~\cite{Hochberg:2016ntt,Geilhufe:2018gry} or carbon nanotubes (CNTs)~\cite{Capparelli:2014lua, Cavoto:2016lqo, Cavoto:2017otc, Cavoto:2019flp} have been carefully investigated in this context.
All of these targets have an energy threshold of the order of a few eV in common and are therefore sensitive to sub-GeV~DM masses.~Furthermore, this class of materials can be inherently sensitive to the direction of the incoming DM particle -- a feature that would facilitate discrimination of a signal associated with the DM wind direction from any isotropic experimental background.~In particular, intrinsically anisotropic materials such as graphene and CNTs are characterised by an enhanced daily modulation of the rate of DM-induced electron ejections -- a pattern that is not expected in typical experimental backgrounds.~This enhanced modulation results from a strong daily change in the overlap between the kinematically allowed values of the momentum transfer vector, $\vec q$, and the values of $\vec q$ that maximise the material ``response function'' (defined below in Eq.~(\ref{eq: response function general})).~Experiments currently in the design or research and development stage that will search for DM-induced electron ejections from graphene sheets or arrays of CNTs include the Princeton Tritium Observatory for Light, Early-Universe, Massive-Neutrino Yield, or PTOLEMY~\cite{Betts:2013uya,PTOLEMY:2018jst, PTOLEMY:2022ldz}, as well as the ``Graphene-FET'' and  ``dark-PMT'' projects~\cite{Apponi:2021lyd}.~The Graphene-FET project focuses on utilizing graphene sheets, while dark-PMT operates arrays of CNTs.\\

\noindent In a companion work (from now onward, Paper I)~\cite{PaperI}, we performed state-of-the-art electronic structure calculations for graphene, and introduced a solid theoretical and computational framework for the accurate modeling of DM-induced electron ejections from graphene-based targets.~Our framework combines effective theory methods -- used to describe the interaction between DM and electrons in a general manner~\cite{Catena:2019gfa} -- with density functional theory (DFT) in order to express the rate of DM-induced electron ejections from graphene-based targets in terms of a single graphene response function.~Remarkably, the latter is directly related to the ``diagonal part'' of the electron momentum density, which by construction is a solid output of DFT~\cite{Feng2019Nov,Trevisanutto2008Nov}.~We then applied this framework to calculate the expected daily modulation of the rate of DM-induced electron ejections from a hypothetical detector using graphene as a target material.  \\

\noindent This work complements Paper I by extending the framework introduced there to the experimentally relevant case of CNTs, and by performing statistically reliable estimates of the expected exclusion limits (assuming a null result) and discovery potential (in the case of a positive signal) for both graphene- and CNT-based detectors.~Our sensitivity studies provide valuable information for the design stage of next-generation experiments such as PTOLEMY, Graphene-FET, and dark-PMT.  \\

This article is organized as follows.~In Sec.~\ref{sec: general formalism} we review our general formalism for modeling the ejection of electrons by the scattering of DM particles in graphene (Paper I), and extend it to the case of CNTs. In Sec.~\ref{sec:exp}, we introduce specific  configurations that a graphene or CNT experiment such as PTOLEMY could operate, and then apply our formalism to investigate the potential of these configurations in Sec.~\ref{sec: sensitivity studies exclusion limits}.~Finally, we conclude in Sec.~\ref{sec:conclusions}.~We complement this work with appendices where we expand on the analytic form of the employed formalism and compare possible setups of CNT-based experiments.

\begin{table}[t]
    \centering
    \begin{tabular*}{\columnwidth}{@{\extracolsep{\fill}}ll@{}}
    \toprule
      $\mathcal{O}_1 = \mathds{1}_{\chi e}$ & $\mathcal{O}_9 = i\mathbf{S}_\chi\cdot\left(\mathbf{S}_e\times\frac{ \mathbf{q}}{m_e}\right)$  \\
        $\mathcal{O}_3 = i\mathbf{S}_e\cdot\left(\frac{ \mathbf{q}}{m_e}\times \mathbf{v}^{\perp}_{\rm el}\right)$ &   $\mathcal{O}_{10} = i\mathbf{S}_e\cdot\frac{ \mathbf{q}}{m_e}$   \\
        $\mathcal{O}_4 = \mathbf{S}_{\chi}\cdot \mathbf{S}_e$ &   $\mathcal{O}_{11} = i\mathbf{S}_\chi\cdot\frac{ \mathbf{q}}{m_e}$   \\                                                                             
        $\mathcal{O}_5 = i\mathbf{S}_\chi\cdot\left(\frac{ \mathbf{q}}{m_e}\times \mathbf{v}^{\perp}_{\rm el}\right)$ &  $\mathcal{O}_{12} = \mathbf{S}_{\chi}\cdot \left(\mathbf{S}_e \times \mathbf{v}^{\perp}_{\rm el} \right)$ \\                                                                                                                 
        $\mathcal{O}_6 = \left(\mathbf{S}_\chi\cdot\frac{ \mathbf{q}}{m_e}\right) \left(\mathbf{S}_e\cdot\frac{{\bf{q}}}{m_e}\right)$ &  $\mathcal{O}_{13} =i \left(\mathbf{S}_{\chi}\cdot  \mathbf{v}^{\perp}_{\rm el}\right)\left(\mathbf{S}_e\cdot \frac{ \mathbf{q}}{m_e}\right)$ \\   
        $\mathcal{O}_7 = \mathbf{S}_e\cdot  \mathbf{v}^{\perp}_{\rm el}$ &  $\mathcal{O}_{14} = i\left(\mathbf{S}_{\chi}\cdot \frac{ \mathbf{q}}{m_e}\right)\left(\mathbf{S}_e\cdot  \mathbf{v}^{\perp}_{\rm el}\right)$  \\
        $\mathcal{O}_8 = \mathbf{S}_{\chi}\cdot  \mathbf{v}^{\perp}_{\rm el}$  & $\mathcal{O}_{15} = i\mathcal{O}_{11}\left[ \left(\mathbf{S}_e\times  \mathbf{v}^{\perp}_{\rm el} \right) \cdot \frac{ \mathbf{q}}{m_e}\right] $ \\       
    \bottomrule
    \end{tabular*}
    \caption{Interaction operators defining the non-relativistic effective theory of spin 0 and 1/2 DM-electron interactions~\cite{Catena:2019gfa}.~$\mathbf{S}_e$ ($\mathbf{S}_\chi$) is the electron (DM) spin, $\mathbf{v}_{\rm el}^\perp=\mathbf{v}-\boldsymbol{\ell}/m_e-\mathbf{q}/(2 \mu_{\chi e})$, where $\mu_{\chi e}$ is the DM-electron reduced mass, $\mathbf{v}_{\rm el}^\perp$ is the relative transverse velocity and $\mathds{1}_{\chi e}$ is the identity in the DM-electron spin space.~In the case of elastic scattering, $\mathbf{v}_{\rm el}^\perp \cdot \mathbf{q}=0$.}
\label{tab:operators}
\end{table}

\section{General formalism for dark matter-induced electron ejections in graphene detectors}
\label{sec: general formalism}

In the complementary work of Paper I, we introduced a general formalism describing DM-induced electron ejections from periodic systems. Unlike in the case of electronic excitations, discussed in our work on semiconductors~\cite{Catena:2021qsr}, in Paper I we modeled the final-state electron as a plane wave, which leads to a reduction of the five material response functions found in~\cite{Catena:2021qsr} into a single one. We achieved this result by using an effective theory framework to list all possible interactions that a non-relativistic scalar or spin-1/2 DM particle can have with the SM~\cite{Catena:2019gfa}. This approach also allowed us to write down explicit expressions for the electron ejection rate for selected benchmark models, that are described in Sec.~II.C of Paper~I. Then we applied the general formalism to the case of two-dimensional targets, with a focus on graphene.

In order to obtain the graphene material response function, two approaches for electronic structure calculation were tested; DFT and tight binding. As has been shown, the tight-binding approach, while working well for capturing crystal effects within a solid, is not internally self-consistent once one aims to obtain a full description of the material on the wavefunction level. In that case, an atomic carbon wavefunction has to be embedded within the formalism, which requires either an inconsistency in the overlap integral values necessary for the correct reconstruction of graphene energy levels, or an unphysical atomic wavefunction. 

While DFT in general provides only an approximation to the individual electronic states, in Paper I we found that the material response function for electronic ejections is proportional to the ``diagonal part'' of the target electron momentum density, an observable well-motivated and described within this framework. This is why, for the scope of this work, we will use DFT as our framework of choice for obtaining predictions and studying various carbon-based detector designs in more detail.

\subsection{Master formula}
In Paper I, we showed that the matrix element describing bound electrons ejected by non-relativistic DM factorizes into free particle and material response functions. The free particle response function $R_\mathrm{free}$ depends on the properties of the free particles such as the momentum and mass of the ejected electron and the DM particle. We give $R_\mathrm{free}$ explicitly in App.~\ref{app: matrix element}. The rate of ejected electrons is given by
\begin{widetext}
\begin{align}
 R&=\frac{n_\chi N_\text{cell}}{32\pi^2m_\chi^2m_e^2}\int\dd^3 \mathbf{k}^\prime\int \mathrm{d}\, E_e\int\dd^3 \mathbf{q}\int\dd^3 \mathbf{v}\,f_\chi(\mathbf{v})\delta\left(\Delta E_e +\frac{q^2}{2m_\chi}- \mathbf{v}\cdot\mathbf{q}\right)
 R_\mathrm{free}(\mathbf{k}^\prime,\mathbf{q},\mathbf{v})
 \;W(\mathbf{k}^\prime-\mathbf{q},E_e)\, ,
 \label{eq:rate_general}
\end{align}
\end{widetext}
where $n_\chi=0.4\,\mathrm{GeV}/\mathrm{cm}^3/m_\chi$ is the DM particle number density, $N_\mathrm{cell}$ is the number of unit cells in the detector, and $m_e$ and $m_\chi$ are the masses of the electron and DM particle, respectively. $\mathbf{k}^\prime$ is the momentum of the final state free electron, $\mathbf{q}$ is the momentum transferred from the DM particle to the electron, and $\mathbf{v}$ is the initial state velocity of the DM particle. We have defined the electron's energy change as~$\Delta E_e\equiv\frac{k^{\prime 2}}{2 m_e}+\Phi- E_e$, with $E_e-\Phi$ being the energy of the initial state electron relative to a free electron at rest. $\Phi=4.3\,\mathrm{eV}$ is the work function of graphene and $E_e\leq 0\,\mathrm{eV}$. The above mentioned material-specific response function is given as
\begin{align}
    W\left(\boldell,E_e\right) &=
    \frac{V_\mathrm{cell}}{(2\pi)^3}\sum_i \int_\text{BZ} \frac{\mathrm{d}^3 \mathbf{k}}{(2\pi)^3} \delta\left(E_e -E_i(\mathbf{k})\right) \nonumber\\
    &\times 
    \left| \widetilde{\psi}_{i\mathbf{k}}(\boldell) \right|^2\,,
\label{eq: response function general}
\end{align}
with $V_\mathrm{cell}$ being the volume of the unit cell, $i$ being the band index, $\mathbf{k}$ being the Brillouin zone momentum (also referred to as crystal momentum), $E_i(\mathbf{k})$ being the energy of an electron in band $i$ with Brillouin zone momentum $\mathbf{k}$, and $\widetilde{\psi}_{i\mathbf{k}}(\boldell)$ being the momentum space electron wave function with linear momentum $\boldell$.
In Eq.~(\ref{eq:rate_general}), the velocity distribution of the DM particles gravitationally bound to the galaxy is assumed to be
\begin{align}
    f_\chi(\mathbf{v})&= \frac{1}{N_{\rm esc}\pi^{3/2}v_0^3}\exp\left[-\frac{(\mathbf{v}+\mathbf{v}_e)^2}{v_0^2} \right]
    \nonumber\\
    &\times \Theta\left(v_{\rm esc}-|\mathbf{v}+\mathbf{v}_e|\right)\,,
    \label{eq:df}
\end{align}
where we use $v_0 = |\mathbf{v}_0| =~238$~km~s$^{-1}$~\cite{Baxter:2021pqo} for the local standard of rest speed, and $v_\mathrm{esc}=544$~km~s$^{-1}$~\cite{Baxter:2021pqo} for the galactic escape speed.~Following~\cite{Geilhufe:2019ndy}, the Earth's velocity with respect to the galactic centre, $\mathbf{v}_e$, is expressed in a coordinate system with $z$-axis in the $\mathbf{v}_0+\mathbf{v}_{\odot}$ direction. 
$\mathbf{v}_{\odot}$ is the Sun's peculiar velocity and $v_e=|\mathbf{v}_0+\mathbf{v}_{\odot}|\simeq 250.5$~km~s$^{-1}$ ~\cite{Baxter:2021pqo} is given as,
\begin{align}
\mathbf{v}_e = v_e \left(
\begin{array}{c}
\sin\alpha_e \sin\beta \nonumber\\
\sin\alpha_e \cos\alpha_e (\cos\beta -1) \nonumber\\
\cos^2\alpha_e + \sin^2\alpha_e\cos\beta
\end{array}
\right) \,,
\end{align}
where $\alpha_e=42^\circ$, $\beta = 2\pi\, t/{\rm day}$, and $t$ is the time variable. Finally, the normalization constant is given by
\begin{align}
N_{\rm esc}&\equiv \erf(v_{\rm esc}/v_0)-{2\over \sqrt{\pi}} {v_{\rm esc}\over v_0}\exp\left(-\frac{v_{\rm esc}^2}{v_0^2}\right)\, .
\end{align}

\subsection{DFT implementation}
\label{sec:graphene}
As discussed above, we use DFT to obtain $W(\boldell, E_e)$,  expanding the target electron Bloch wave-functions, $\psi_{i\mathbf{k}}(\mathbf{x})$ in plane waves,
\begin{equation}
    \psi_{i\mathbf{k}}(\mathbf{x})=\frac{1}{\sqrt{V}}\sum_\mathbf{G} u_i(\mathbf{k}+\mathbf{G}) e^{i(\mathbf{k}+\mathbf{G})\cdot\mathbf{x}}\,.
    \label{eq: Bloch wave function}
\end{equation}
These Bloch wave functions are normalized over a finite volume $V$, while the $u_i$ coefficients obey $\sum_\mathbf{G}\left|u_i(\mathbf{k}+\mathbf{G})\right|=1$ for all bands $i$ and all Brillouin zone momenta $\mathbf{k}$. Fourier transforming and squaring Eq.~(\ref{eq: Bloch wave function}) and inserting it in Eq.~(\ref{eq: response function general}) gives
\begin{align}
    W\left(\boldell, E_e\right)  &= V_\mathrm{cell}\sum_i \int_\text{BZ} \frac{\mathrm{d}^3 \mathbf{k}}{(2\pi)^3} \delta\left(E_e - E_i(\mathbf{k})\right)\nonumber\\
    &\times\sum_\mathbf{G}|u_i(\mathbf{k}+\mathbf{G})|^2\delta^{(3)}(\mathbf{k}+\mathbf{G}-\boldell)\, .
\label{eq: response function dft}    
\end{align}
The Bloch coefficients are obtained in a self-consistent field DFT calculation with \texttt{QuantumEspresso v.6.4.1}~\cite{Giannozzi_2009, Giannozzi_2017, doi:10.1063/5.0005082}, and the sum over band indices and integral over the Brillouin zone momentum are carried out in \texttt{QEdark-EFT}~\cite{QEdark-EFT}. 

\begin{figure}
    \centering
    \includegraphics[width=0.48\textwidth]{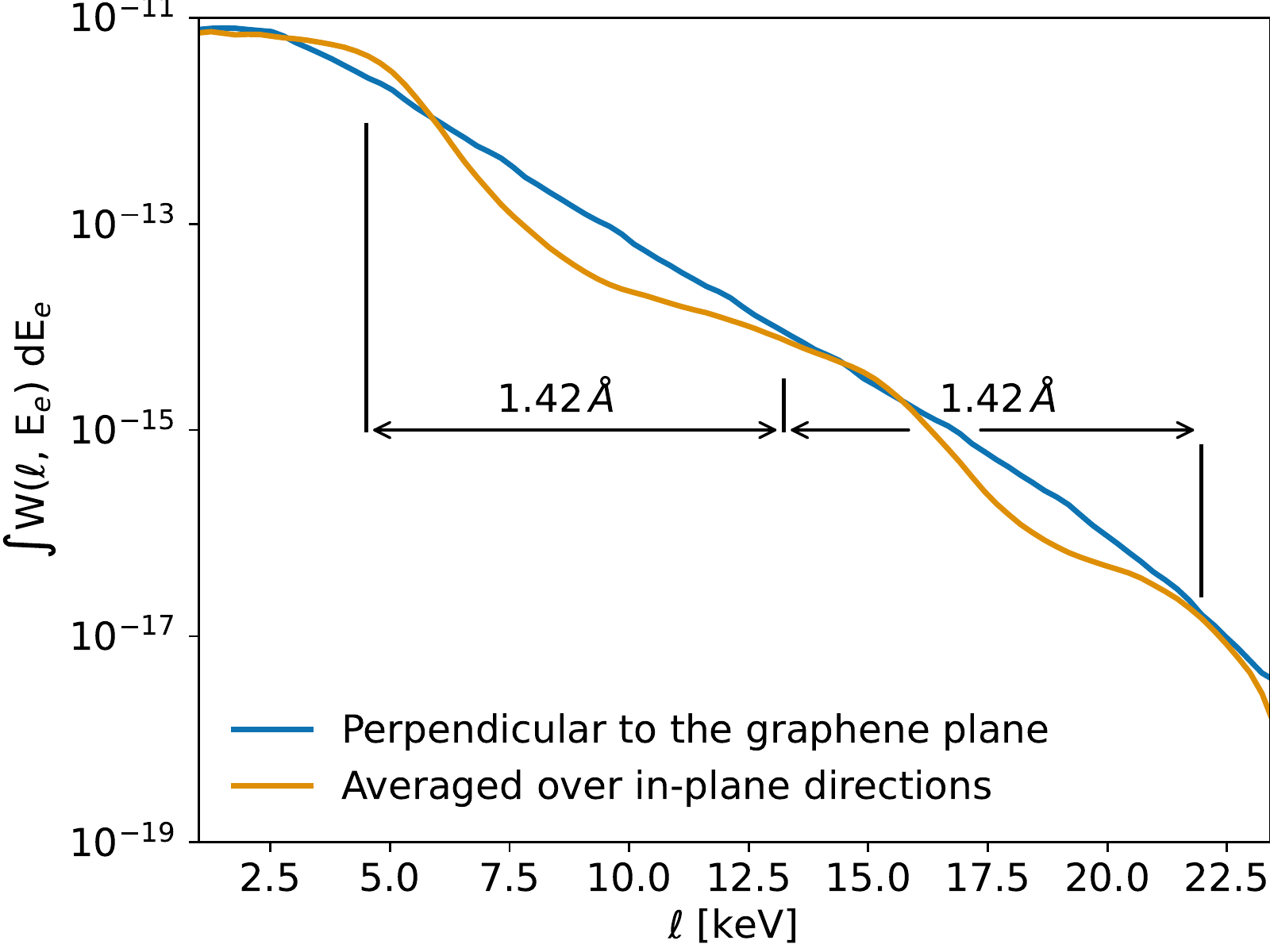}
    \caption{Graphene response function integrated over energy as a function of the initial electron momentum, $\ell$. The blue line represents response function fall-off with momenta that are perpendicular to the graphene sheet, whereas the orange line represents averaged in-sheet momenta. We can see that the in-plane  crystal momentum of the lattice with atomic spacing $\sim 1.42\,\mathrm{\AA}$ inflates the low-momentum electrons within the sheet.}
    \label{fig: W fall off}
\end{figure}

DFT directly and self-consistently computes the ground state electron density, which, as we argued in Paper I, is closely related to $W(\boldell,E_e)$, the latter being proportional to the ground state electron momentum density when band mixing effects can be neglected. DFT is therefore exceptionally well suited to obtain $W(\boldell,E_e)$ for the case of electron ejections from materials.
~The electron momentum distributions in the directions perpendicular and parallel to the graphene sheet are shown in Fig.~\ref{fig: W fall off}, where we plot the integrated graphene response as a function of the in-plane and out-of-plane initial electron momentum, $\ell$. Note that above approximately $5\,\mathrm{keV}$ there are more electrons with a momentum perpendicular to the graphene sheet than parallel to it. As we will see, this observation has an important impact on our predictions.

\subsection{Extension to carbon nanotubes}
\label{sec:nanotubes}
In this section, we extend the above formalism to the important case of electron ejections by DM scattering in carbon nanotubes.~This study is relevant for the development of PTOLEMY-CNT, which will search for DM with arrays of single- or multi-wall metallic carbon nanotubes. Furthermore, this detector design has been shown to outperform non-curved graphene sheets for the complementary cosmic neutrino background search due to the Heisenberg uncertainty final-state spectrum smearing~\cite{PTOLEMY:2022ldz}.

Carbon nanotubes are cylindrical tubes with walls of single or a few atom thickness and a radius of $R_\mathrm{tube}\approx 10\,\mathrm{nm}$. The electronic structure of carbon nanotubes can be modelled using DFT, 
taking effects of interactions between tubes into account. However, we leave this DFT calculation 

for future work, and use an approximate expression for the response function of carbon nanotubes that we obtain from that of graphene as explained below.

We start by noticing that a carbon nanotube can be approximated by the superposition of $n_s$ tangential graphene sheets.~This geometrical approximation is kinematically well-motivated as the typical momentum of the initial state electrons is around few $\mathrm{keV}$, which corresponds to a de-Broglie wavelength of $\lambda\lesssim 1\,\mathrm{nm}$.~Since $\lambda\ll R_\mathrm{tube}$, the target electrons do not resolve the curvature of the tubes, and segments of the tube walls can be approximated as locally flat and thus described by tangential planes.~As a result, the rate of electron ejections by DM scattering in carbon nanotubes can be approximated by the sum of contributions from $n_s$ tangential planes, each of mass $1/n_s$ times the mass of the nanotube.

The $n_s$ tangential planes used in this approximation can be obtained through subsequent, active rotations of a reference graphene sheet.~Let us denote by $\Omega_{(i)}$, $i=1,\dots,n_s$ the matrix that represents the rotation associated with the $i$-th tangential plane.~Under $\Omega_{(i)}$ the Bloch wave function $\psi_{i\mathbf{x}}(\mathbf{x})$ in Eq.~(\ref{eq: Bloch wave function}) transforms as follows\\
\begin{align}
    \psi_{i\mathbf{k}}(\mathbf{x})  \longrightarrow \psi'_{i\mathbf{k}}(\mathbf{x})  =\psi_{i\mathbf{k}}\left(\Omega_{(i)}^{-1}\mathbf{x}\right) \,,
\end{align}
which implies 
\begin{align}
 |\widetilde{\psi}_{i\mathbf{k}}(\boldsymbol{\ell})|^2 \longrightarrow 
 |\widetilde{\psi}'_{i\mathbf{k}}(\boldsymbol{\ell})|^2 &= (2\pi)^3 \sum_{\mathbf{G}} |u_i(\mathbf{k}+\mathbf{G})|^2 \nonumber\\
 &\times \delta^{(3)} \left(\Omega_{(i)}(\mathbf{k}+\mathbf{G})-\boldsymbol{\ell}\right)\,.
\end{align}
By using
\begin{align}
\delta^{(3)} \left(\Omega_{(i)}(\mathbf{k}+\mathbf{G})-\boldsymbol{\ell}\right)
= \delta^{(3)} \left(\mathbf{k}+\mathbf{G}-\Omega^{-1}_{(i)}\boldsymbol{\ell}\right)\,,
\end{align}
we can write the response function of a rotated graphene sheet as
\begin{align}
W(\boldsymbol{\ell},E_e) \longrightarrow W'_{(i)}(\boldsymbol{\ell},E_e) = W\left(\Omega^{-1}_{(i)}\boldsymbol{\ell},E_e\right)\,,
\end{align}
and approximate the response function of a single nanotube as
\begin{equation}
    W_\mathrm{tube}(\boldell,E_e)\approx\sum_{i=1}^{n_s} \frac{W\left(\Omega^{-1}_{(i)}\boldsymbol{\ell},E_e\right)}{n_s}\,,
\end{equation}
where the sum runs over the number of tangential planes used in the approximation.~In our calculations, we take the limit of an infinite number of planes, $n_s\rightarrow \infty$, which implies
\begin{equation}
    W_\mathrm{tube}(\boldell,E_e)\approx \frac{1}{2\pi} \int{\rm d}\phi \,W\left(|\boldsymbol{\ell}|\hat{\boldsymbol{n}},E_e\right)\,,
    \label{eq:Wtube}
\end{equation}
where, $\hat{\boldsymbol{n}}=(\cos\phi\sin\theta,\sin\phi\sin\theta,\cos\theta$),  $\cos\theta=\boldsymbol{\ell}\cdot \hat{\mathbf{z}}/|\boldsymbol{\ell}|$, $\hat{\mathbf{z}}$ is a unit vector in the direction of the symmetry axis of the nanotube and $\phi$ is the associated azimuthal angle.~In practice, we perform the integral in Eq.~(\ref{eq:Wtube}) together with the velocity and momentum transfer integrals in Eq.~(\ref{eq:rate_general}) by random sampling $\phi$ with Monte Carlo methods.

\begin{figure*}
\centering
\includegraphics[scale=0.4]{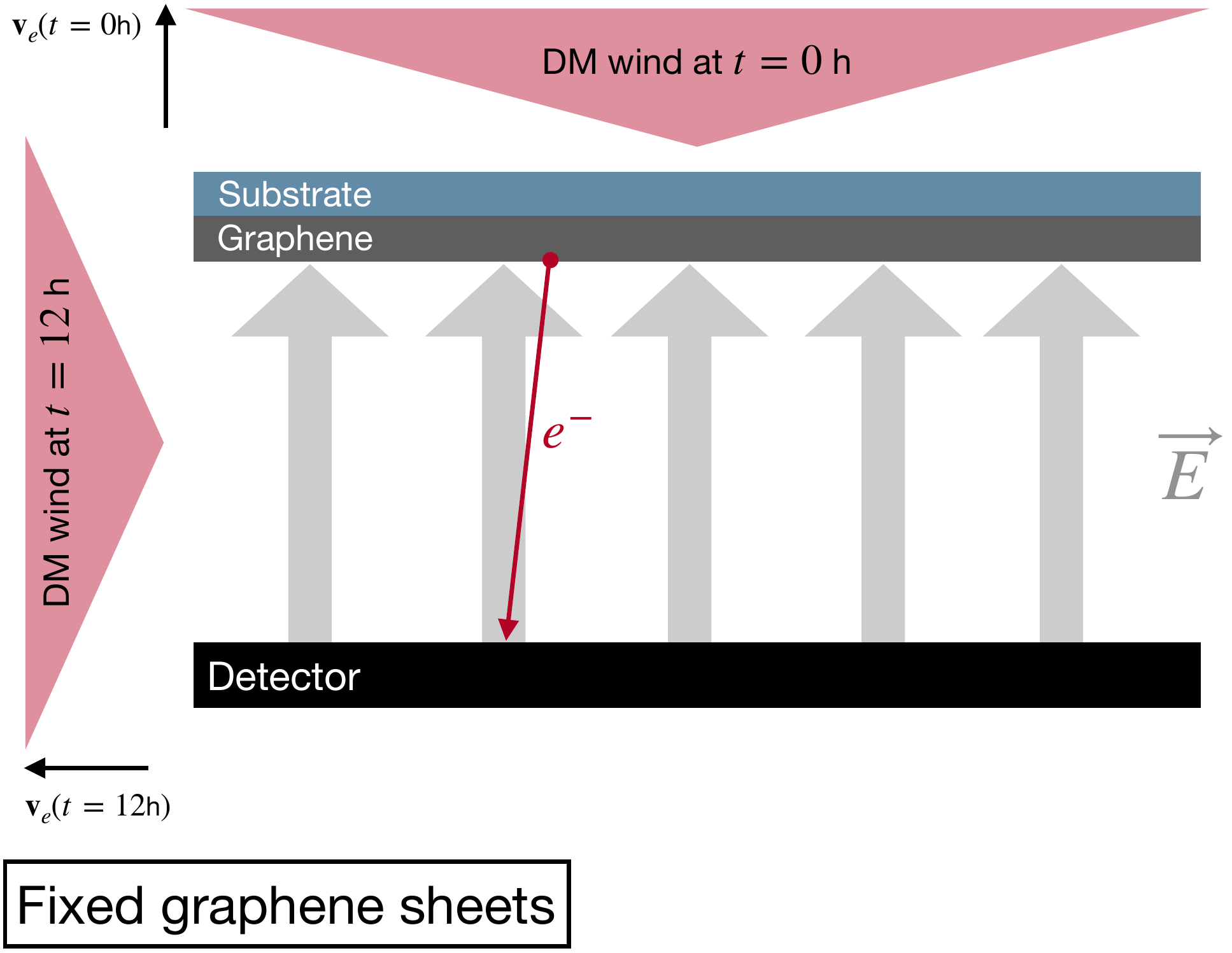}
\includegraphics[scale=0.4]{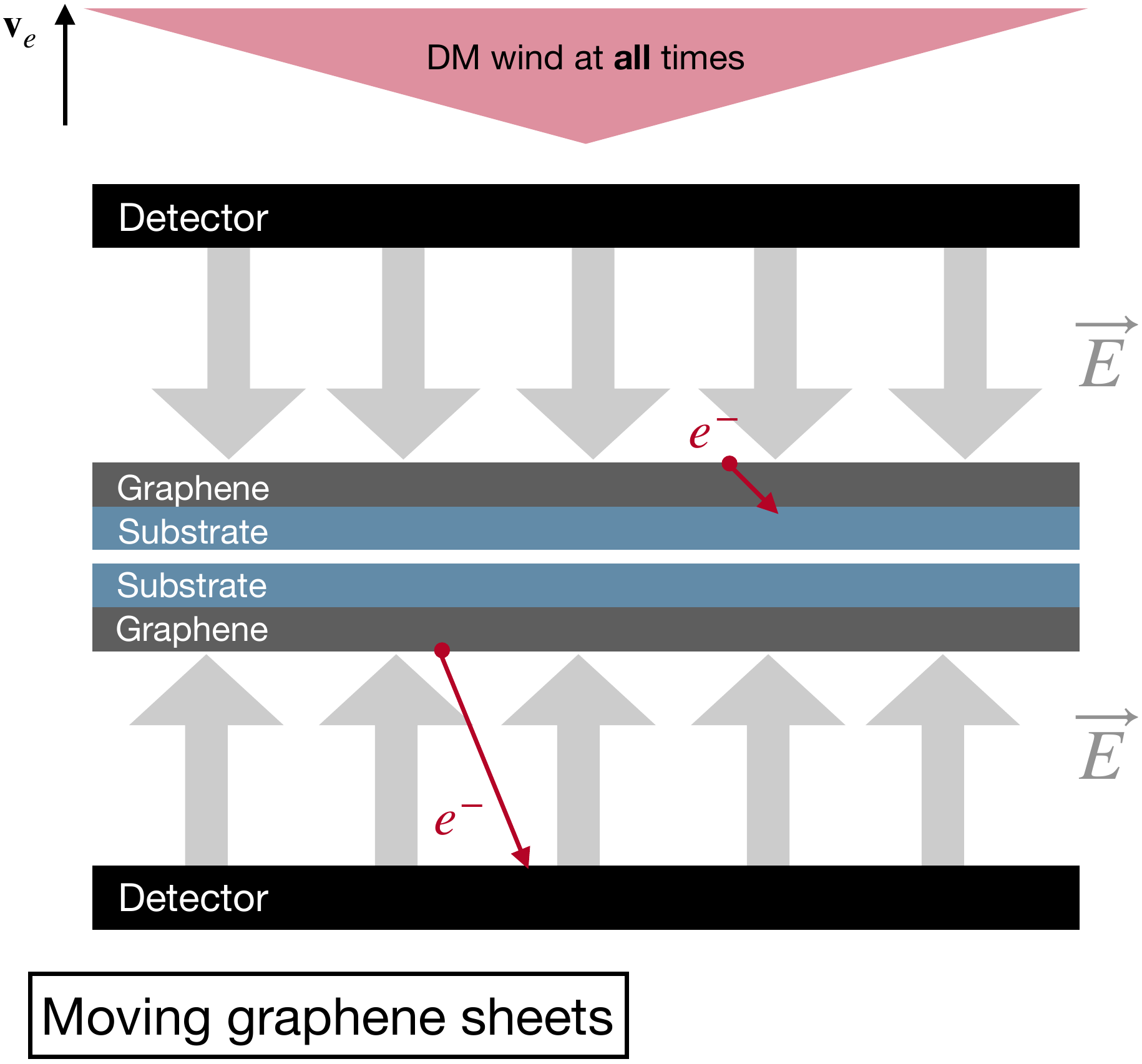}
\caption{Schematics of experimental setups with graphene sheet targets. The graphene sheet (dark gray) is grown on a substrate (blue). The electric field in the vacuum surrounding the sheet is shown with gray arrows. Electrons ejected from the sheet are accelerated by the electric field as indicated by the red arrow. \textbf{Left:} Setup in which graphene sheets are fixed in the lab. \textbf{Right:} Setup in which graphene sheets are tracking $\mathbf{v}_e$.}
\label{fig: graphene setups}
\end{figure*}

\section{Experimental settings for graphene-based dark matter detectors}
\label{sec:exp}

In the lab.\ frame, the flux of DM particles through the Earth, that is the so-called DM wind, 
peaks at $\mathbf{v}=-\mathbf{v}_e$, as one can see from Eq.~(\ref{eq:df}).~Since $\mathbf{v}_e$ varies during the day, the rate of electron ejections by DM scattering is expected to modulate correspondingly over a period of 24 hours.~For a given $\mathbf{k}'$, the amplitude of this daily modulation is large when the kinematically allowed regions for $\mathbf{q}$, found by solving the delta function constraint in Eq.~(\ref{eq:rate_general}), and the regions where $\mathbf{k}'-\mathbf{q}$ maximises $W(\mathbf{k}'-\mathbf{q})$ have a non trivial overlap at a given time, and a marginal overlap 12 hours later.~Whether the overlap between peaks of $W$ and kinematically allowed regions in $\mathbf{q}$-space follows this time evolution depends on the initial angle between $\mathbf{v}_e$ and the symmetry axis of the assumed target material.
~Consequently, anisotropic materials characterised by 

an angle-dependent response function can be used to amplify the expected daily modulation of the electron ejection rate by an appropriate choice of detector orientation.~Importantly, none of the known experimental backgrounds is expected to exhibit a similar daily modulation, which would therefore be a smoking gun for DM discovery.
~In this article, we consider two experimental setups that are in principle capable of detecting this
daily modulation.
~The two setups can be realised by using multi-layer graphene or carbon nanotube array detectors.~Both targets have an anisotropic response function (see Secs.~\ref{sec:graphene} and  \ref{sec:nanotubes}).

\begin{figure*}
\centering
\includegraphics[scale=0.4]{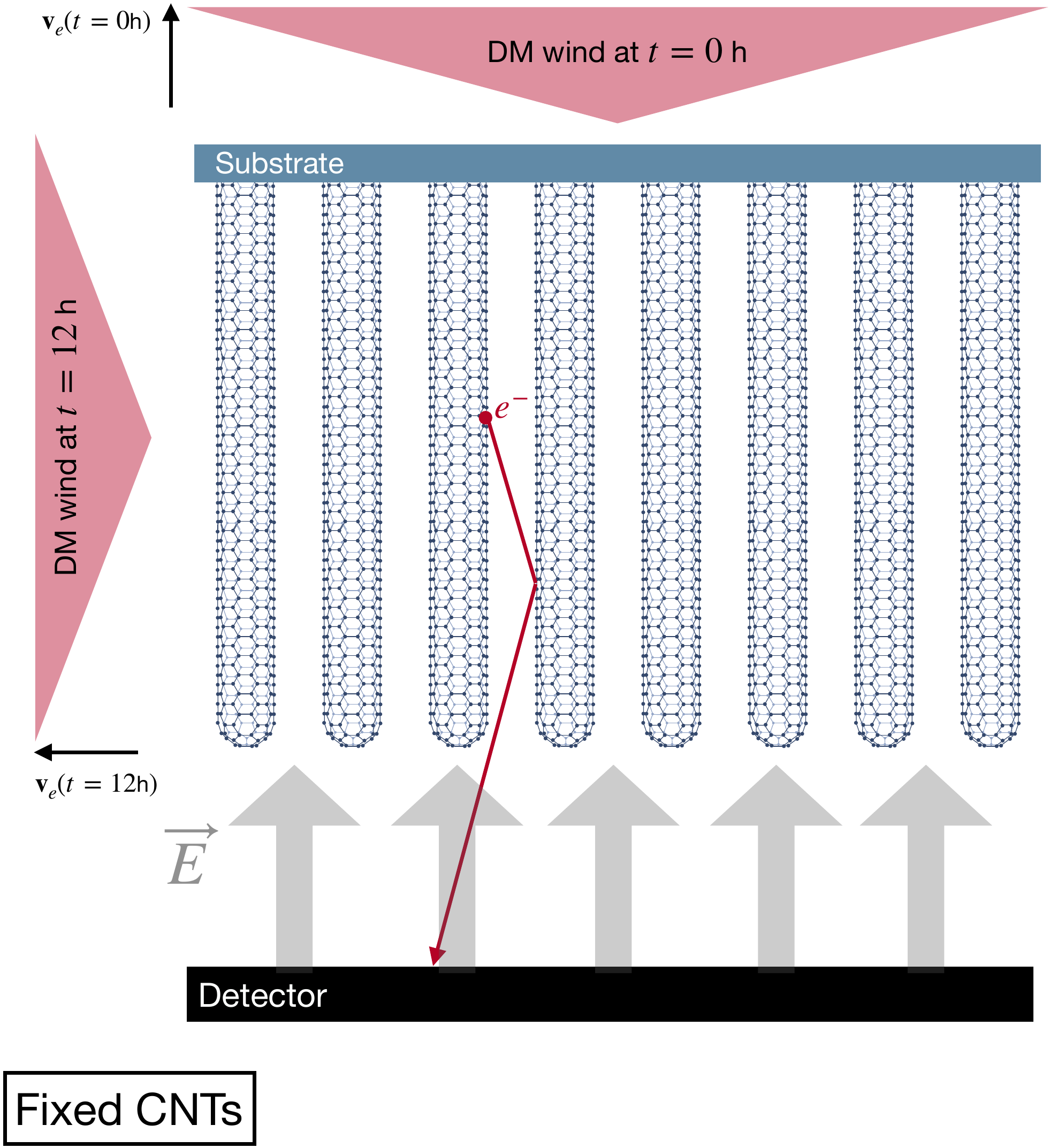}
\includegraphics[scale=0.4]{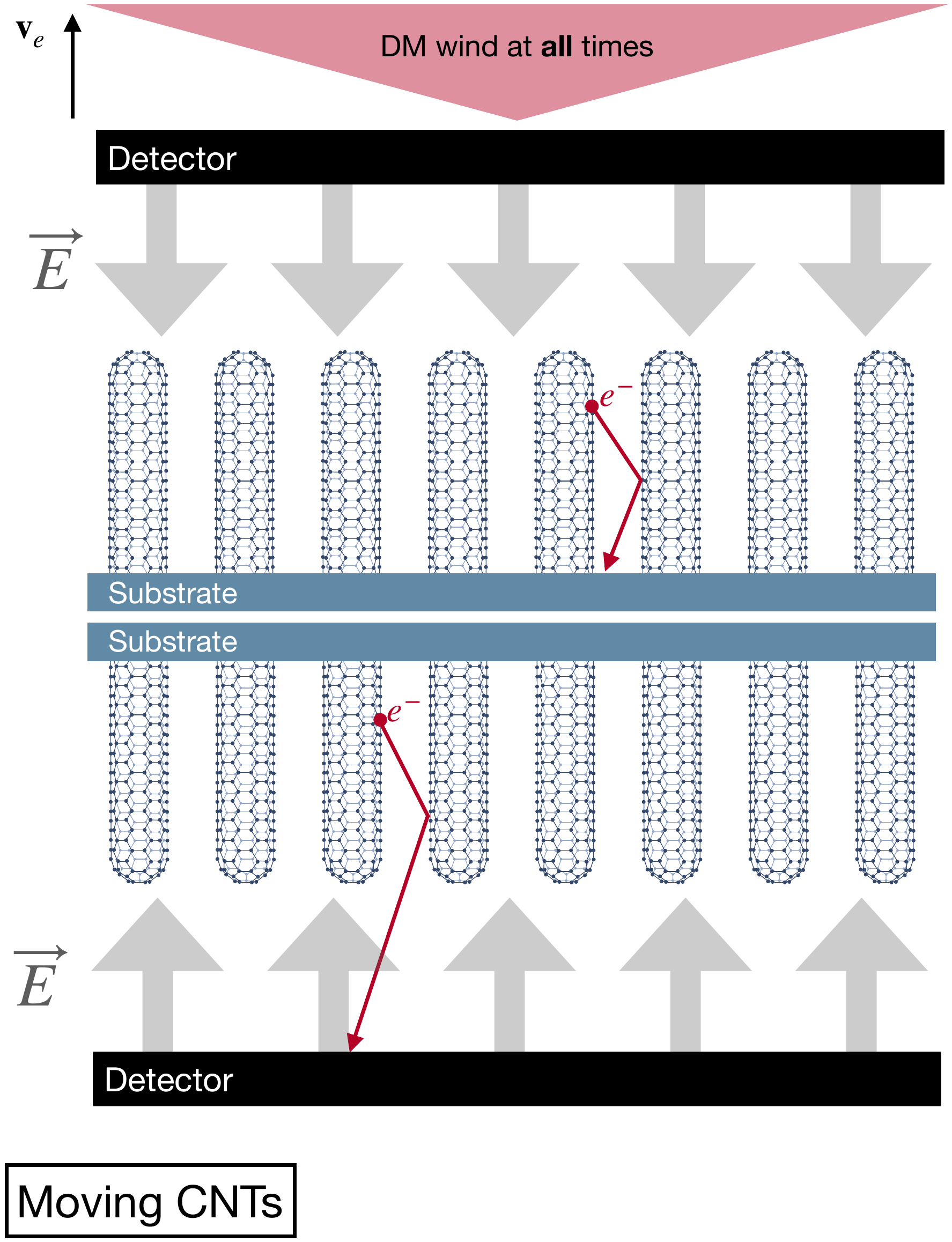}
\caption{Schematics of experimental setups with carbon nanotubes (CNTs) as targets. The CNTs are grown on a substrate (blue). The electric field in the vacuum surrounding the sheet is shown with gray arrows. Electrons ejected into the field are indicated by the red arrow. \textbf{Left:} Setup in which the CNTs are fixed in the lab. \textbf{Right:} Setup in which the CNTs are tracking $\mathbf{v}_e$.}
\label{fig: CNTs setups}
\end{figure*}

In the first setup, we consider an anisotropic detector fixed in the lab frame, so that the angle between the detector's symmetry axis, e.g.~the $z$-axis $\hat{\mathbf{z}}$, and the direction of $\mathbf{v}_e$ varies during the day, producing a daily modulation in the DM induced signal. Both graphene and carbon nanotubes can be used to realise this setup, as shown in the left panels of Fig.~\ref{fig: graphene setups} and Fig.~\ref{fig: CNTs setups} for graphene and carbon nanotubes, respectively.

In a second setup, we consider a pair of identical anisotropic detectors mounted with opposite orientations onto a platform which tracks the DM wind. For this setup, a smoking gun signal for DM discovery would be a statistically significant difference in the number of events in the two detectors.~The right panel of Fig.~\ref{fig: graphene setups} shows a realisation of this setup based on graphene sheets, while the right panel in Fig.~\ref{fig: CNTs setups} illustrates a realisation of the same setup based on carbon nanotubes.~The latter will be employed by PTOLEMY-CNT~\cite{Cavoto:2019flp}.

As far as the detection of the ejected electrons is concerned, we assume that this can be achieved by drifting the ejected electrons in an external electric field.~When electrons are ejected from graphene sheets (solid gray stripe in Fig.~\ref{fig: graphene setups}), we assume that all electrons ejected into the electric field will drift to the detector, whereas the electrons ejected into the substrate (light blue band in Fig.~\ref{fig: graphene setups}) will be absorbed by the substrate and not detected. When electrons are ejected from CNTs (vertical honeycombed structures attached to the light blue substrate in Fig.~\ref{fig: CNTs setups}), they need to propagate through the CNT array in order to be detected. In spherical coordinates, with the CNTs (graphene sheets) aligned with (perpendicular to) the z-axis and electrons leaving the CNTs (graphene sheets) in the negative z-direction are detected whereas the electrons leaving the CNTs (graphene sheets) in the positive z-direction are not, the rate of detected electrons, $\mathscr{R}$, can finally be written as follows
\begin{equation}
\label{eq: angular rate}
    \mathscr{R}    = \int \mathrm{d}\,k^\prime \mathrm{d}\,\theta^\prime \mathrm{d}\,\phi^\prime \sin\theta^\prime \frac{\ddd R}{\mathrm{d}k^\prime\mathrm{d}\cos\theta^\prime\,\mathrm{d}\phi^\prime } \,f\left(k^\prime, \theta^\prime, \phi^\prime \right)\,,
\end{equation}
where
\begin{widetext}
\begin{align}
\label{eq: differential rate}
    \frac{\ddd R}{\mathrm{d}k^\prime\mathrm{d}\cos\theta^\prime\,\mathrm{d}\phi^\prime }(\mathbf{k}^\prime)=\frac{n_\chi \left(k^\prime\right)^2 N_\text{cell}}{32\pi^2m_\chi^2m_e^2}\int \mathrm{d}\, E_e\int\dd^3 \mathbf{q}\int\dd^3 \mathbf{v}\,f_\chi(\mathbf{v})\delta\left(\Delta E_e +\frac{q^2}{2m_\chi}- \mathbf{v}\cdot\mathbf{q}\right)
  R_\mathrm{free}(\mathbf{k}^\prime,\mathbf{q},\mathbf{v})
 \;W(\mathbf{k}^\prime-\mathbf{q},E_e)\, ,
\end{align}
\end{widetext}
is the triple differential electron ejection rate, and $\left\{k^\prime,\theta^\prime,\phi^\prime\right\}$ are the spherical coordinates of the final state electron momentum.
$f\left(k^\prime, \theta^\prime, \phi^\prime \right)$ is the probability of an electron ejected with momentum $k^\prime$ and propagating in direction $\{\theta^\prime, \phi^\prime\}$ being detected. Obtaining an accurate $f\left(k^\prime, \theta^\prime, \phi^\prime \right)$ is beyond the scope of this paper, and for our analysis we assume 
\begin{equation}
\label{eq: angular cutoff}
    f\left(k^\prime, \theta^\prime, \phi^\prime \right)=\Theta\left(  \theta^\prime-\theta^* \right)
\end{equation}
where $\theta^*$ is the cutoff angle below which an electron is not detected. 
In this work we use $\theta^*=100^\circ$ for CNTs to obtain an asymmetry in the number of events reported by the two detectors and a differential rate comparable to that found in Ref.~\cite{Cavoto:2017otc}. For graphene-based experiments, we use $\theta^*=90^\circ$.

\section{Sensitivity studies: statistical framework}
\label{sec:statistics}
In this section we introduce the statistical framework we use to assess the expected performance of the two experimental setups described in the previous section. For each setup, we are interested in two possible experimental outcomes. In the first, the experiment reports the detection of DM particles with a statistical significance corresponding to three standard deviations, i.e.~``$3\sigma$''. In the second, we assume that the number of observed events is too small to report a DM particle discovery, and should therefore be interpreted as a ``null result''. We explore the implications of the two experimental outcomes by using different test statistics, as we briefly explain below.

Let us start by focusing on the scenario in which DM particles have actually been detected. The experimental setups discussed above would establish such a discovery through the observation of an asymmetry in the number of events reported in different detector components, or in different periods of the day. 
For the fixed target setup, the detection of DM particles would imply an asymmetry in the number of events recorded in the 12 hours of the day around the expected times of maximum and minimum rate~\footnote{In practice this requires recording events in 24 one-hour wide bins, sorting them and taking the 12 bins with the highest number of events to contribute to $n_+$ and the 12 bins with the lowest number of events to contribute to $n_-$.}. 
For the moving twin experiment setup, the discovery of DM would imply an asymmetry in the number of events observed in the two detector components. 
Let us now denote by $n_{+}$ and $n_{-}$, $n_{+}\ge n_{-}$, the number of counts recorded in the two ``regions of interests'' i.e. in the time periods/detector components introduced above, and by $n_{+}- n_{-}$ the associated asymmetry. Furthermore, let $E[n_{\pm}]=\mu s_{\pm}+\theta_{\pm}$ be the expected value of $n_{\pm}$, where $\theta_{\pm}$ ($\mu s_{\pm}$) is the expected number of background (DM signal) events in the given region of interest, while $\mu$ is the ``strength parameter'', i.e.~$c_i^2$ for the operators in Tab.~\ref{tab:operators} and $g/\Lambda$ or $g/\Lambda^2$ in the case of dipole and anapole interactions, respectively. For all interactions, we calculate $\mu s_{\pm}$ by integrating the rate formula we obtained in the previous sections. With this notation, we can write the probability to record $n_{\pm}$ counts given the expectation $E[n_{\pm}]$, i.e.~given $(\mu,\theta_{\pm})$, as follows\\
\begin{align}
\mathscr{P}_{\pm}(\mu,\theta_{\pm}) &=\frac{ e^{-(\mu s_{\pm}+\theta_{\pm})} }{n_{\pm}!}
\, (\mu s_{\pm}+\theta_{\pm})^{n_{\pm}} \nonumber\\
&\rightarrow \frac{1}{\sqrt{2 \pi \sigma^2_{\pm}}} \exp\left[-\frac{1}{2} \left( \frac{\mu s_{\pm}+\theta_{\pm}-n_{\pm}}{\sigma_{\pm}} \right)^2\right]
\,,     
\label{eq:gauss}
\end{align}
where in the second line $E[n_{\pm}]\gg 1$ and $\sigma_{\pm}^2=E[n_{\pm}]$. To eliminate $n_{\pm}$ from the equations, we now introduce a specific value for the strength parameter, $\mu'$, implicitly defined via
\begin{align}
n_{+}-n_{-} \equiv \mu' (s_{+}-s_{-})\,. 
\label{eq:mup}
\end{align}
In an actual experiment, $\mu'$ would be the unknown value of $\mu$ which underlies the data, whereas in a Monte Carlo simulation, $\mu'$ is the benchmark value of $\mu$ from which data are sampled. Since the probability density function of the difference of the two normal random variables $n_+$ and $n_-$ is a normal random variable with expectation value $E[n_+]-E[n_-]$ and variance $\sigma^2=\sigma_{+}^2+\sigma_{-}^2$, the probability of observing an asymmetry $n_{+}-n_{-}$ given $\mu$ and $\theta\equiv\theta_{+}+\theta_{-}$, $\theta_{+}=\theta_{-}$, is proportional to the likelihood function
\begin{align}
\mathscr{L}(\mu,\theta) = \exp\left[-\frac{1}{2} \left( \mu - \mu'\right)^2\left( \frac{s_{+}-s_{-}}{\sigma}\right)^2\right] \,.
\label{eq:L}
\end{align}
We use Eq.~(\ref{eq:L}) to introduce the test statistics $q_0$ defined as
\begin{align}
q_0 &= \left\{
\begin{array}{ll}
- 2\ln \frac{\mathscr{L}(0,\theta)}{\mathscr{L}(\mu',\theta)} = \frac{(s_+ - s_{-})^2 \mu^{\prime 2}}{\theta} & \textrm{for~$\mu'\ge0$} \\
0& \textrm{for~$\mu'< 0$}
\end{array}\right.\,.
\label{eq:q0}
\end{align}
Here, we neglect statistical fluctuations in the number of background events, and set $\theta$ to its expectation value. 
For $\mu'=0$, $q_0$ obeys a ``half chi-square" distribution with one degree of freedom, whereas for $\mu'\neq0$ it follows a non-central chi-square distribution~\cite{Cowan:2010js}. Consequently, we can use $q_0$ to express the statistical significance for DM particle discovery\footnote{Significance, $Z$, and $p$-value are related by $Z\equiv \Phi^{-1}(1-p)$, where $\Phi$ is the cumulative distribution function of a normal distribution with mean 0 and variance 1~\cite{Cowan:2010js} (i.e.~the standard normal distribution).} as $Z=\sqrt{q_0}$~\cite{Cowan:2010js}. Notice that $\mathscr{L}(0,\theta)<\mathscr{L}(\mu',\theta)$ when $\mu'\neq0$ underlies the data. Furthermore, the larger $\mu'$, the larger $Z$ and, therefore, the better one can reject the null hypothesis, i.e.~$\mu=0$ in favour of the alternative hypothesis, $\mu=\mu'$. By obtaining the $\mu'$ for which $\sqrt{q_0}=3$, we find the smallest strength parameter, or coupling constant value, that can be measured with a statistical significance corresponding to ``$3\sigma$''.

As a second experimental outcome, we consider the one in which the observed asymmetry $n_{+} - n_{-}$ is too small to report a DM particle discovery. If an asymmetry in the total number of recorded events, $n_{-}+n_{+}$, cannot be established, the experimental data can still be used to exclude values of the coupling constants $c_i^2$, $g/\Lambda$ or $g/\Lambda^2$ that would imply $E[n_{-}+n_{+}]> n_{-}+n_{+}$. Conservatively, here we assume that all recorded events are due to DM, i.e. $E[n_{-}+n_{+}]=\mu(s_{-}+s_{+})$, and do not perform any background subtraction. We then compute 90\% confidence level (C.L.) exclusion limits on $\mu$, or, equivalently, on $c_i^2$, $g/\Lambda$ or $g/\Lambda^2$, by imposing that the probability of observing $n_{-}+n_{+}$ or fewer events when our expectation is $E[n_{-}+n_{+}]=\mu(s_{-}+s_{+})$ is 10\%. We therefore solve 
\begin{align}
    &e^{-\mu(s_{-}+s_{+})}\sum_{i=0}^{n_{-}+n_{+}} \frac{\mu(s_{-}+s_{+})^i}{i!} = 0.1
    \label{eq: cdf}
\end{align}
for $\mu$ for each of the four experimental setups introduced above. In this analysis, we assume two reference data samples:~1) $n_{-}+n_{+}=0$, and 2) $n_{-}+n_{+} = \mu'(s_{-}+s_{+})+\theta$, where $\mu'$ and $\theta$ solve the equation $\sqrt{q_0}=3$, as explained above. The first sample implies the strongest exclusion limits one can expect, whereas the second one corresponds to a scenario where DM is at the threshold of discovery.

\section{Sensitivity studies: expected ejection rates, exclusion limits and discovery potential} \label{sec: sensitivity studies exclusion limits}
In this section, we numerically evaluate the ejection rates from graphene sheets and carbon nanotubes for various models of DM and determine the sensitivity of selected detector setups within these models.

\subsection{Ejection rate for graphene sheets}
\begin{figure}
    \centering
    \includegraphics[width=0.485\textwidth]{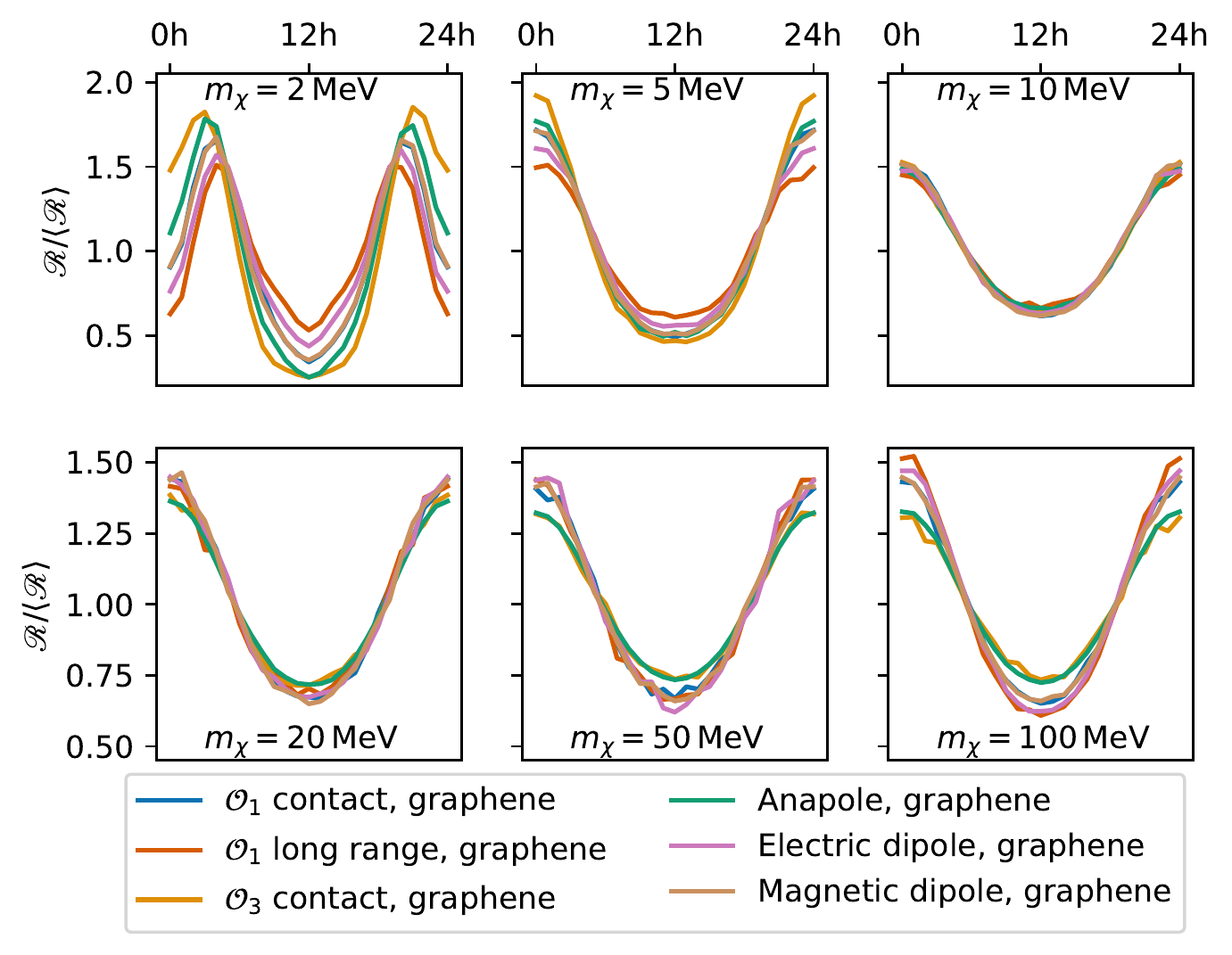}
    \caption{Daily modulation pattern of detectable electrons in the fixed graphene sheets setup (Fig.~\ref{fig: graphene setups}, left panel) for $\mathcal{O}_1$ contact interaction (blue), $\mathcal{O}_1$ long range interaction (red), $\mathcal{O}_3$ contact interaction (yellow), the anapole interaction (green), electric dipole interaction (magenta) and the magnetic dipole interaction (brown). The mass of the DM particle increases from $2\,\mathrm{MeV}$ in the top left corner, to $100\,\mathrm{MeV}$ in the bottom right corner. 
    When compared to other detector setups, it is easier to recognize a DM signal (since all DM candidates provide a similar signal) but harder to distinguish different interaction models once a signal is detected.}
 \label{fig: daily modulation sheets}
\end{figure}
\begin{figure*}
    \centering
    \includegraphics[width=0.48\textwidth]{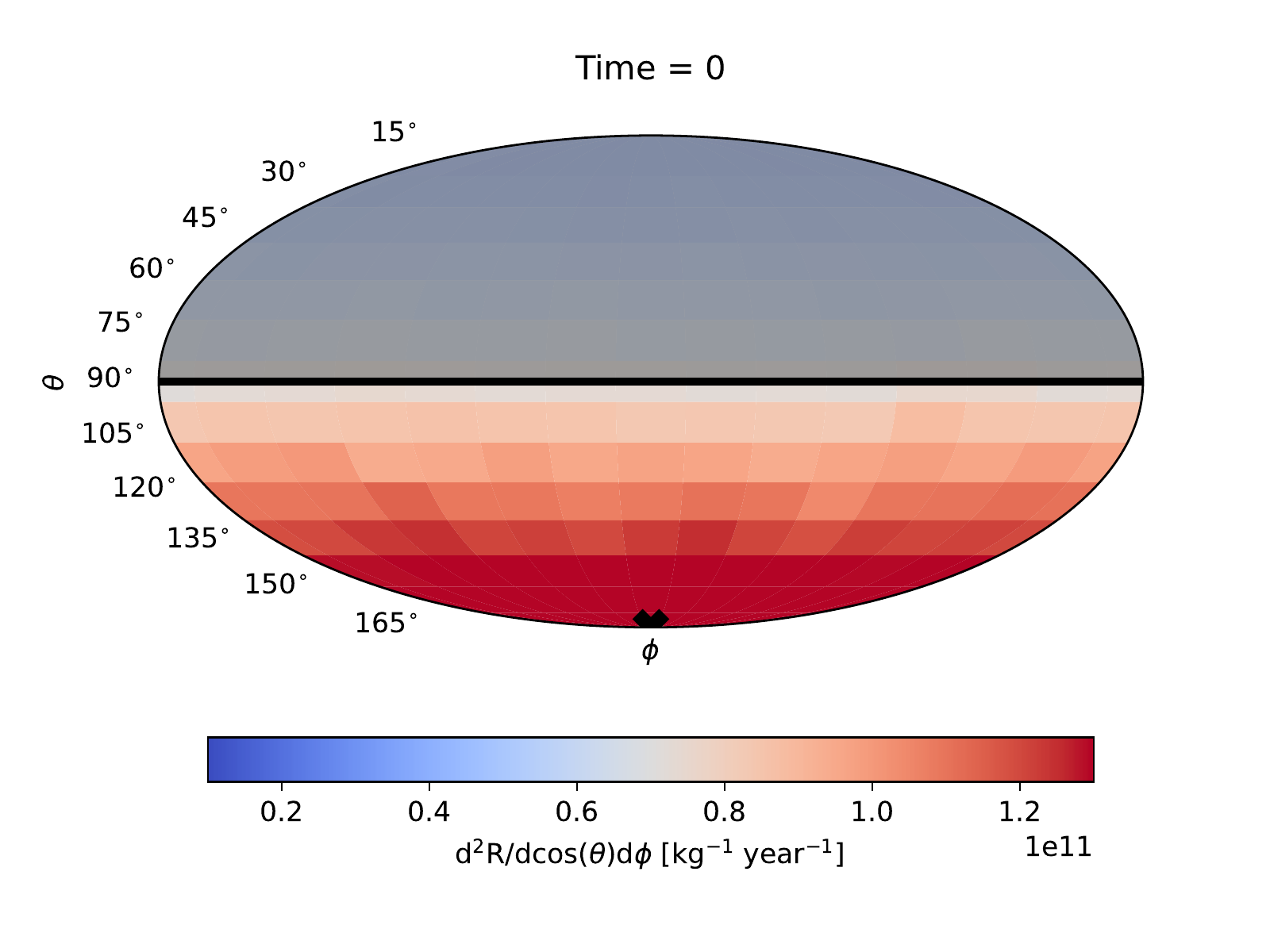}
    \includegraphics[width=0.48\textwidth]{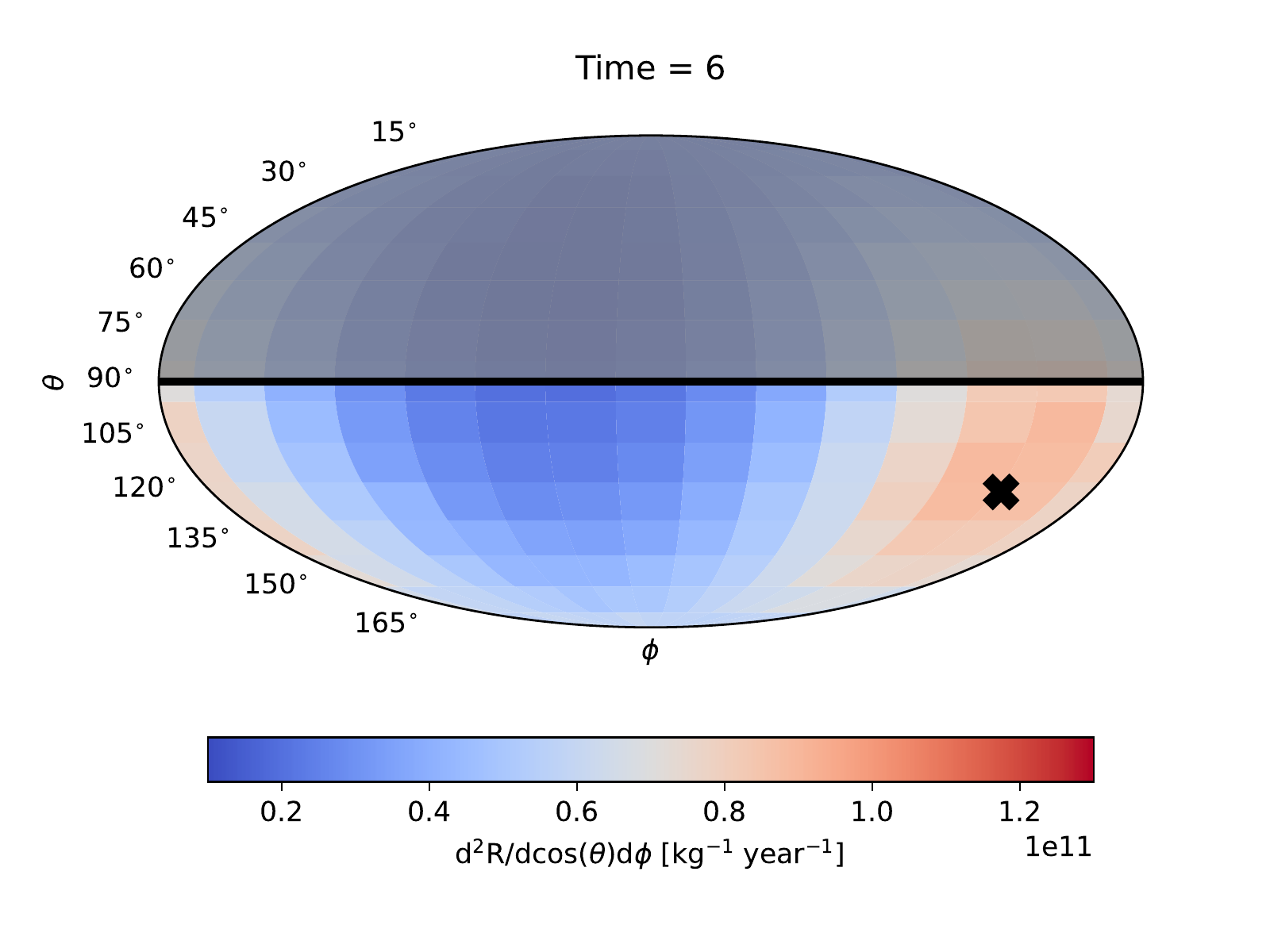}
    \includegraphics[width=0.48\textwidth]{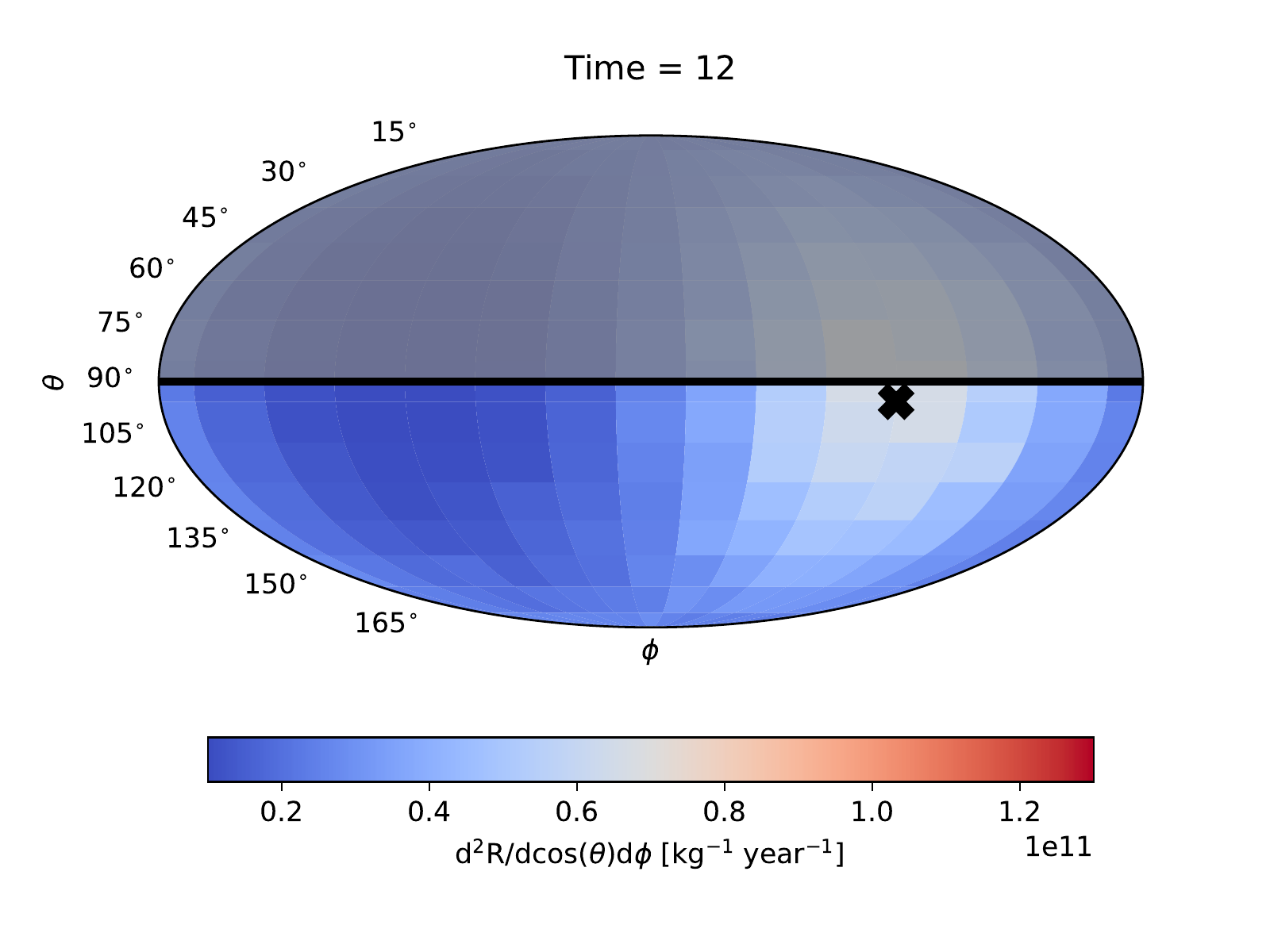}
    \includegraphics[width=0.48\textwidth]{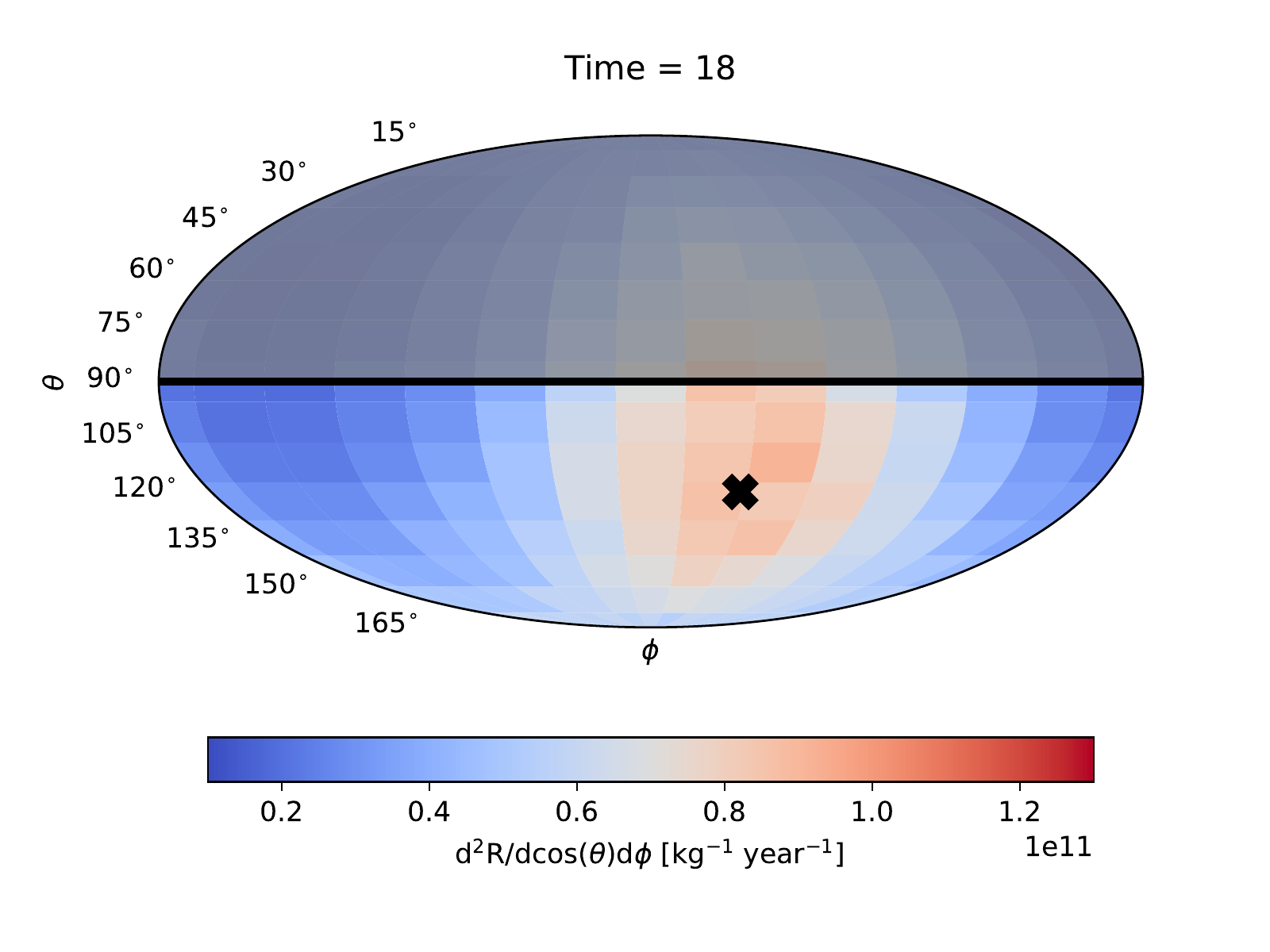}
    \caption{Projection plot for the magnetic dipole interaction for graphene sheets, $m_\chi =$ 5\,MeV, for different times (in hours) throughout the day. At time = 0\,h, the DM wind (denoted as a black cross) is perpendicular to the graphene sheet, which is shown as a solid black line at $\theta =$ 90$^\circ$. The shaded region denotes the substrate where ejected electrons will not be detected. This plot represents a family of solutions for which the operator does not contain the term $\frac{ \mathbf{q}}{m_e}\times \mathbf{v}^{\perp}_{\rm el}\big|_{\boldell=0} = \frac{ \mathbf{q}}{m_e}\times \mathbf{v}$, and therefore $\mathbf{q}$ in the direction of $\mathbf{v}$ is not suppressed. The DM wind direction and ejected electron momentum are therefore similar.}
    \label{fig: projections magnetic dipole}
\end{figure*}

\begin{figure*}
    \centering
    \includegraphics[width=0.48\textwidth]{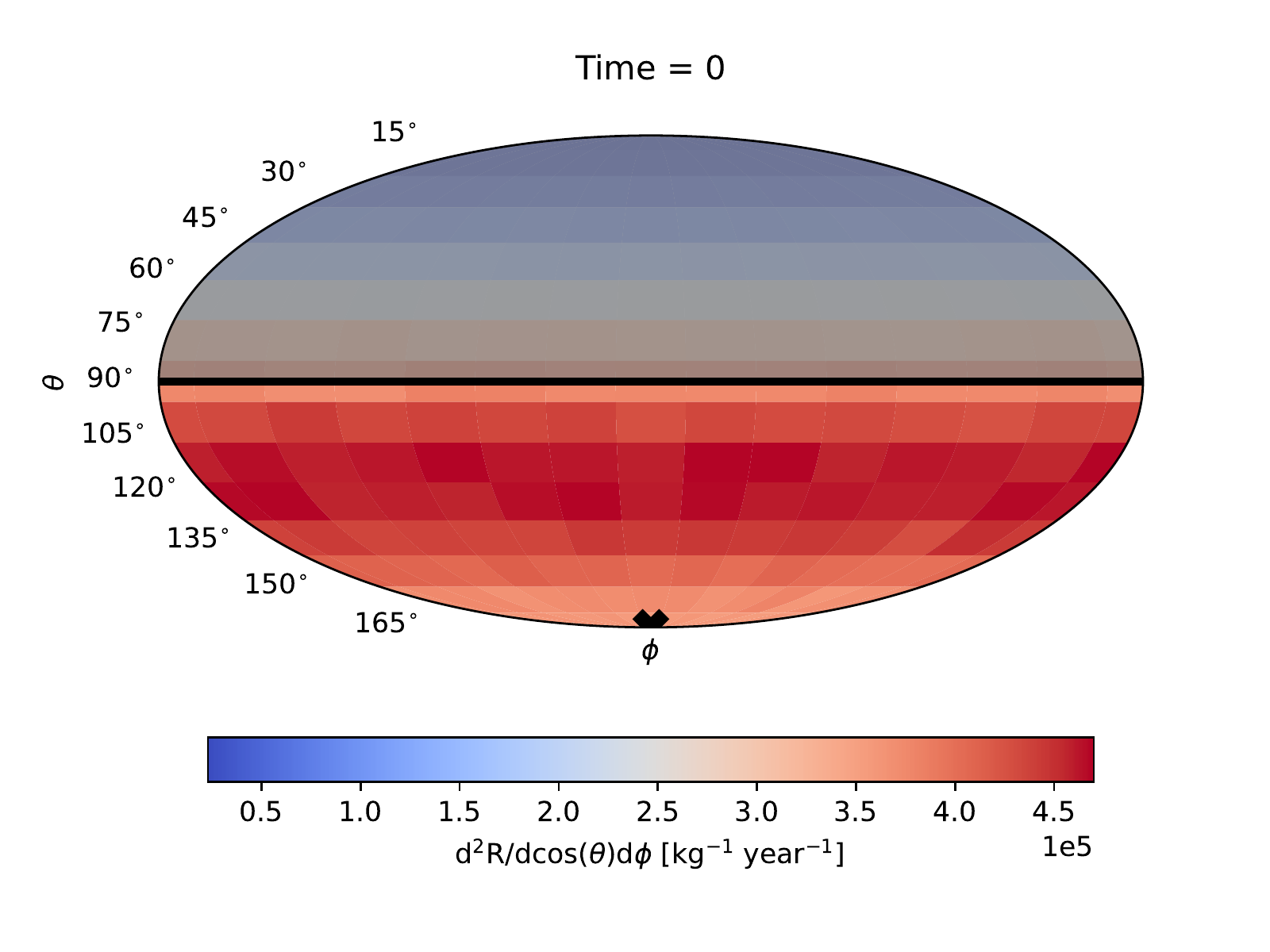}
    \includegraphics[width=0.48\textwidth]{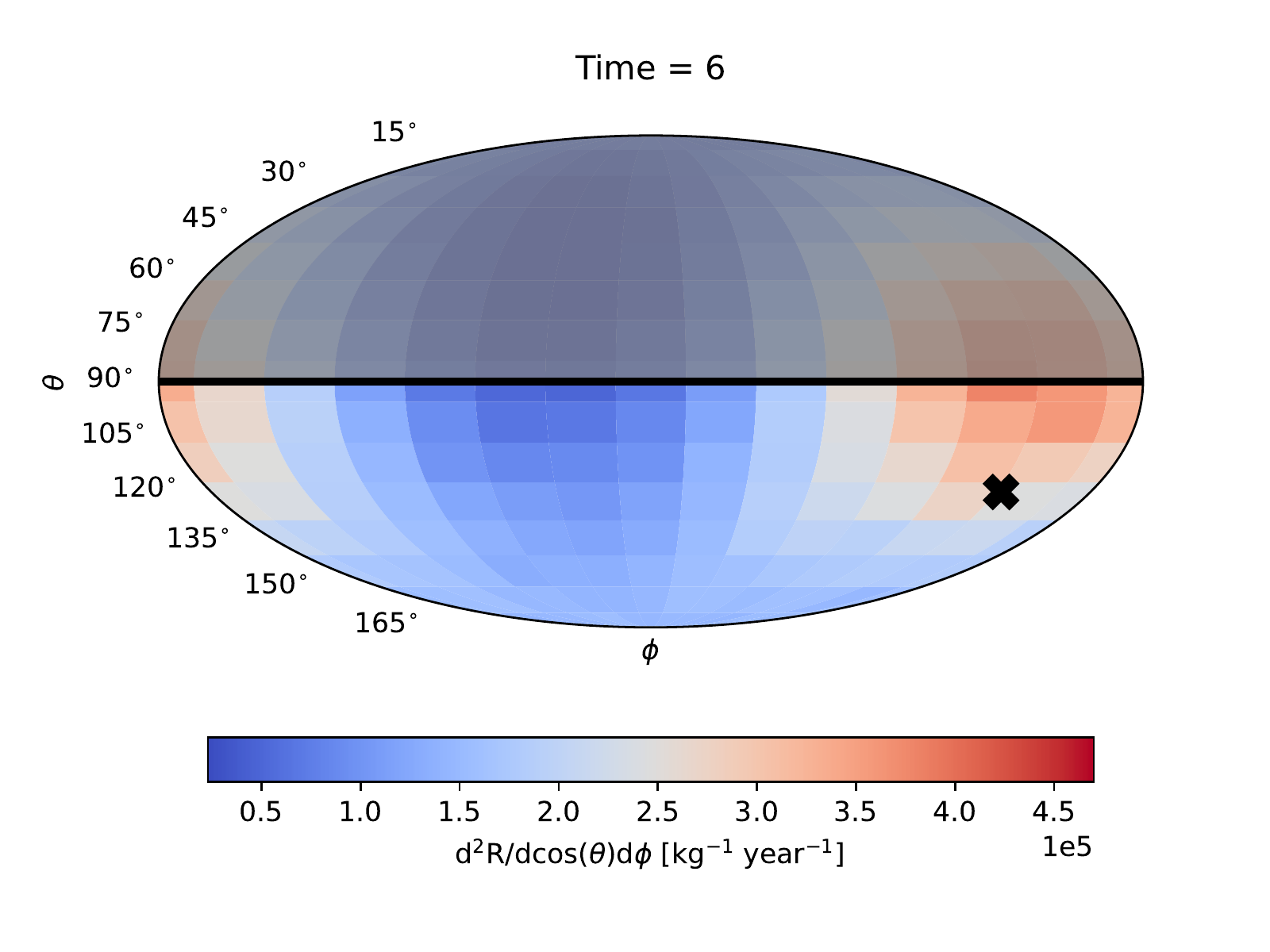}
    \includegraphics[width=0.48\textwidth]{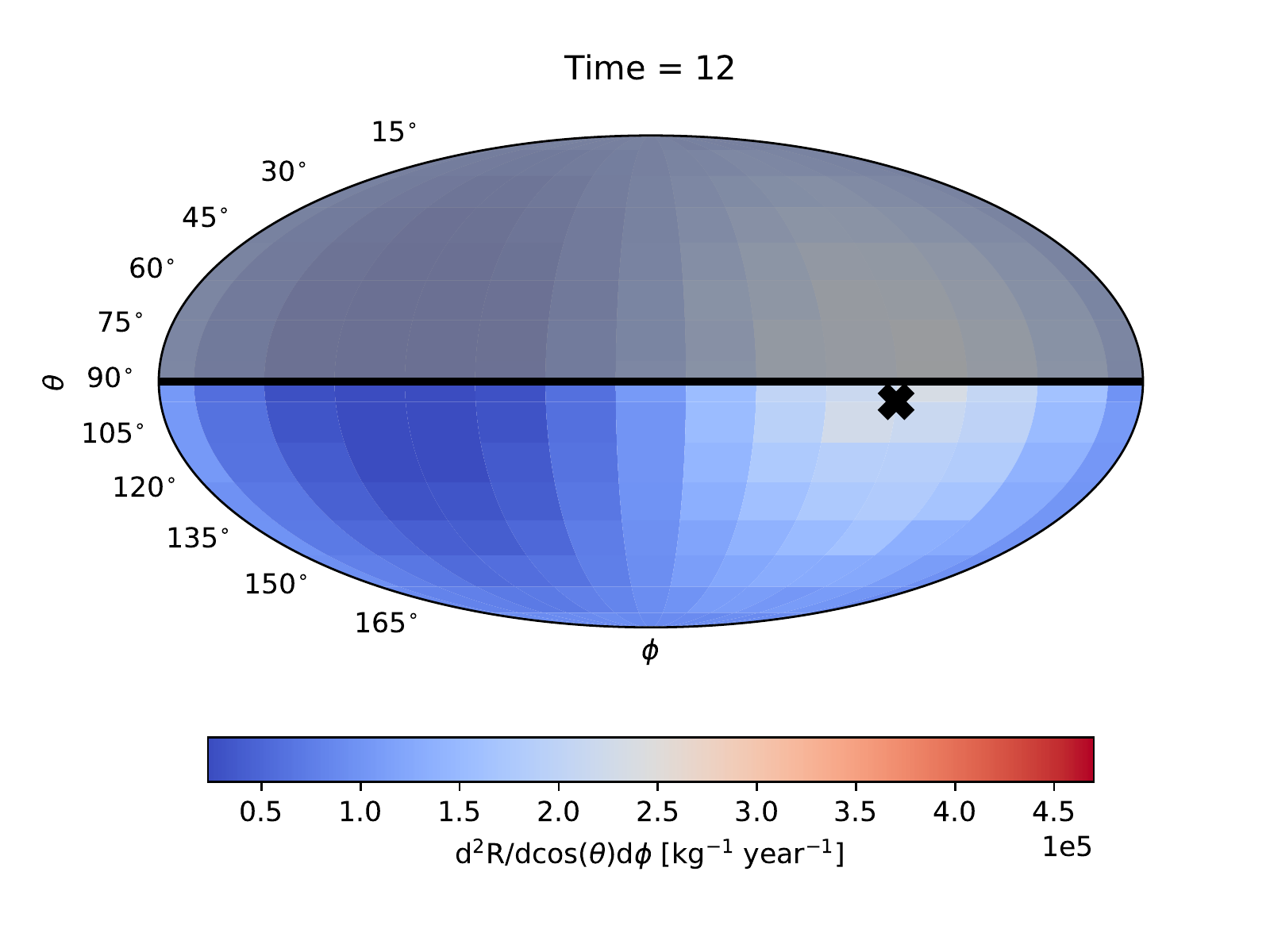}
    \includegraphics[width=0.48\textwidth]{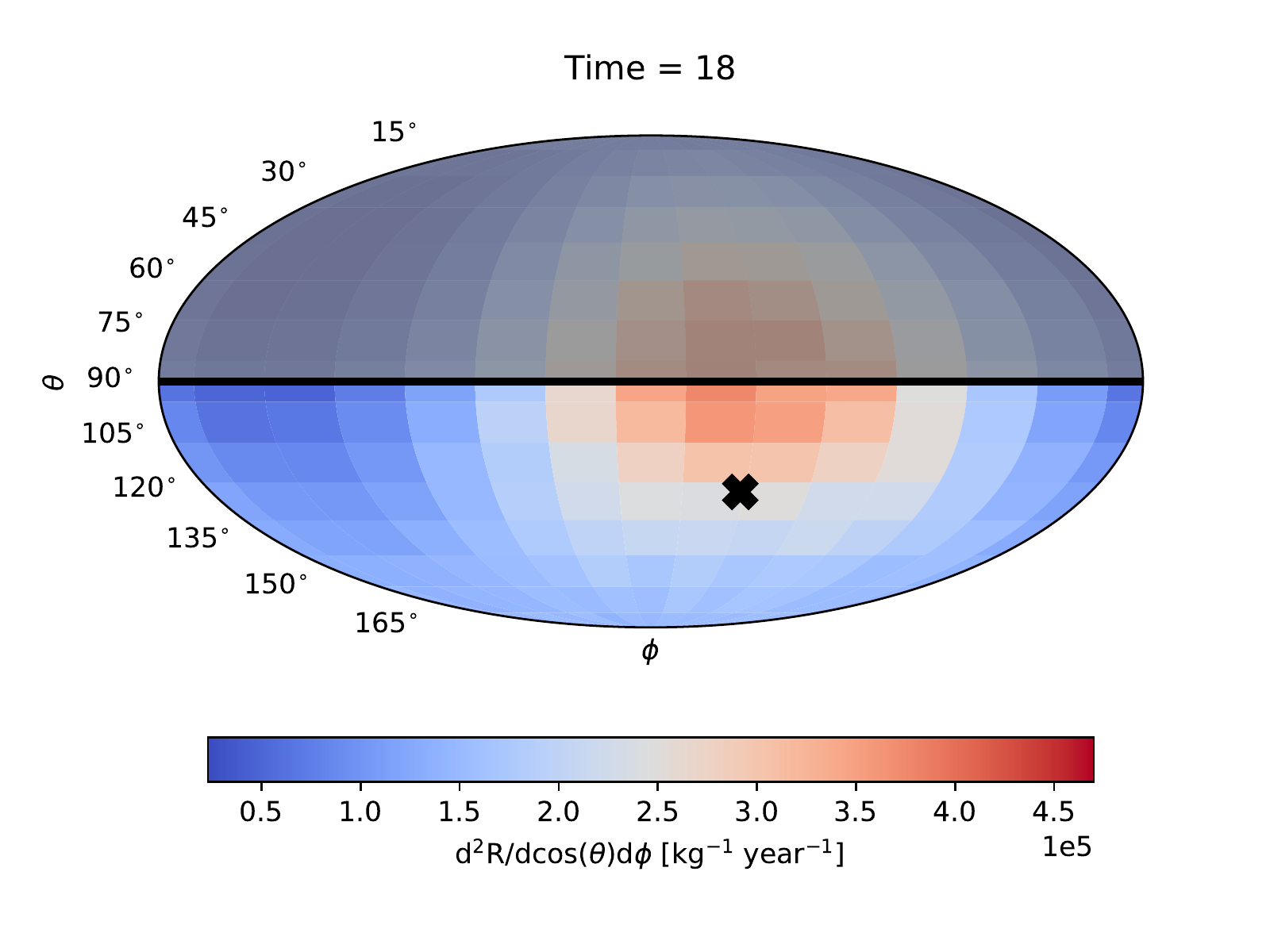}
    \caption{Projection plot for $\mathcal{O}_3$ contact interaction for graphene sheets, $m_\chi =$ 5\,MeV, for different times throughout the day. At time = 0\,h, DM wind (denoted as a black cross) is perpendicular to the graphene sheet, which is shown as a solid black line at $\theta =$ 90$^\circ$. The shaded region denotes the substrate where ejected electrons will not be detected. This plot represents a family of solutions for which the operator contains the term $\frac{ \mathbf{q}}{m_e}\times \mathbf{v}^{\perp}_{\rm el}\big|_{\boldell=0} = \frac{ \mathbf{q}}{m_e}\times \mathbf{v}$, suppressing $\mathbf{q}$ in the direction of $\mathbf{v}$ and offsetting the direction of the ejected electron momentum from that of the DM wind.}
    \label{fig: projections c3s}
\end{figure*}

For the case of a static graphene sheet fixed in the lab (Fig.~\ref{fig: graphene setups} (left)), the rate of ejected electrons varies throughout the day as the orientation of the DM wind changes. As we showed already in Paper I, this change depends on the DM mass and on the form of the DM-electron interaction; we reproduce the results of Paper I in Fig.~\ref{fig: daily modulation sheets}. The displayed interactions are $\mathcal{O}_1$ (contact and long range interaction), $\mathcal{O}_3$ (contact interaction) as well as the anapole, electric dipole, and magnetic dipole interactions. DM mass ranges from $2\,\mathrm{MeV}$ to $100\,\mathrm{MeV}$. We see that the modulation curves are qualitatively similar for the considered interactions and for most DM masses with a maximum at time~=~$0$h and a minimum at time~=~$12$h. However, the low DM mass of $2\,\mathrm{MeV}$ we see that the peaks are displaced from time~=~$0$h to around time~=~$3$h and that the precise location of the peak depends somewhat on the interaction type. As discussed in the associated Paper I, the displacement of the peak at $2\,\mathrm{MeV}$ is due to the momentum distribution of the most loosely bound electrons in the graphene sheet. 

Figs.~\ref{fig: projections magnetic dipole} and \ref{fig: projections c3s} show a two-dimensional projection of the rate integrated over all final-state electron energies as a function of time for the magnetic dipole and the $\mathcal{O}_3$ interactions. From these plots, we can see a preferential direction of the ejected electrons as well as the overall rate modulation for various angular orientations along with the direction of the DM wind (denoted as a black marker).

As can be seen from Eq.~(\ref{eq: differential rate}), the final differential rate depends on four independent contributing factors, namely the distribution of the DM velocities, kinematic constraints imposed by energy conservation, the properties of the material entering through the material response function $W$, and the form of the interaction entering through the free particle response function $R_\mathrm{free}$.

The DM velocity distribution in Eq.~(\ref{eq:df}) is the truncated Maxwell-Boltzmann distribution typically assumed within the Standard Halo Model (SHM)~\cite{Baxter:2021pqo}. It peaks at $\mathbf{v}=-\mathbf{v}_e$ and is exponentially suppressed for velocities of $\mathbf{v}$ away from $-\mathbf{v}_e$. The associated speed distribution is found by integrating Eq.~(\ref{eq:df}) over angles, and multiplying by $v^2=|\mathbf{v}|^2$.~The minimum speed a DM particle must have in order to transfer a momentum $q$ and deposit an energy $\Delta E_e$ is
\begin{align}
v_{\rm min} = \frac{\Delta E_e}{q} + \frac{q}{2 m_\chi} \,, 
\label{eq:vmin}
\end{align}
as one can see by setting to zero the argument of the Dirac $\delta$ function in Eq.~(\ref{eq:rate_general}). The same $\delta$ function also implies that the integrand in the electron ejection rate peaks at $v=v_{\rm min}$, and hence for $\vec{v}$ parallel to $\mathbf{q}$, as long as $v_{\rm min}$ is larger than the most probable speed for all allowed $q $ and $\Delta E_e$.~This is the case for $m_\chi$ below a few MeV.

As can be seen from Fig.~\ref{fig: W fall off}, the properties of the electronic wavefunction in the material favor states where the modulus of the  initial state electron momentum, $|\boldsymbol{\ell}|=|\mathbf{k}'-\mathbf{q}|$, is around $4$~keV.~Let us now denote by $\zeta\sim 4$~keV the ``typical'' value of $|\boldsymbol{\ell}|$ preferred by the electronic properties of graphene. For $m_\chi> \zeta/ v_{\rm max}$, we find that $\zeta$ is smaller than the ``typical'' momentum transferred in a DM-electron interaction\footnote{By imposing $v_{\rm min}<v_{\rm max}$, one finds that $q_{\rm min}<q<q_{\rm max}$, where
\begin{align}
q_{\rm min}&=m_\chi v_{\rm max} - \sqrt{m_\chi^2 v_{\rm max}^2 - 2m_\chi(\Phi - E_e) } \,,\nonumber \\
q_{\rm max}&=m_\chi v_{\rm max} + \sqrt{m_\chi^2 v_{\rm max}^2 - 2m_\chi(\Phi - E_e) }\,. \nonumber
\end{align}}, $q_{\rm typ}\equiv m_\chi v_{\rm max}$, where $v_{\rm max}$ is the maximum possible speed of a DM particle gravitationally bound to the Milky Way.~For $\zeta < q_{\rm typ}$, we also find that the angle between $\mathbf{k}'$ and $\mathbf{q}$ is expected to be smaller than $\pi/2$.\footnote{The vectors $\boldsymbol{\ell}$, $\mathbf{q}$ and $\mathbf{k}'$ must form a triangle.~It is instructive to draw this triangle placing the vertex at the intersection of $\boldsymbol{\ell}$ and $\mathbf{q}$ at the centre of a circle of radius $|\boldsymbol{\ell}|$.~Doing so, we also set $|\boldsymbol{\ell}|$ and $|\mathbf{q}|$ to their representative values, $\zeta$ and $q_{\rm typ}$, respectively. For $\zeta<q_{\rm typ}$, the angle between $\mathbf{k}'$ and $\mathbf{q}$ (with $|\mathbf{q}|=q_{\rm typ}$)  is maximum when $\mathbf{k}'$ is tangent to the above circle of radius $|\boldsymbol{\ell}|=\zeta$.~This occurs when the angle between the vectors $\mathbf{k}'$ and $\boldsymbol{\ell}$ is equal to $\pi/2$, and therefore the angle between $\mathbf{q}$ and $\mathbf{k}'$ is less than $\pi/2$.} This in turn implies that the angle between $\mathbf{v}$ and $\mathbf{k}'$ is smaller than $\pi/2$ when the initial velocity $\mathbf{v}$ is aligned with $\mathbf{q}$, i.e.~for DM masses below a few MeV.~As a result, we predict a forward-backward asymmetry in the rate of DM-induced electron ejections, and therefore a directional sensitivity to the incoming DM wind for graphene detectors.~An illustration of this is provided in Fig.~\ref{fig: projections magnetic dipole}.

Of course, this feature holds only if its effect is not countered by the fourth term in Eq.~(\ref{eq: differential rate}), the DM response function $R_\mathrm{free}$. Almost all the effective operators shown in Tab.~\ref{tab:operators} do not suppress the ejection of electrons along the direction of the DM wind with the exception of the operators $\mathcal{O}_3$ and $\mathcal{O}_5$. These contain the term $\frac{ \mathbf{q}}{m_e}\times \mathbf{v}^{\perp}_{\rm el}\big|_{\boldell=0}=\frac{ \mathbf{q}}{m_e}\times \mathbf{v}$, which disfavors a parallel orientation of $\mathbf{v}$ and $\mathbf{q}$ and, in combination with the previously discussed terms, favors a rate offset from the direction of the DM wind (as can be seen in Fig.~\ref{fig: projections c3s}).

\subsection{Ejection rate for carbon nanotubes}
\begin{figure}
    \centering
    \includegraphics[width=0.45\textwidth]{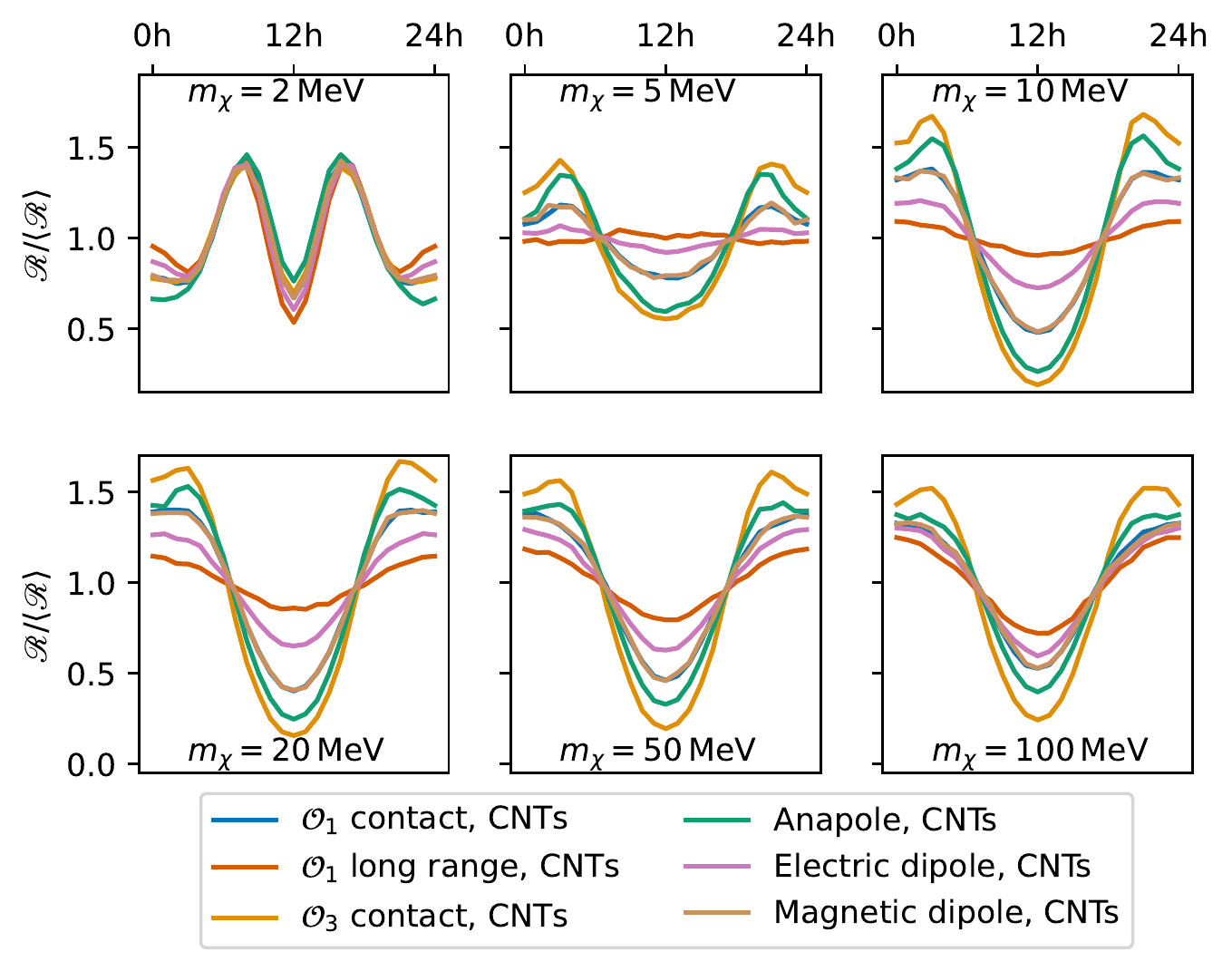}
    \caption{Same as Fig.~\ref{fig: daily modulation sheets}, but for the case of fixed carbon nanotubes. We see that for $m_\chi = 2\,\mathrm{MeV}$ the rate is maximal around time~=~$6$h and time~=~$18$h for all interactions. For larger masses, however, the daily modulation pattern strongly differs between different interaction types. Interestingly, the $\mathcal{O}_1$ long range and electric dipole interactions produce a daily modulation pattern that is largely flat at $m_\chi=5\,\mathrm{MeV}$, making establishing a daily modulation difficult.} 
    \label{fig: daily modulation CNTs}
\end{figure}
\begin{figure*}
    \centering
    \includegraphics[width=0.48\textwidth]{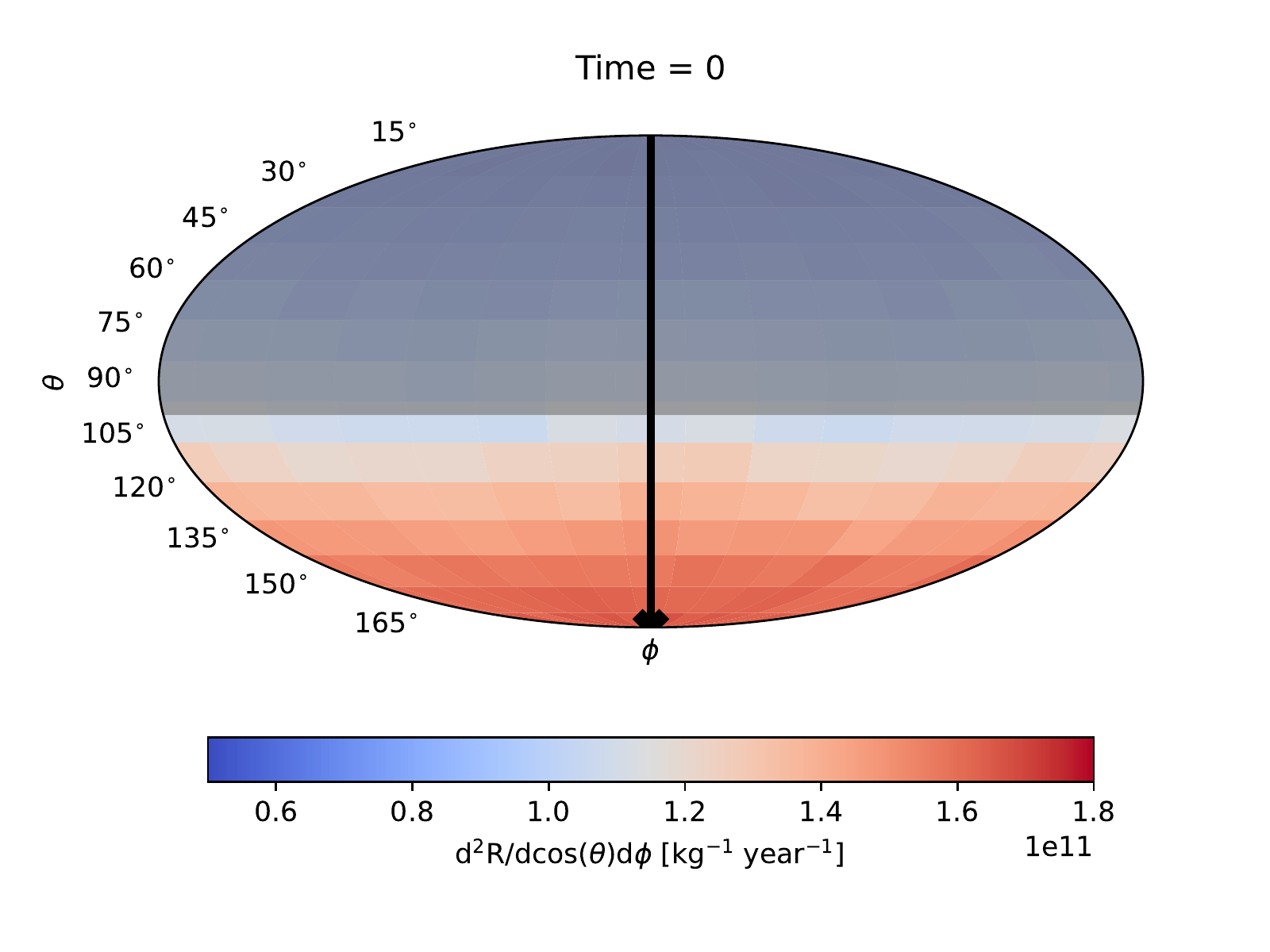}
    \includegraphics[width=0.48\textwidth]{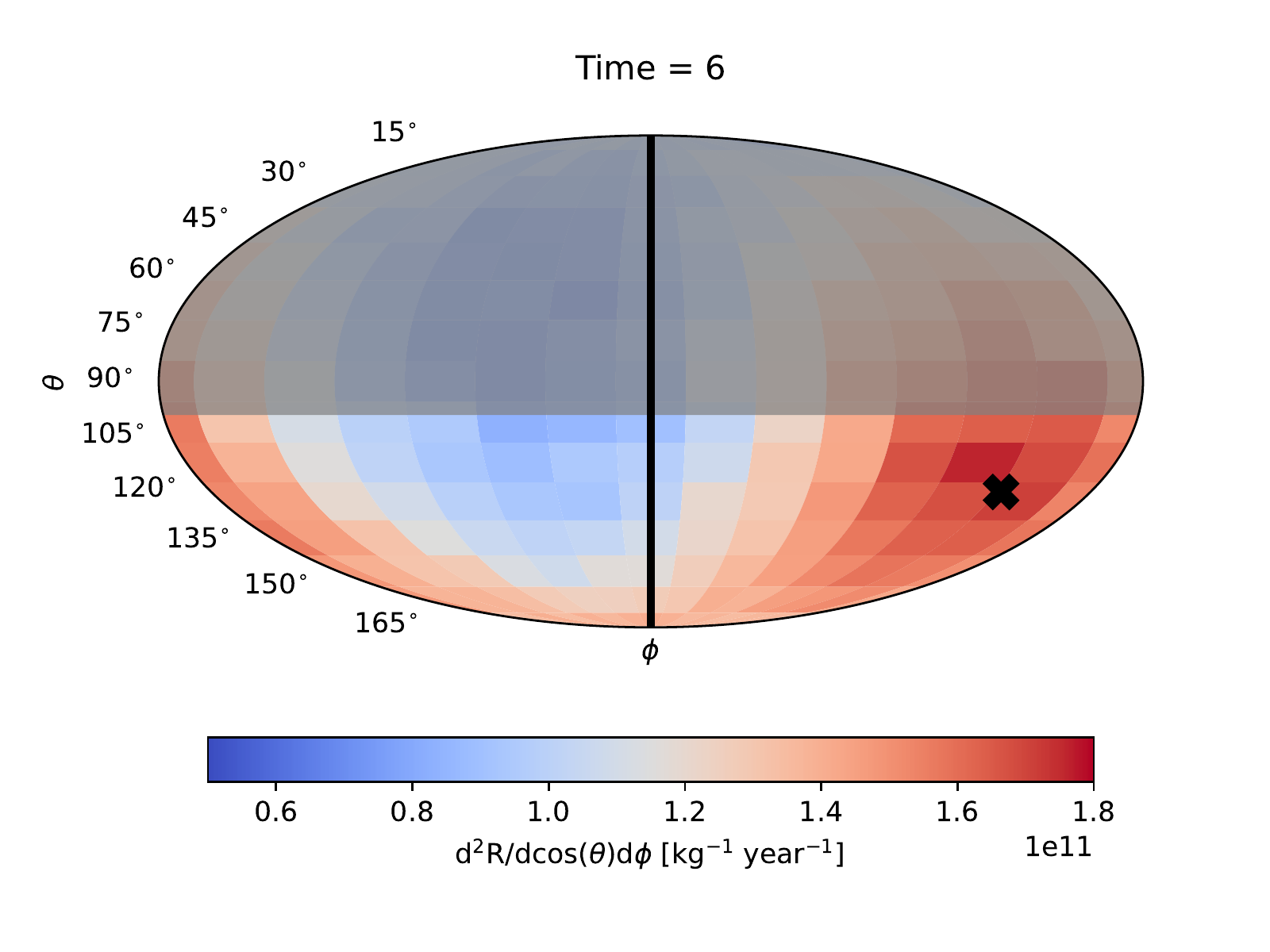}
    \includegraphics[width=0.48\textwidth]{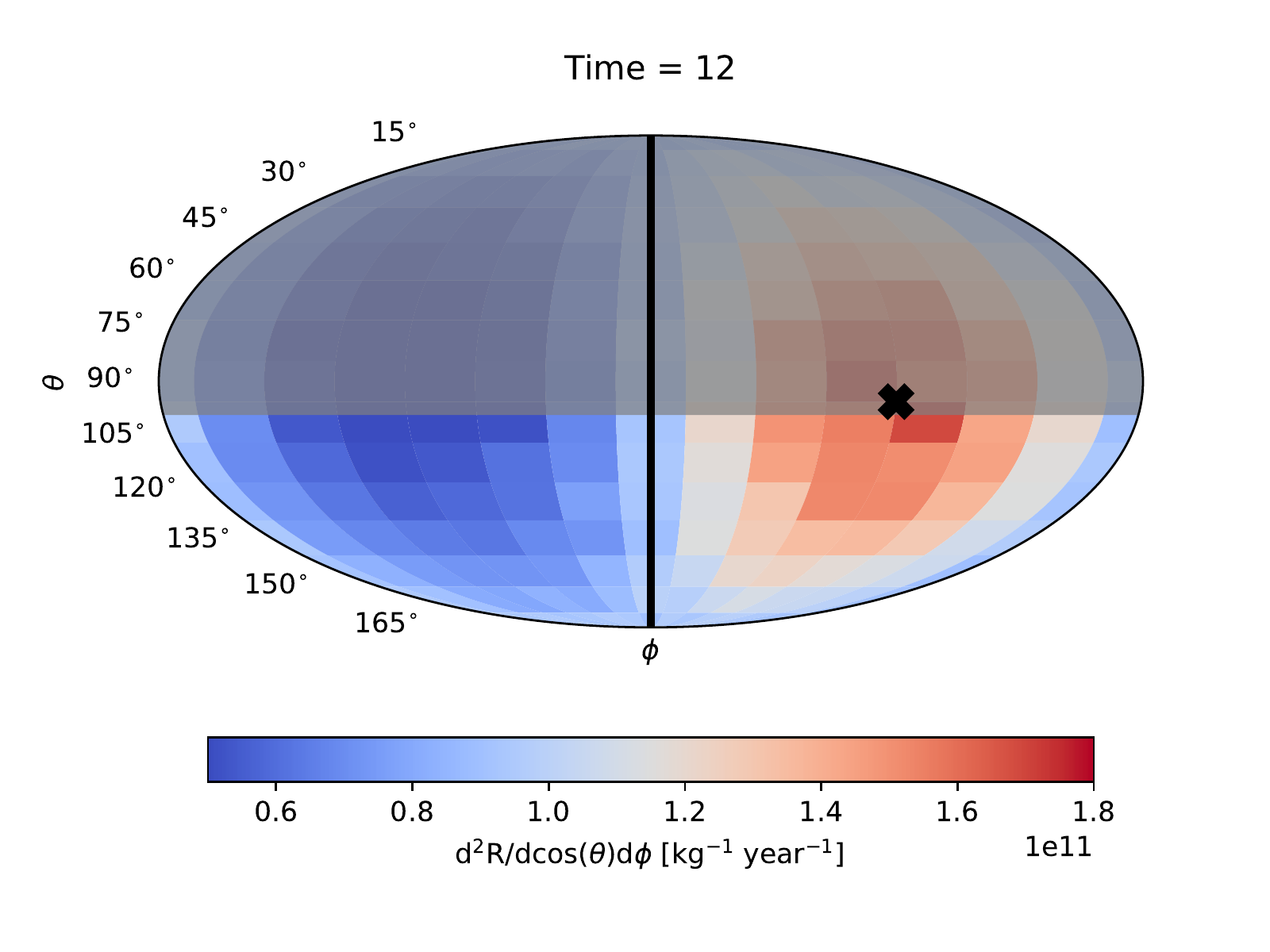}
    \includegraphics[width=0.48\textwidth]{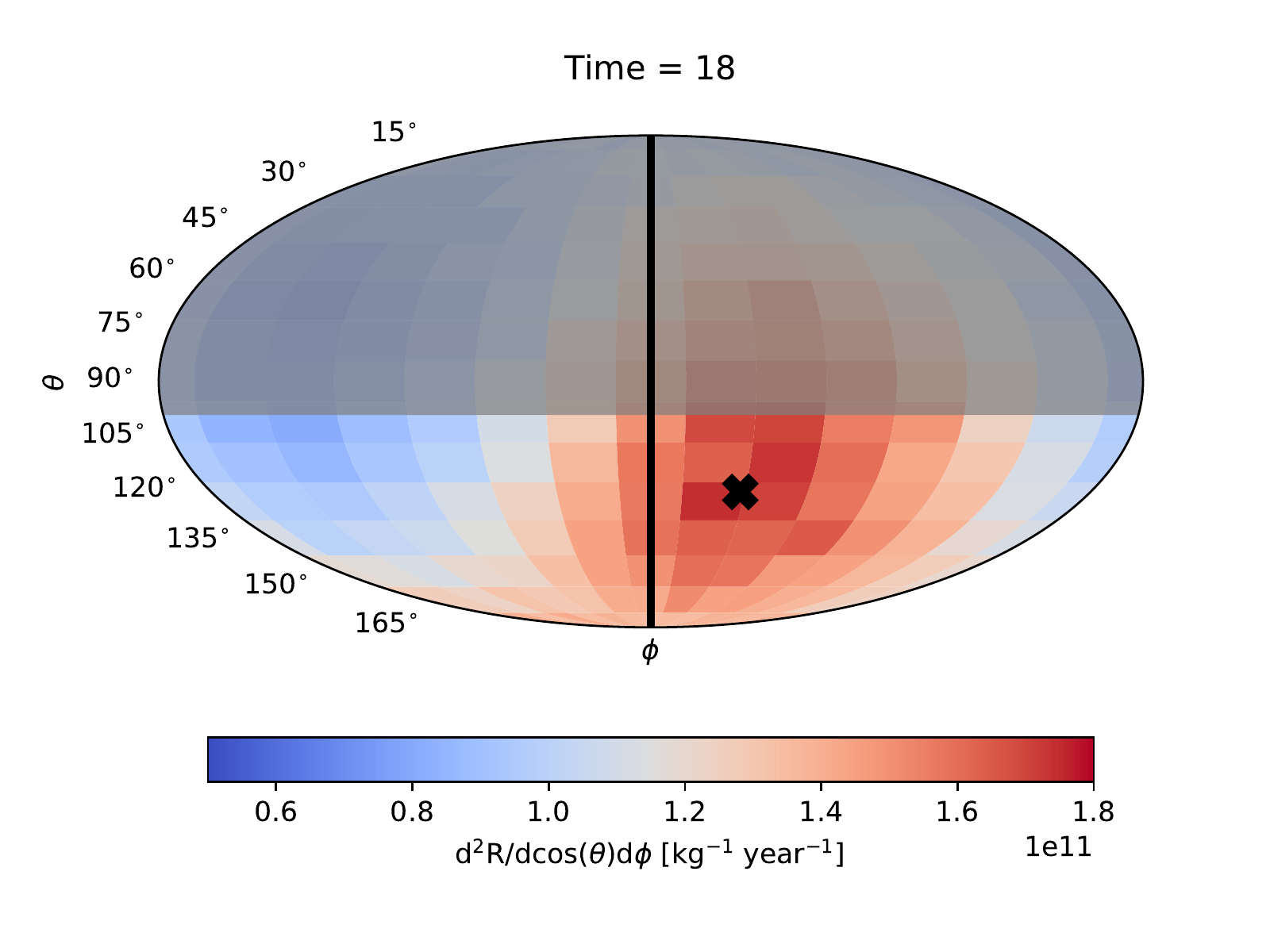}
    \caption{Same as Fig.~\ref{fig: projections magnetic dipole}, but for a detector made of CNTs. The vertical black line corresponds to the orientation of the tube axis.}
    \label{fig: projections magnetic dipole tubes}
\end{figure*}

\begin{figure*}
    \centering
    \includegraphics[width=0.48\textwidth]{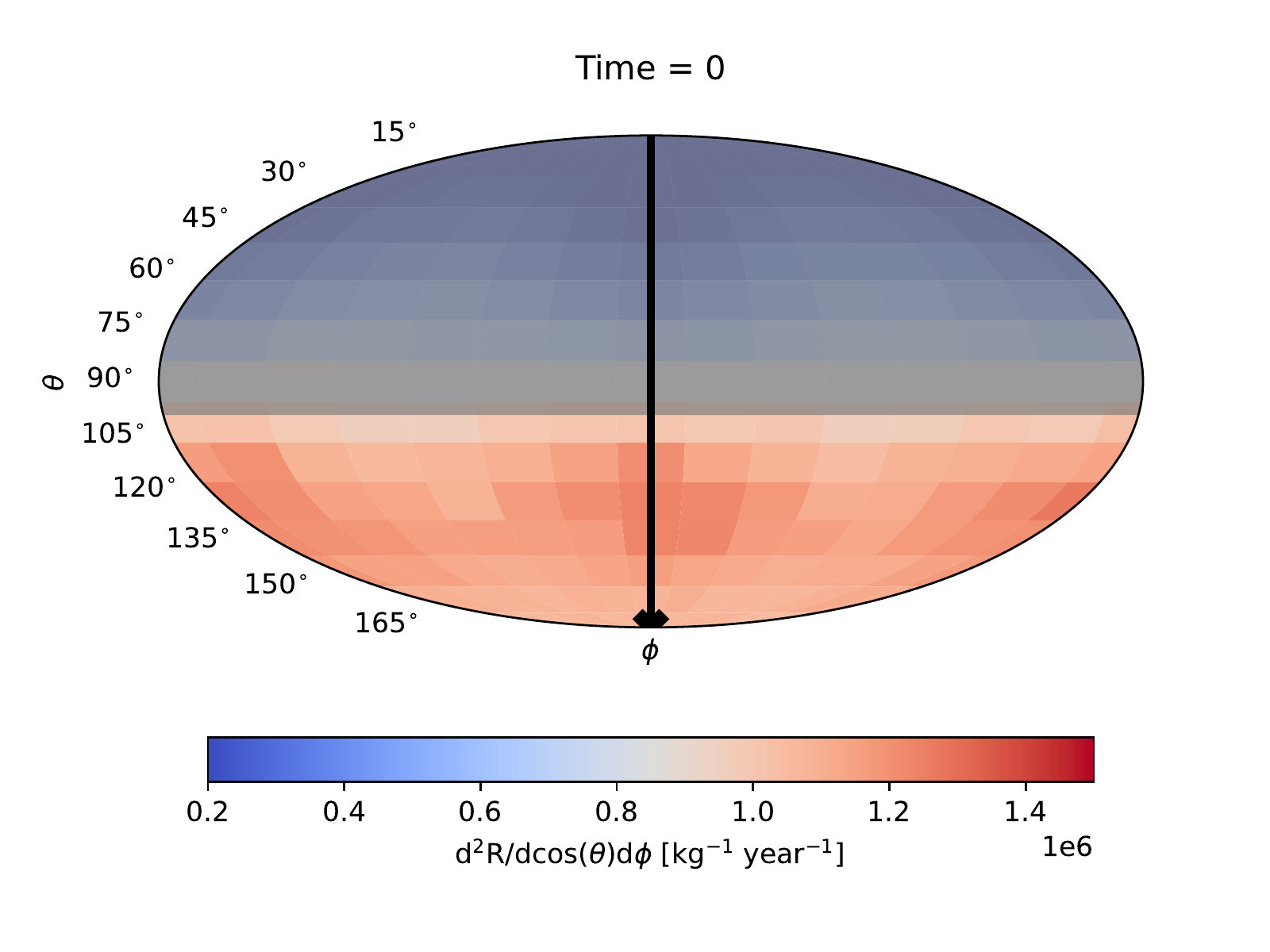}
    \includegraphics[width=0.48\textwidth]{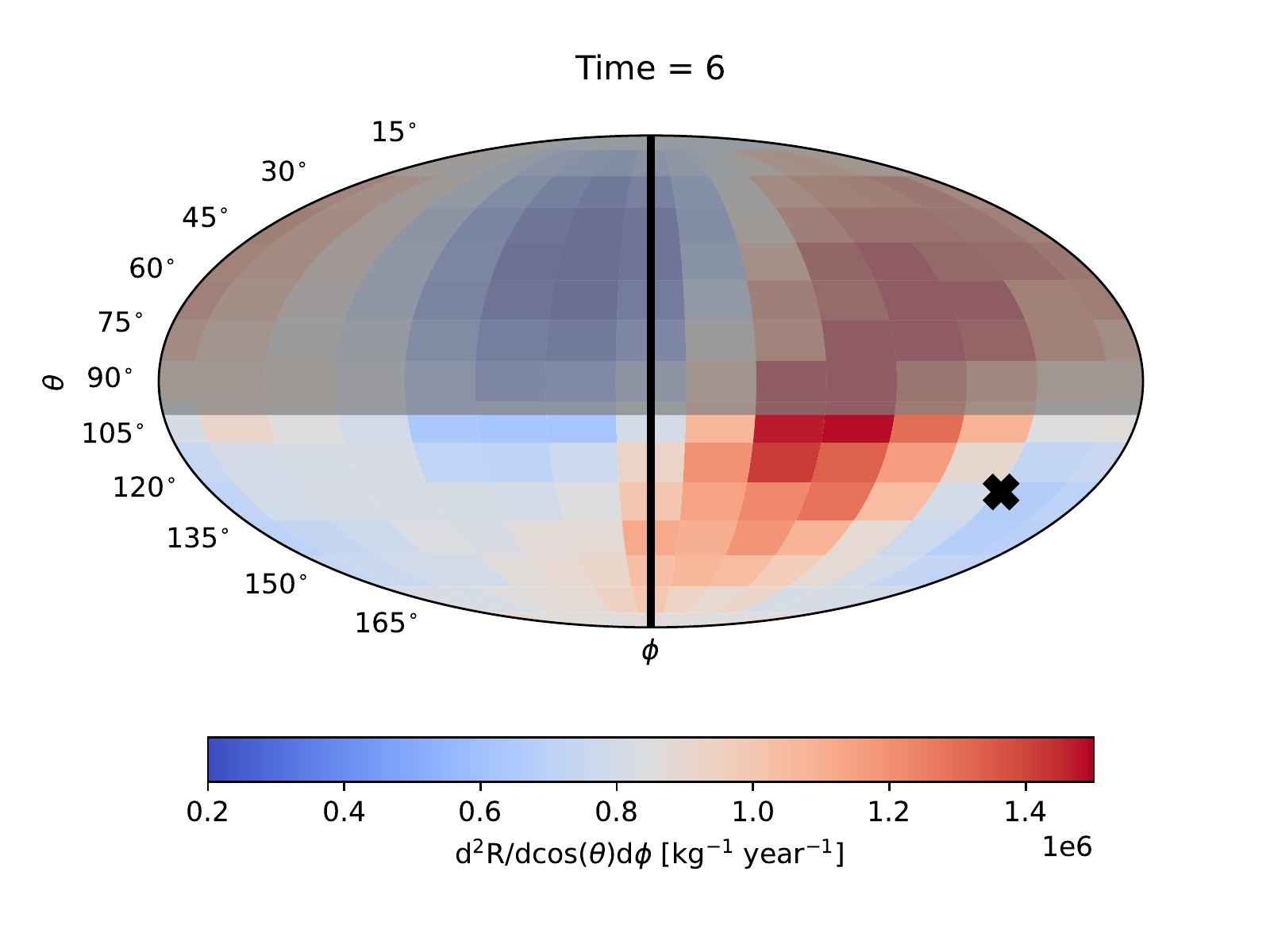}
    \includegraphics[width=0.48\textwidth]{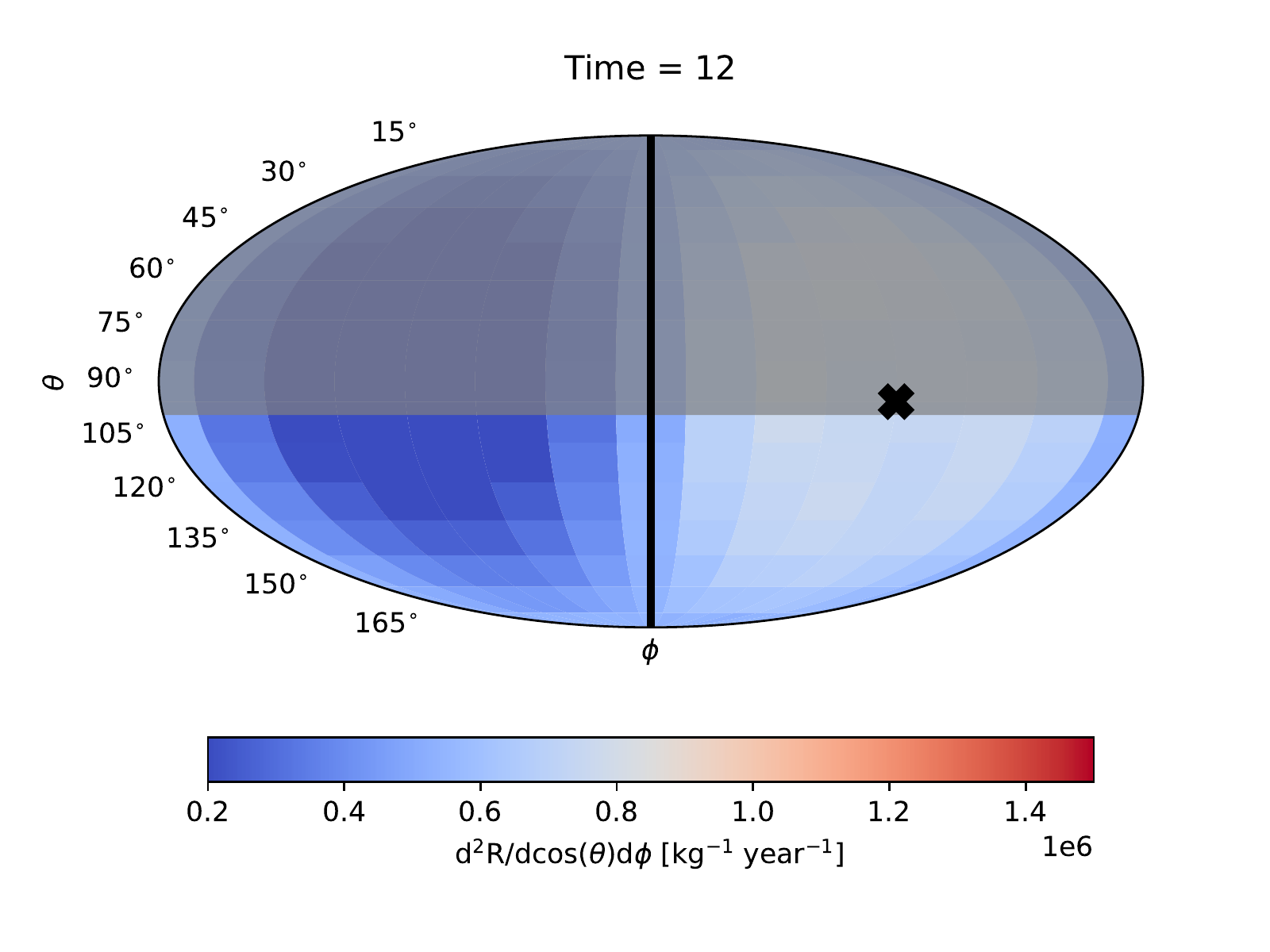}
    \includegraphics[width=0.48\textwidth]{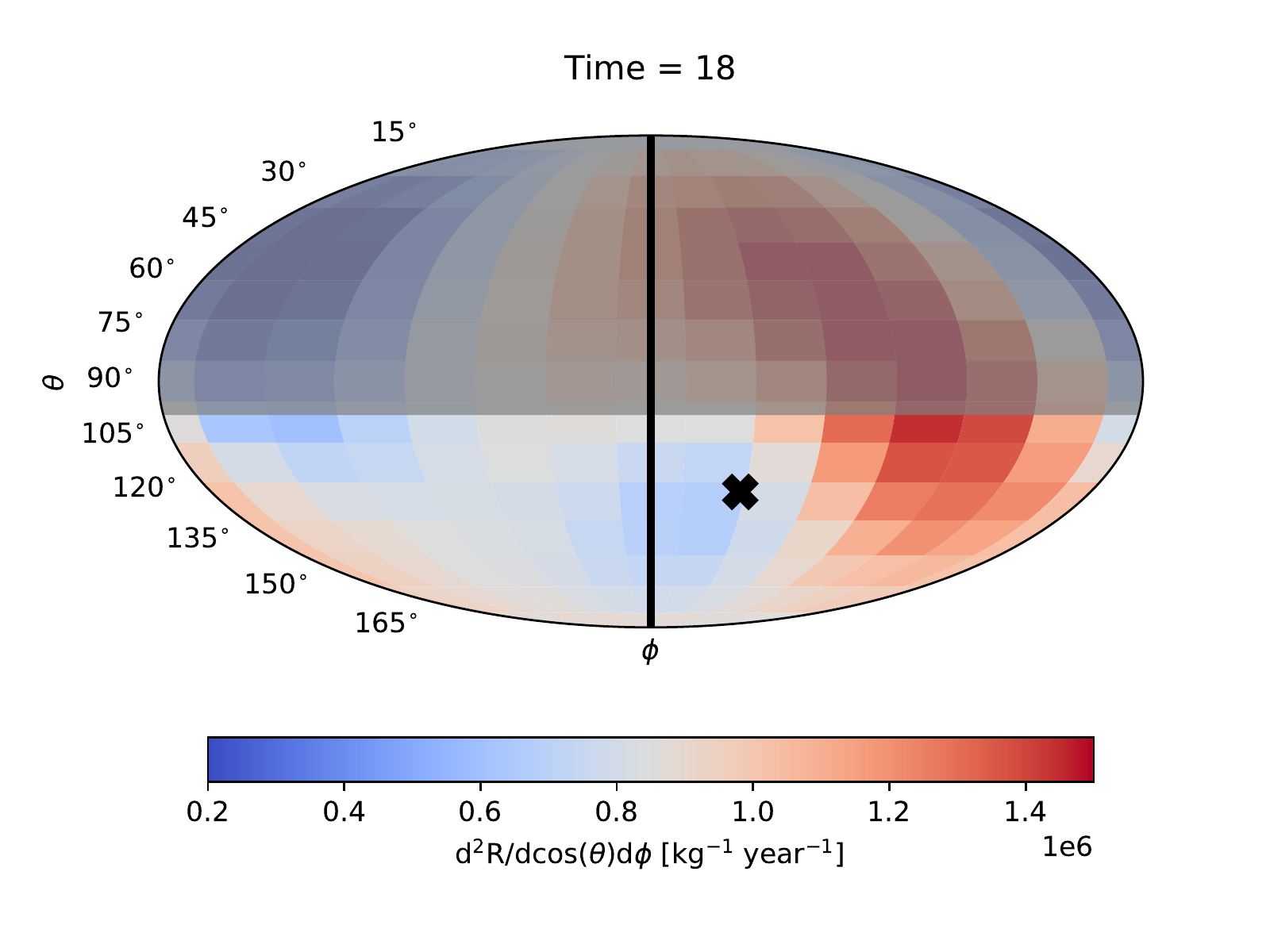}
    \caption{Same as Fig.~\ref{fig: projections c3s}, but for a detector made of CNTs. The vertical black line corresponds to the orientation of the tube axis.}
    \label{fig: projections c3s tubes}
\end{figure*}
In this subsection, we discuss the numerical results obtained for the carbon-nanotube experiments, both in a static setup observing daily modulation as well as in a second, moving setup with two identical detector arms in which one tracks the direction of the DM wind and the second is aimed  in a different direction to establish a rate difference.

The angular dependence of the ejected electrons is more complicated for CNTs than for graphene sheets as in this case, there are two competing effects, each dominating a different interaction energy regime. On one hand, due to the additional crystal momentum within the graphene (CNT) sheet, there are more electron states with large out-of-plane lying momentum than in-plane (as discussed in the previous section). So, generally, interactions in which DM is oriented perpendicularly to the sheet, are preferred over those in which DM arrives parallel to the sheet. The preference is particularly strong for low DM candidate masses, since they have low energy and so, due to energy-momentum conservation, require an electron with a large momentum in order to overcome the work function. Large-mass candidates carry enough kinetic energy on their own and rely less on the initial momentum of the target electron. However, this effect is countered in the case of CNTs by the fact that the acceptance of the detector limits the allowed angles of ejected electrons, which must be oriented close to along the direction of the tube walls to escape. This, together with the fact that long range interactions carry an additional factor of $\sim \frac{1}{q^2}$ suppressing large momentum transfers causes a non-trivial pattern to emerge in the daily modulation plots for various masses and interaction types.

In Fig.~\ref{fig: daily modulation CNTs}, we show the daily modulation plot for the $\mathcal{O}_1$ operator (both contact and long range interactions), $\mathcal{O}_3$ operator (contact interaction), and the anapole, electric and magnetic dipole interactions. For $m_\chi=2\,\mathrm{MeV}$, the modulation curves are qualitatively similar for all considered interactions, with maxima around time~=~$6$h and time~=~$18$h, when the DM wind forms a roughly $35^\circ$ degree angle with the tube axis. For $m_\chi=100\,\mathrm{MeV}$, the modulation curves are also mostly qualitatively similar with minima around time~=~$12$h for all plotted interactions and maxima at time~=~$0$h for all the interactions with the exception of $\mathcal{O}_3$ contact, whose produced rate-maxima are displaced. For masses between 2 and 100 $\mathrm{MeV}$, we see qualitative differences between the interactions. Note in particular that for the $\mathcal{O}_1$ long range and the electric dipole interactions, the curves flatten out around $m_\chi=5\,\mathrm{MeV}$ (due to the balancing of the two competing directional effects discussed in the previous paragraph). As will be discussed in Sec.~\ref{sec: exclusion limits}, this causes an experiment searching for a daily modulation signal to lose sensitivity at this mass, and is seen as peaks in the yellow curves around $m_\chi=5\,\mathrm{MeV}$ in the top right and bottom left panels of Fig.~\ref{fig: sensitivity plot CNTs}.

From the comparison of Fig.~\ref{fig: daily modulation sheets} and Fig.~\ref{fig: daily modulation CNTs}, it would seem that the graphene sheet setup is less sensitive to differentiating between various interaction models. This is however due to the fact that in this approach, we integrate over all outgoing electrons that are ejected under the graphene sheet plane and the effects of various interaction types wash out. The detector acceptance of carbon nanotubes cuts out a narrower window in the direction of the outgoing electrons making different couplings distinguishable. Therefore, if one would impose a similar directional cut on the outgoing electrons from the graphene sheets, one would be able to better distinguish various interaction models as well (as can be seen from Figs.~\ref{fig: projections magnetic dipole} and \ref{fig: projections c3s}).

Figs.~\ref{fig: projections magnetic dipole tubes} and~\ref{fig: projections c3s tubes} show the angular dependence of the total event rate integrated over all electron energies for the magnetic dipole and the $\mathcal{O}_3$ contact interaction respectively. As for the case of graphene, we see that the direction of electron ejections largely tracks the DM wind for the magnetic dipole interaction, whereas it is displaced from the direction of the DM wind for $\mathcal{O}_3$ contact interaction.

\subsection{Exclusion limits and discovery potential}
\label{sec: exclusion limits}
\begin{figure*}
    \centering
    \includegraphics[width=0.48\textwidth]{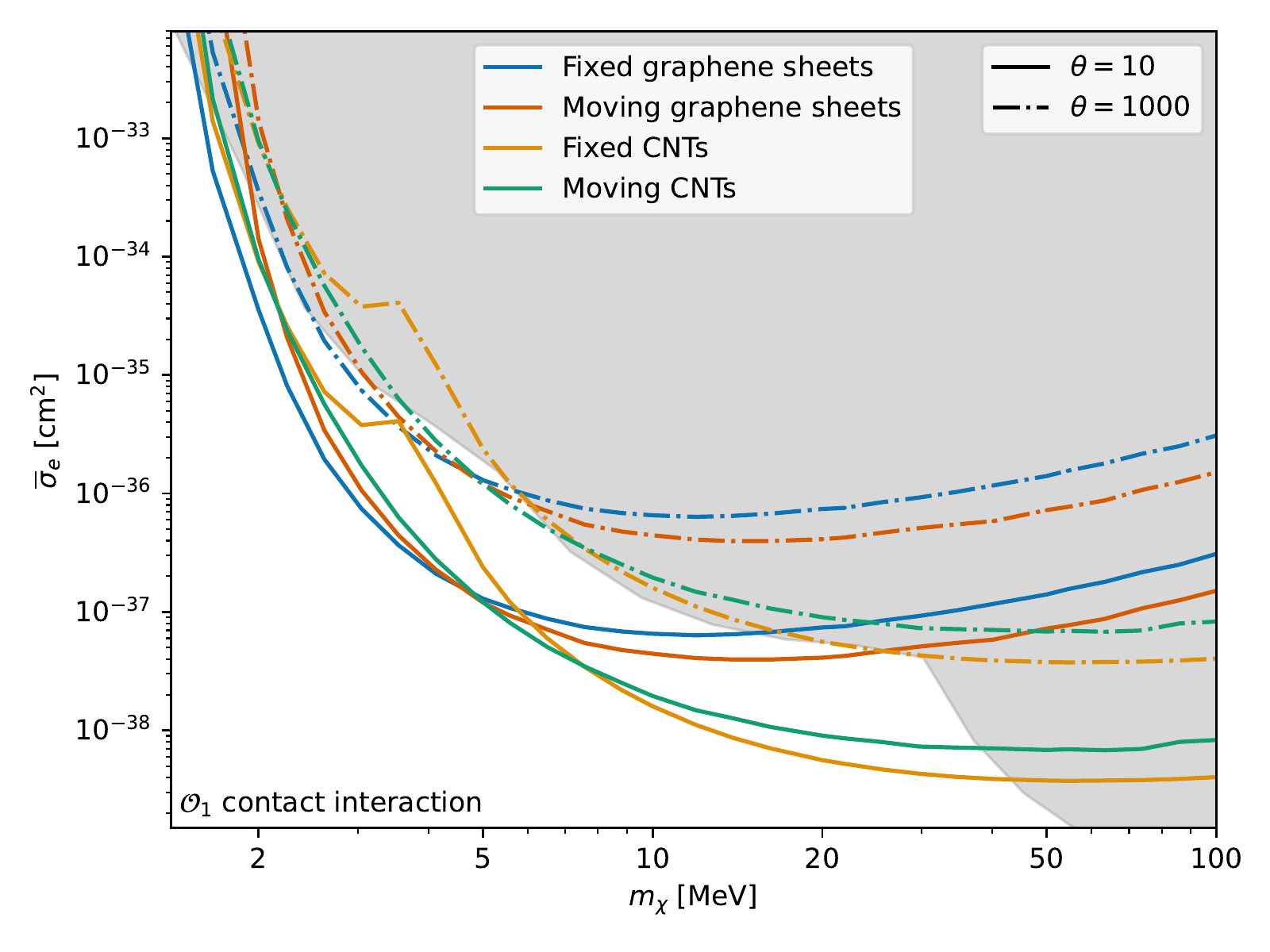}
    \includegraphics[width=0.48\textwidth]{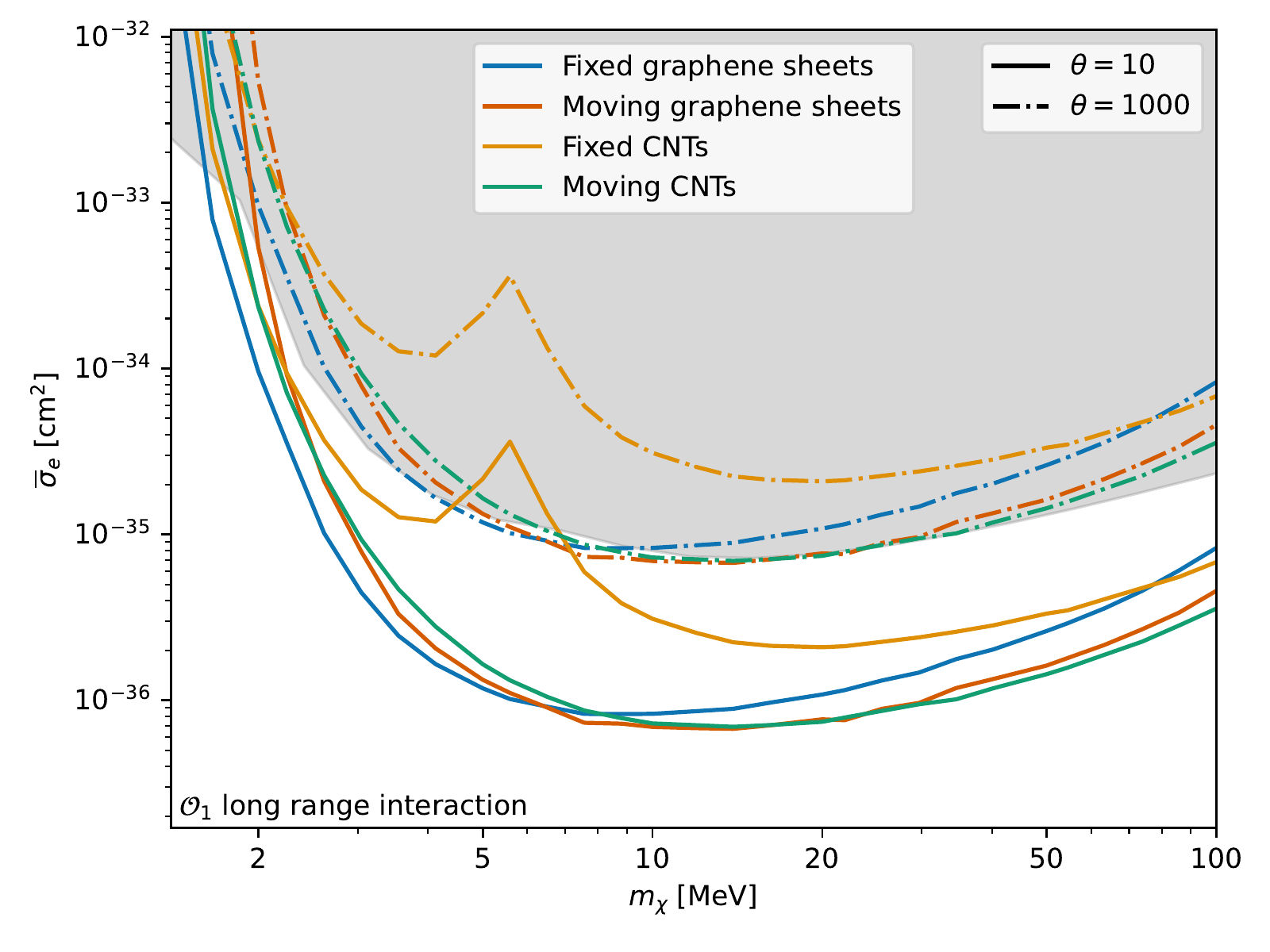}
    \includegraphics[width=0.48\textwidth]{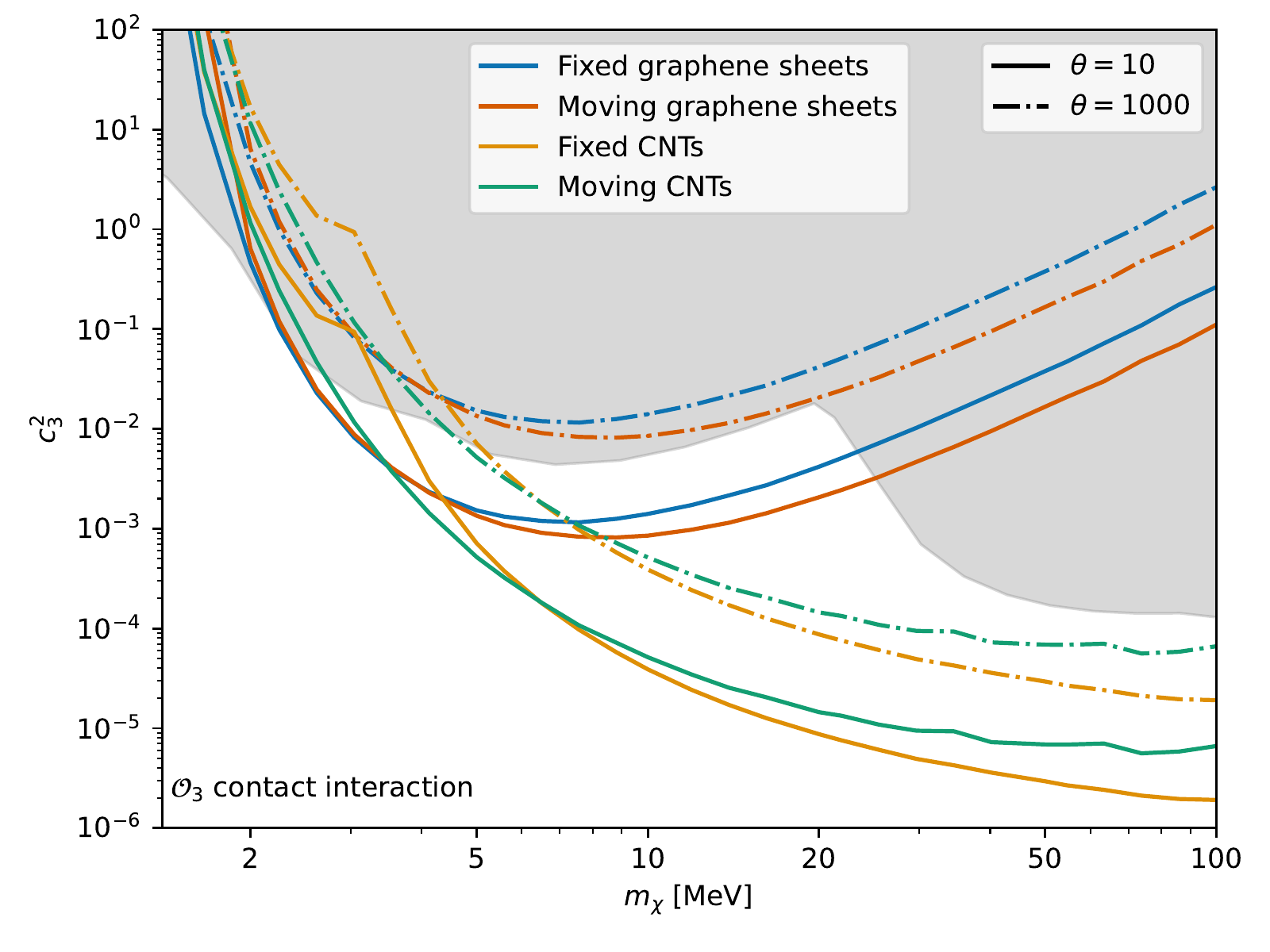}
    \includegraphics[width=0.48\textwidth]{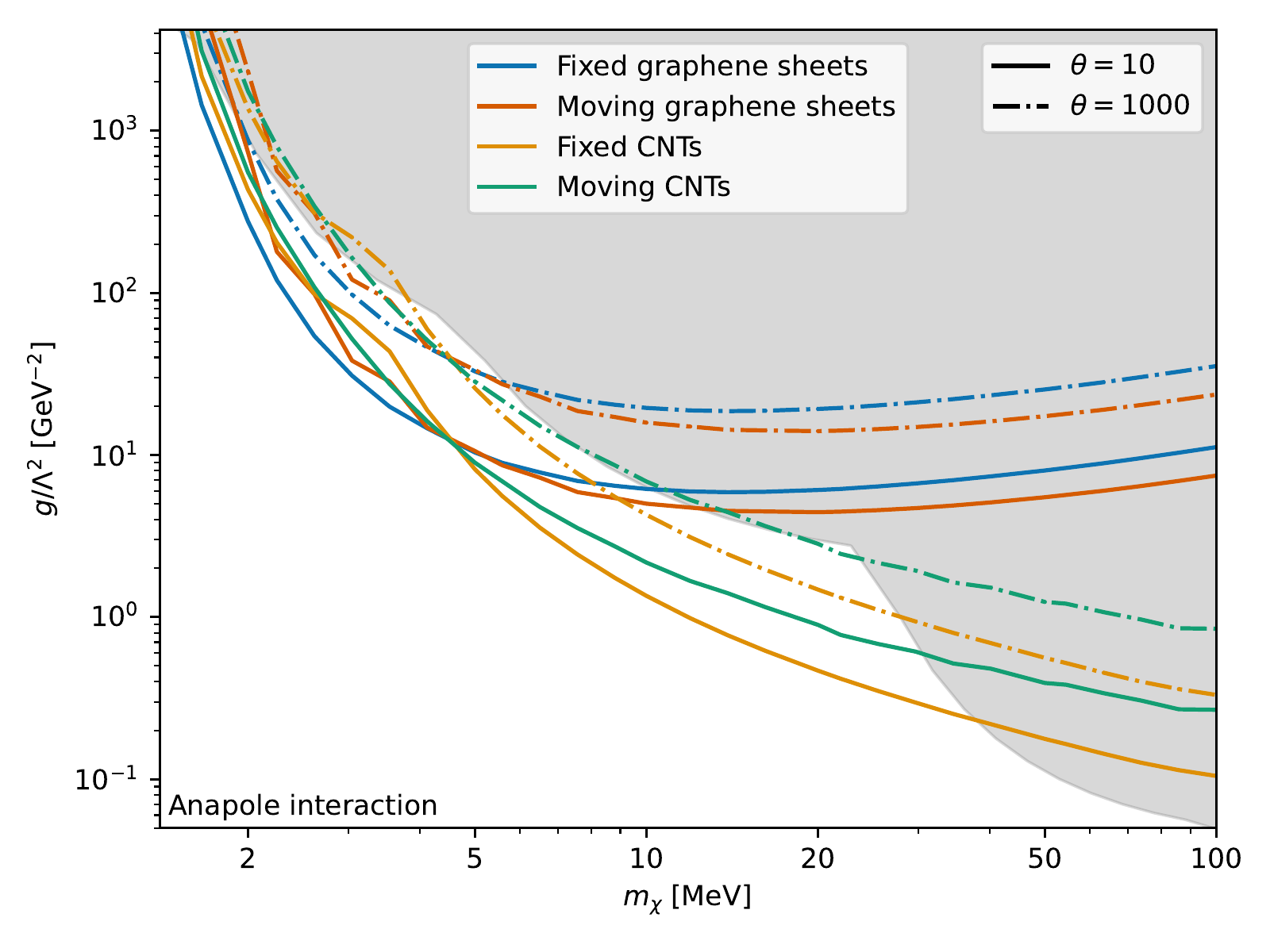}
    \includegraphics[width=0.48\textwidth]{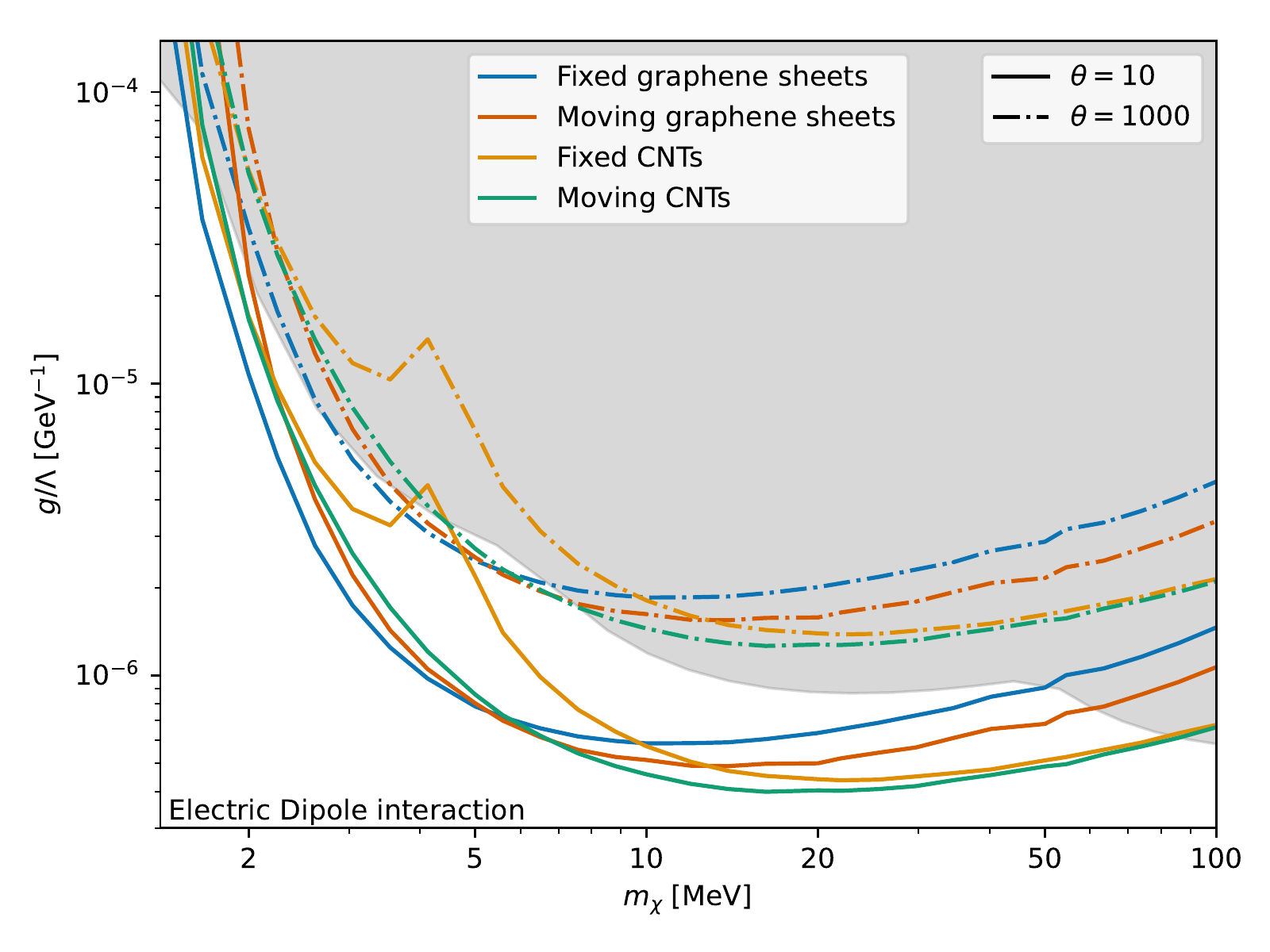}
    \includegraphics[width=0.48\textwidth]{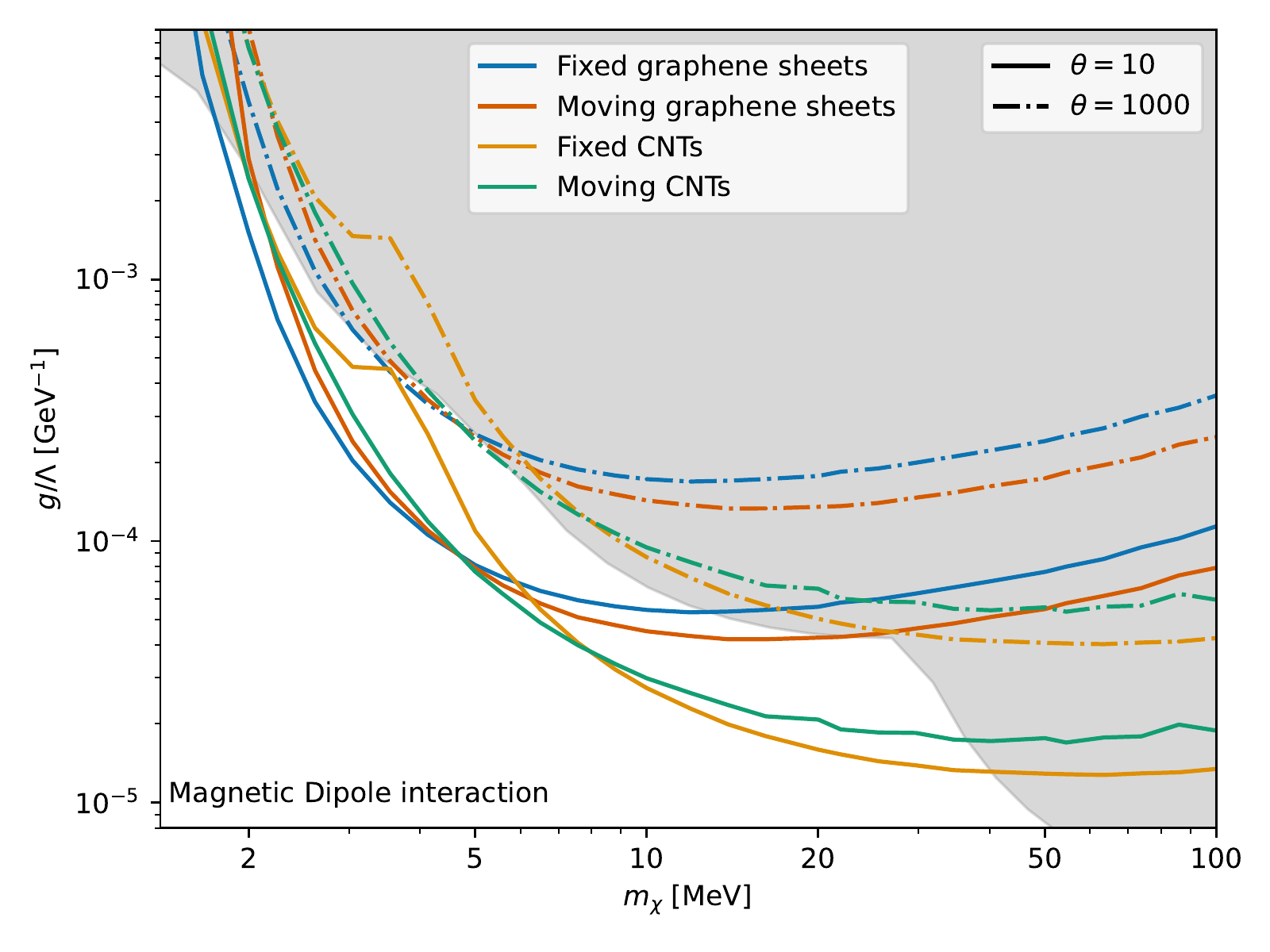}
    \caption{Sensitivities of carbon nanotube detection setups with a $10\,\mathrm{g}\,\mathrm{yr}$ exposure for $\mathcal{O}_1$ short range (top left), $\mathcal{O}_1$ long range (top right), $\mathcal{O}_3$ short range (center left), anapole (center right), electric dipole (bottom left) and magnetic dipole (bottom right) interactions. The lines correspond to $3\sigma$ discovery significance, where parameter points above the line are expected to be discovered with more than $3\sigma$ significance in the event of $\theta$ number of background events. The line styles correspond to different numbers of background events, with the solid line corresponding to $\theta=10$ background events, and the dash-dotted line corresponding to $\theta=1000$ background events. The different colors correspond to the different experimental setups shown in Figs.~\ref{fig: graphene setups} and~\ref{fig: CNTs setups}. The gray region is already excluded by other experiments~\cite{SENSEI:2020dpa,Catena:2021qsr,QEdark-EFT,XENON10:2011prx,Essig:2012yx,XENON:2019gfn,XENONCollaborationSS:2021sgk,Catena:2019gfa}. The bumps in the sensitivity curves for some of the experimental setups are due to $s_+$ and $s_-$ becoming similar at these masses.}
    \label{fig: sensitivity plot CNTs}
\end{figure*}
\begin{figure*}
    \centering
    \includegraphics[width=0.48\textwidth]{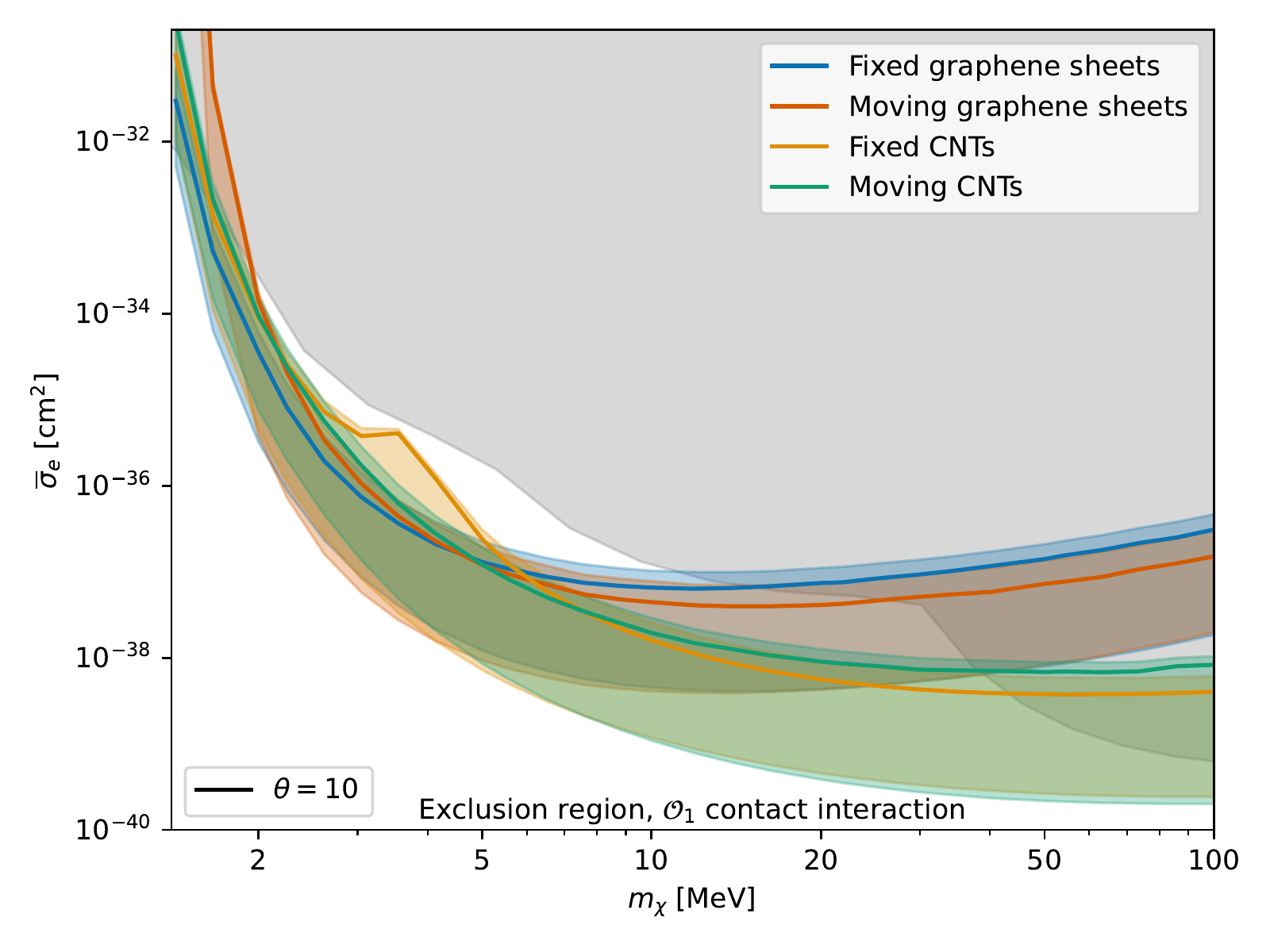}
    \includegraphics[width=0.48\textwidth]{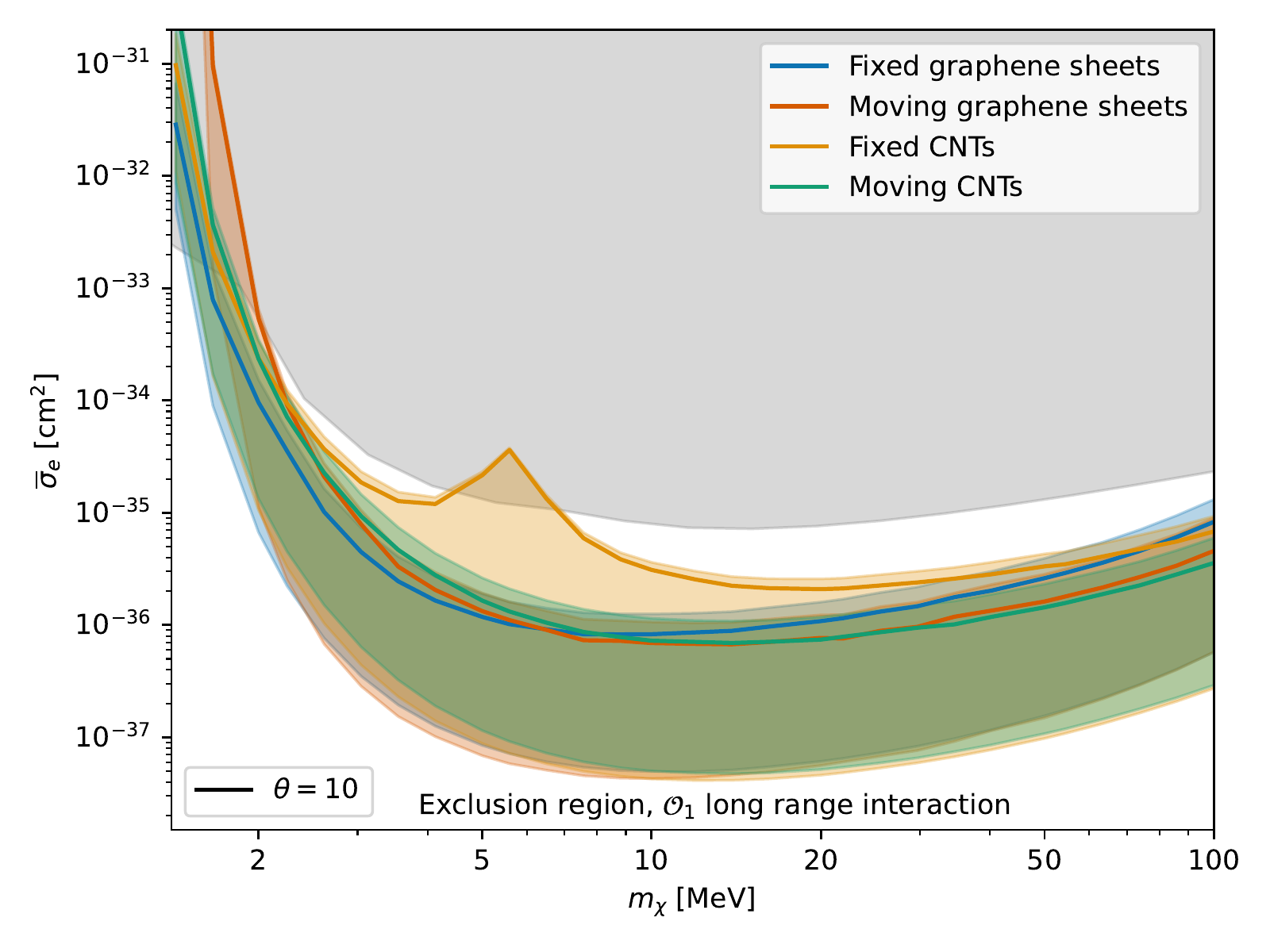}
    \includegraphics[width=0.48\textwidth]{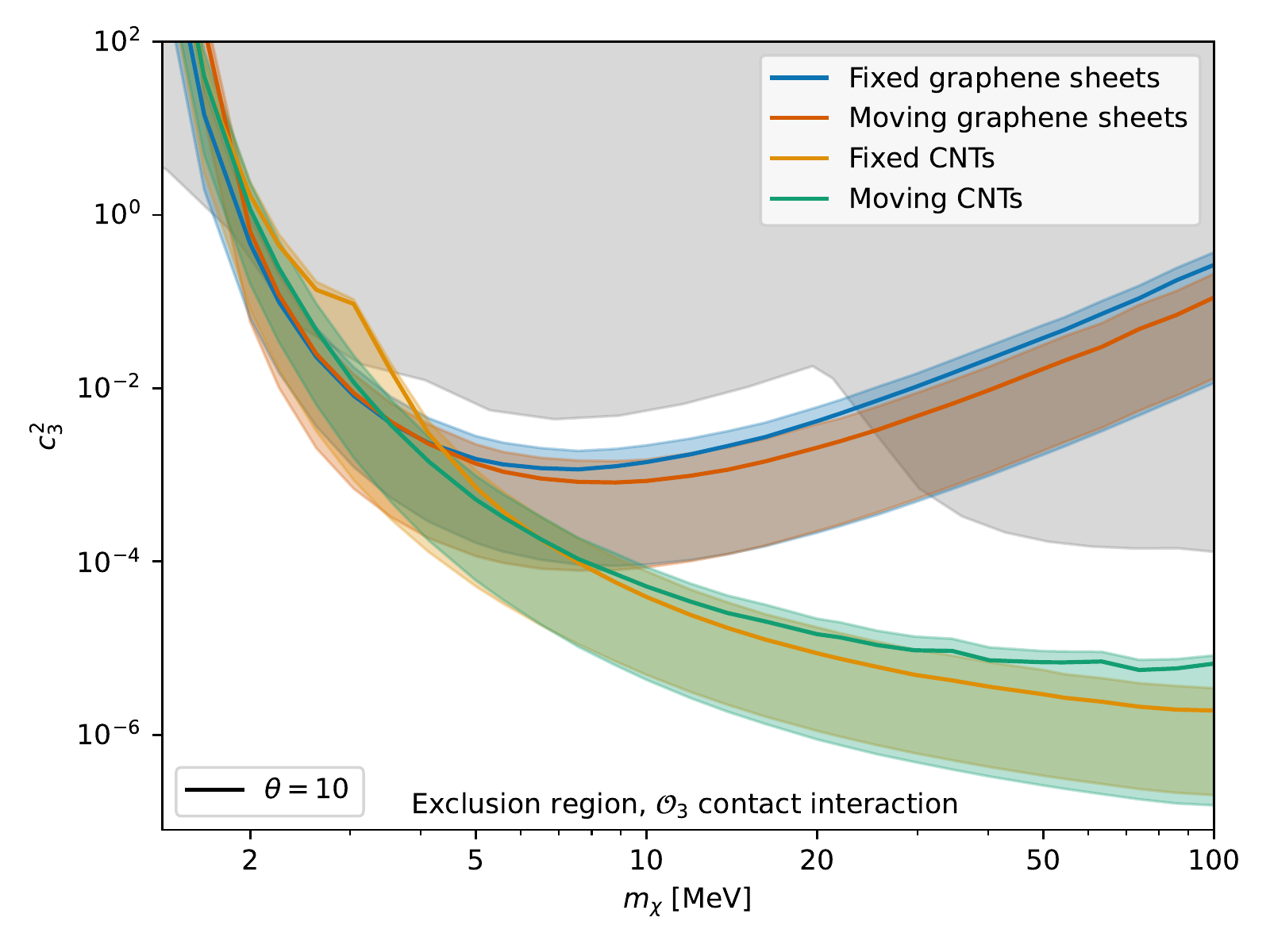}
    \includegraphics[width=0.48\textwidth]{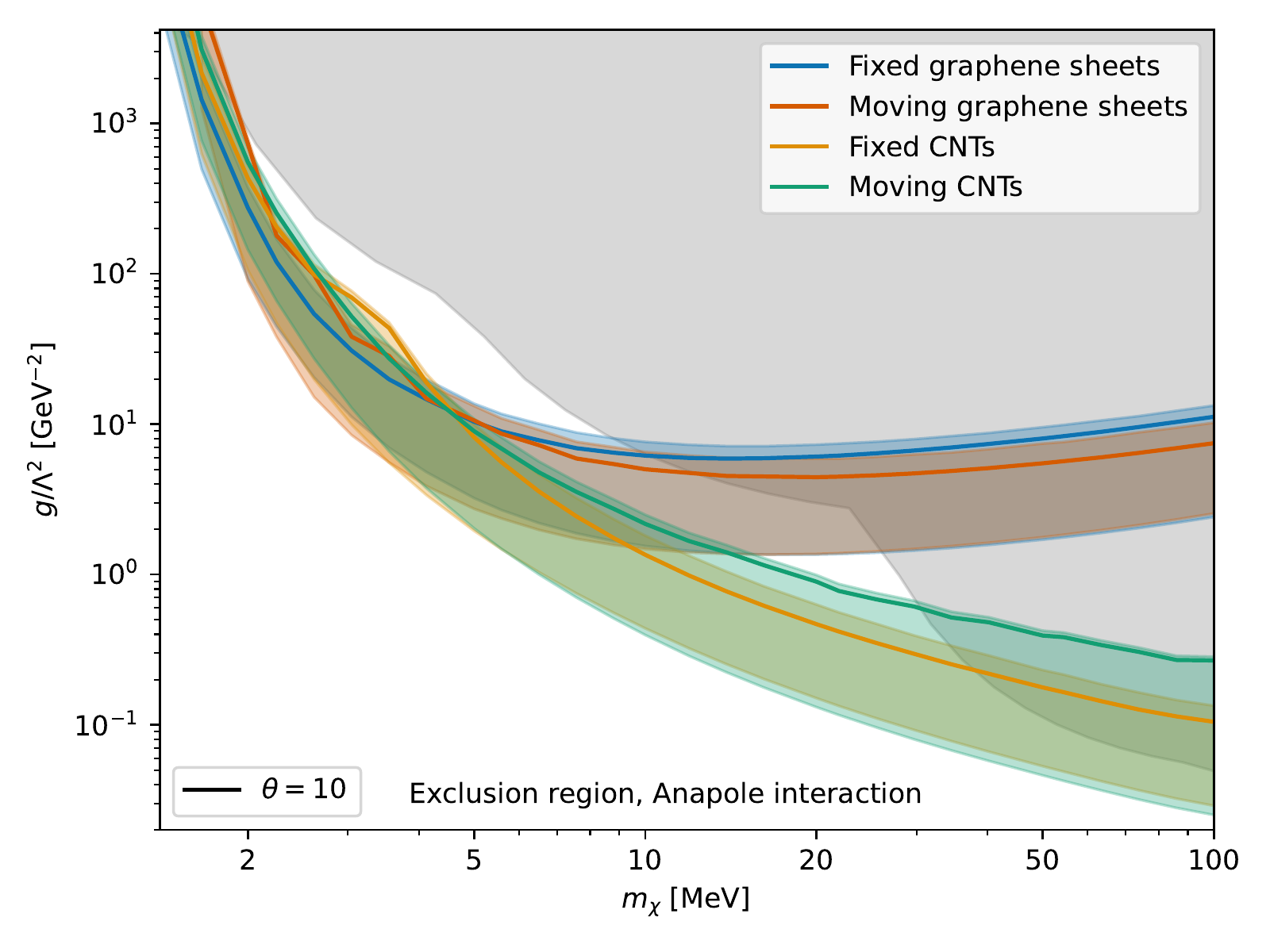}
    \includegraphics[width=0.48\textwidth]{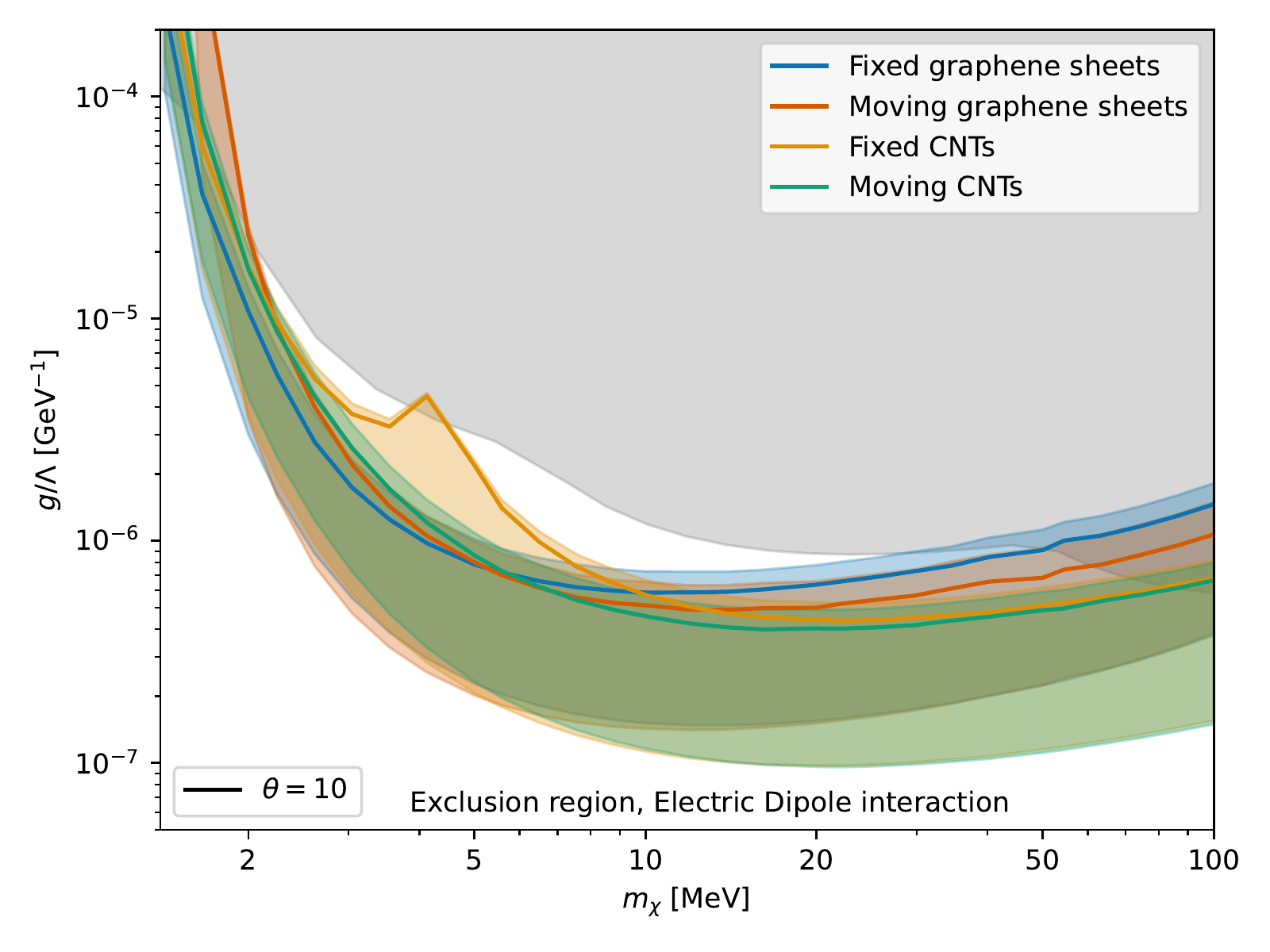}
    \includegraphics[width=0.48\textwidth]{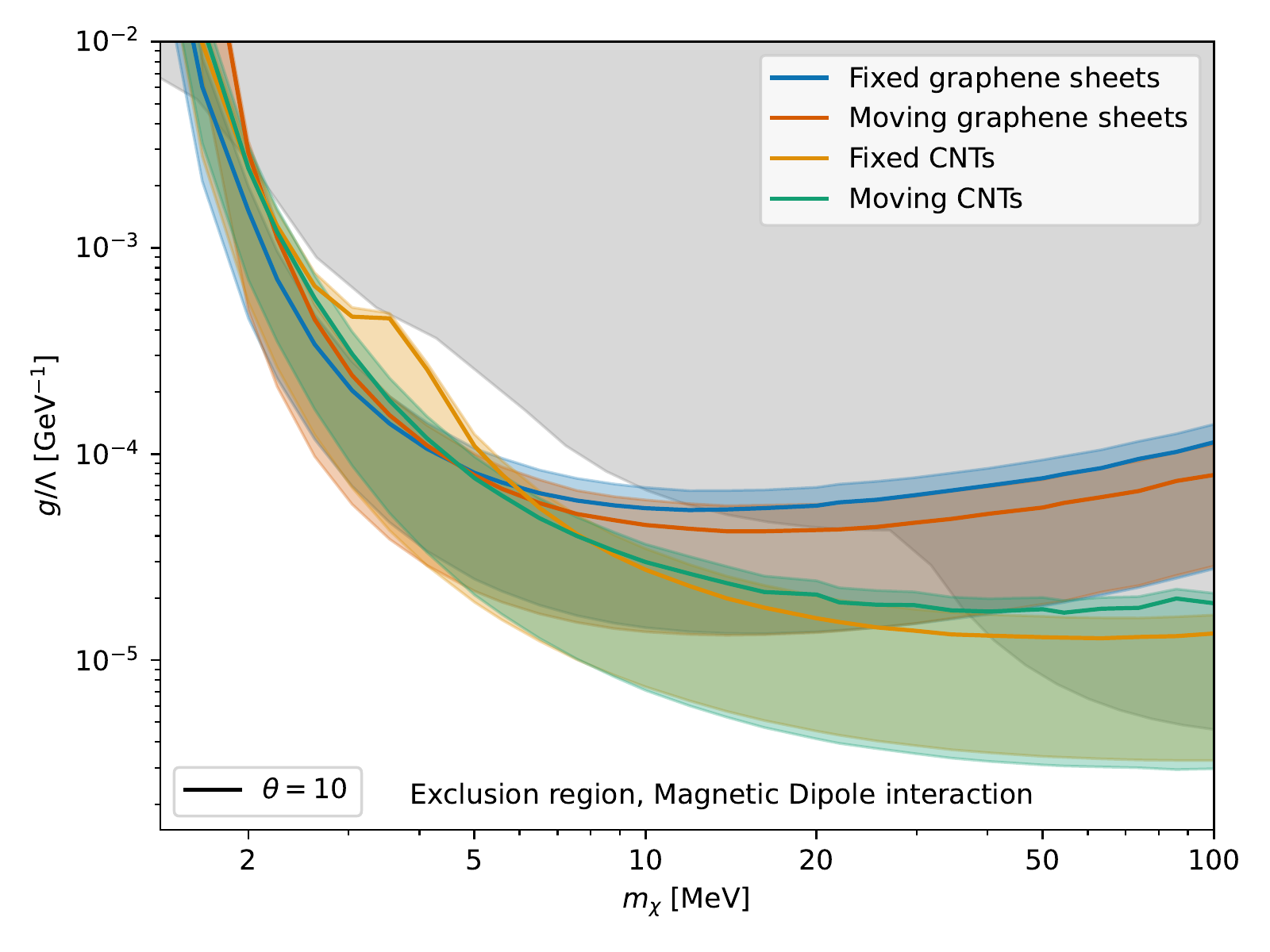}
    
    \caption{As Fig.~\ref{fig: sensitivity plot CNTs}, with lines corresponding to where daily modulation can be established at $3\sigma$ significance for the experimental setups shown in Figs.~\ref{fig: graphene setups} and~\ref{fig: CNTs setups}, together with an envelope region whose lower border is obtained from Eq.~\ref{eq: cdf} with $n_-+n_+=0$, and the upper border with $n_{-}+n_{+} = \mu'(s_{-}+s_{+})+\theta$, where $\theta=10$ and $\mu'$ is the strength parameter corresponding to the $3\sigma$ significance line. The exposure is assumed to be $10\,\mathrm{g}\,\mathrm{yr}$.}
    \label{fig: Exclusion}
\end{figure*}

To compare the expected performance of CNT and graphene sheet-based experiments, we give the expected sensitivities for discovering a daily modulation signal and expected exclusion limits in the case of a null result for a 10\,g-yr exposure. Fig.~\ref{fig: sensitivity plot CNTs} shows the expected 3$\sigma$ sensitivity for the four experiments illustrated in Figs.~\ref{fig: graphene setups} and~\ref{fig: CNTs setups} obtained with the likelihood defined in Eq.~(\ref{eq:q0}). Different colors correspond to different experimental setups, and various line styles correspond to a different number of background events. The solid line shows the 3$\sigma$ sensitivity of discovery for ten background events $\theta$, whereas the dashed line represents the same calculation with $\theta=1000$ background events. The panels correspond to different forms of DM-electron interactions, and the gray area is excluded by other direct-detection experiments. 

We see that for the contact-like interactions, $\mathcal{O}_1$ and $\mathcal{O}_3$ contact, as well as for the anapole and magnetic dipole interactions, the CNT-based experiments perform considerably better than the graphene sheet-based ones. This is due to contact interactions exhibiting a considerably stronger daily modulation pattern in CNTs than in graphene sheets. Comparing the $m_\chi=100\,\mathrm{MeV}$ panels in Figs.~\ref{fig: daily modulation sheets} and~\ref{fig: daily modulation CNTs}, one sees that while the electric dipole and magnetic dipole have minima between $\mathscr{R}/\langle\mathscr{R}\rangle = 0.5$ and $\mathscr{R}/\langle\mathscr{R}\rangle = 0.7$ for both CNTs and graphene sheets, the contact-like interactions have a minimum well below $\mathscr{R}/\langle\mathscr{R}\rangle = 0.5$ for CNTs and well above $\mathscr{R}/\langle\mathscr{R}\rangle = 0.5$ for graphene sheets. 

At lower DM masses, graphene sheets outperform CNTs, but the difference, when compared to moving-CNTs setups, is small. The fixed-CNTs setup does however suffer a notable loss of sensitivity at a region between $m_\chi=3\,\mathrm{MeV}$ and $m_\chi=6\,\mathrm{MeV}$, depending on the interaction type. The loss of sensitivity in this mass range is due to the flattening of the daily modulation plot in Fig.~\ref{fig: daily modulation CNTs} (given by the balancing of the two competing directional effects discussed above), where the daily modulation pattern is flatter for $m_\chi=5\,\mathrm{MeV}$ than for $m_\chi=2\,\mathrm{MeV}$ and $m_\chi=10\,\mathrm{MeV}$. For the electric dipole and $\mathcal{O}_1$ long range interactions these curves are almost entirely flat, explaining why the regions of insensitivity of the fixed CNTs setup in Fig.~\ref{fig: sensitivity plot CNTs} are more pronounced for these interactions. 

Finally, we see that with a 10\,g-yr exposure, there is a sizeable non-excluded parameter space in which DM can be discovered at 3$\sigma$ confidence limit in the event of $\theta=10$ background events. This is in particular true for the moving-CNTs setup. Increasing the number of background events to $\theta=1000$ shrinks this parameter space severely, still leaving tiny patches of allowed parameter space in which DM can be discovered. An exception to this is the $\mathcal{O}_3$ contact interaction, which induces a very strong daily modulation signal in CNTs at DM masses above $5\,\mathrm{MeV}$, leaving a vast allowed parameter space in which DM can be discovered at 3$\sigma$ even with $\theta=1000$ background events. 

In Fig.~\ref{fig: Exclusion}, we show a contour around the $\theta=10$ reaches shown previously in Fig.~\ref{fig: sensitivity plot CNTs}. This contour is obtained from Eq.~(\ref{eq: cdf}) and represents the region in which the exclusion limits of DM can lie. The lower edge of the contour is given by the exclusion limit in the scenario of no observed events, i.e. $n_-+n_+=0$, whereas the upper edge is the ``worst case'' scenario for $\theta=10$, where there are still not enough DM events for discovery, and one has to resort to excluding DM having observed $n_{-}+n_{+} = \mu'(s_{-}+s_{+})+\theta$ events, where $\mu'$ is the value of the strength parameter at which DM can be detected at 3$\sigma$. We see that the CNT-based setups have the potential to considerably constrain the parameter space for all the considered models in the case of a null result. 

\begin{figure*}
    \centering
    \includegraphics[scale=0.32]{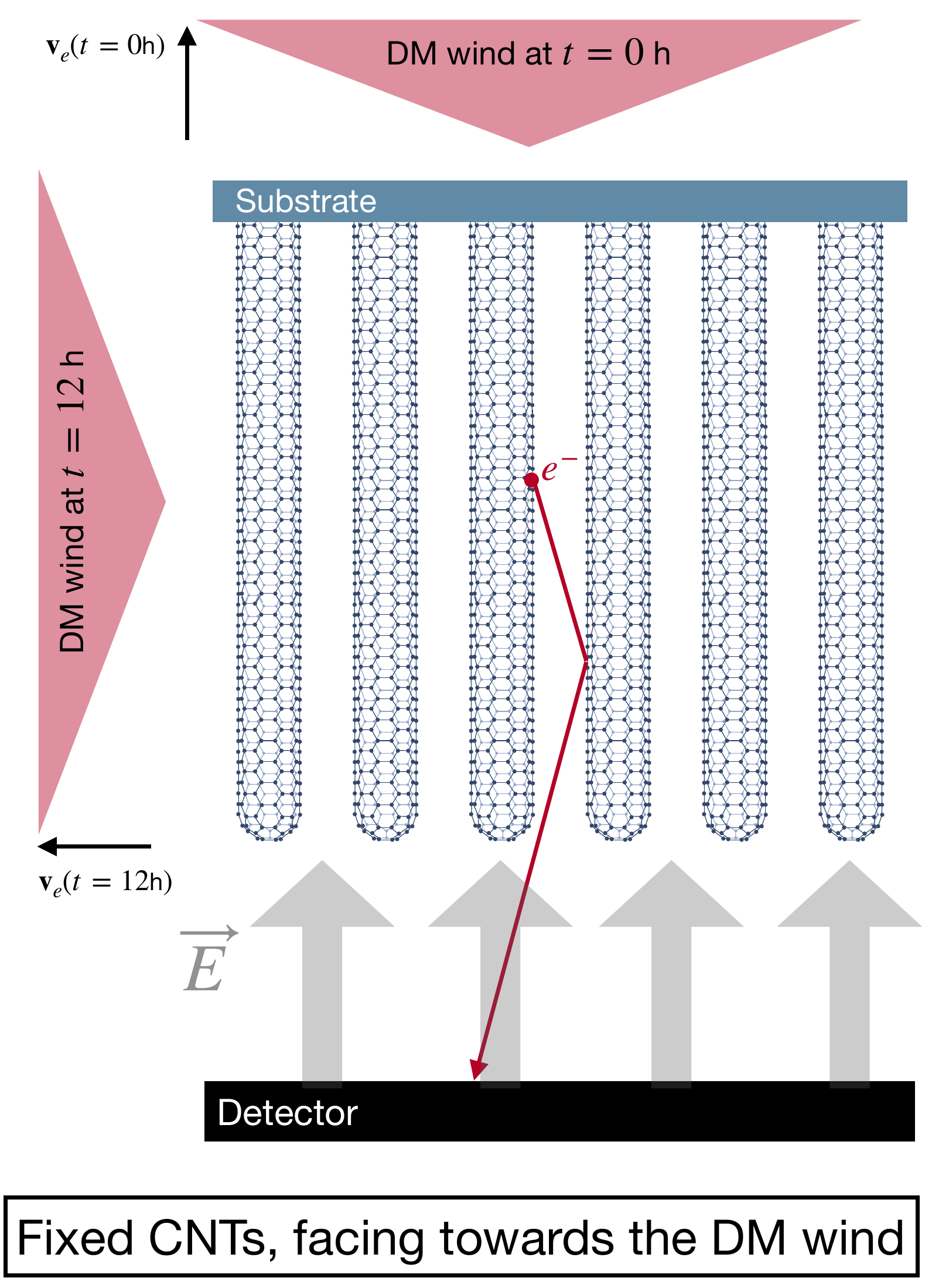}
  \qquad\qquad  \includegraphics[scale=0.32]{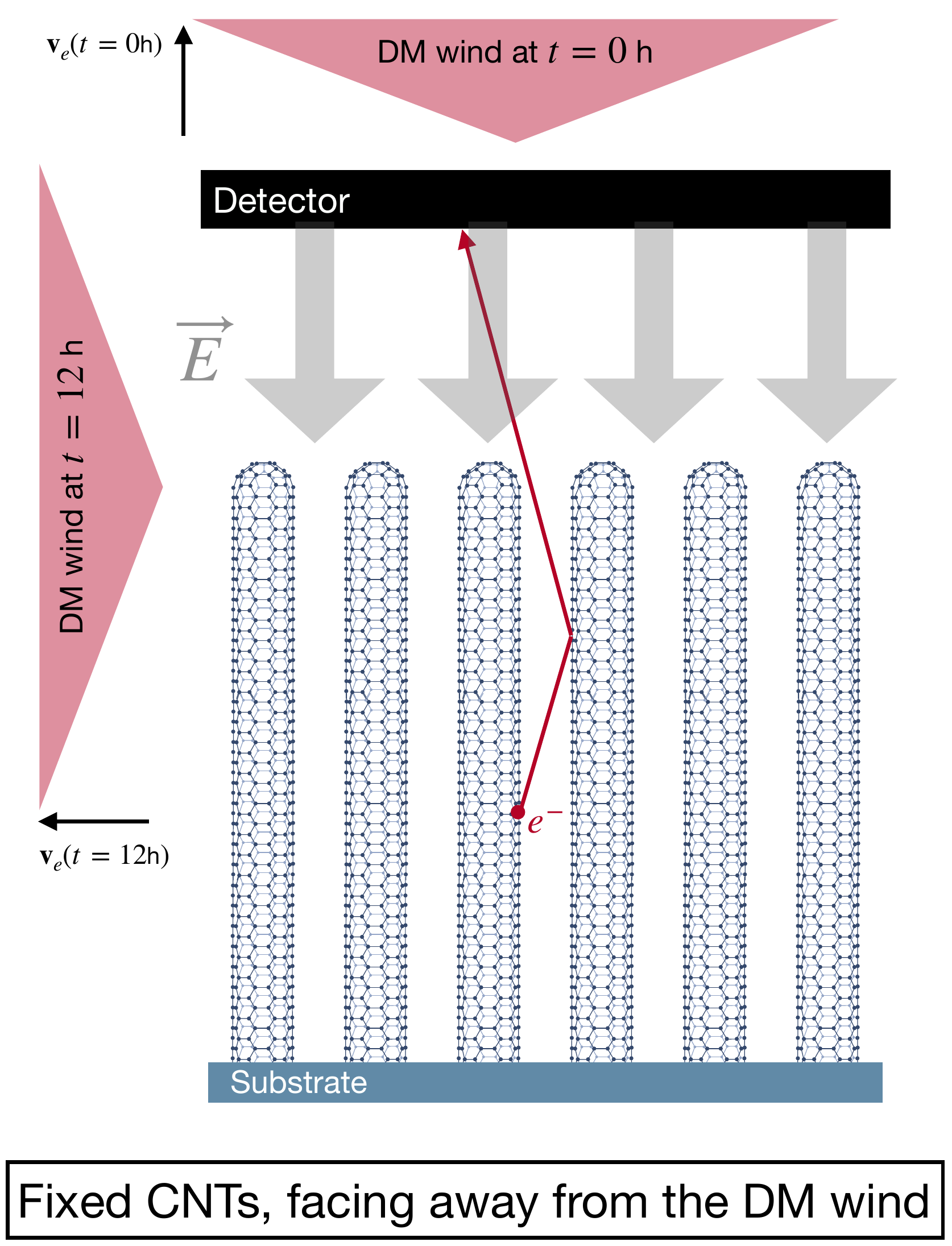}
    \caption{Illustrations of fixed CNT setups. The left panel was already shown in Fig.~\ref{fig: CNTs setups} (left) and is repeated here for comparison.}
    \label{fig: Fixed  CNT several setups}
\end{figure*}
\begin{figure*}
    \centering
    \includegraphics[scale=0.27]{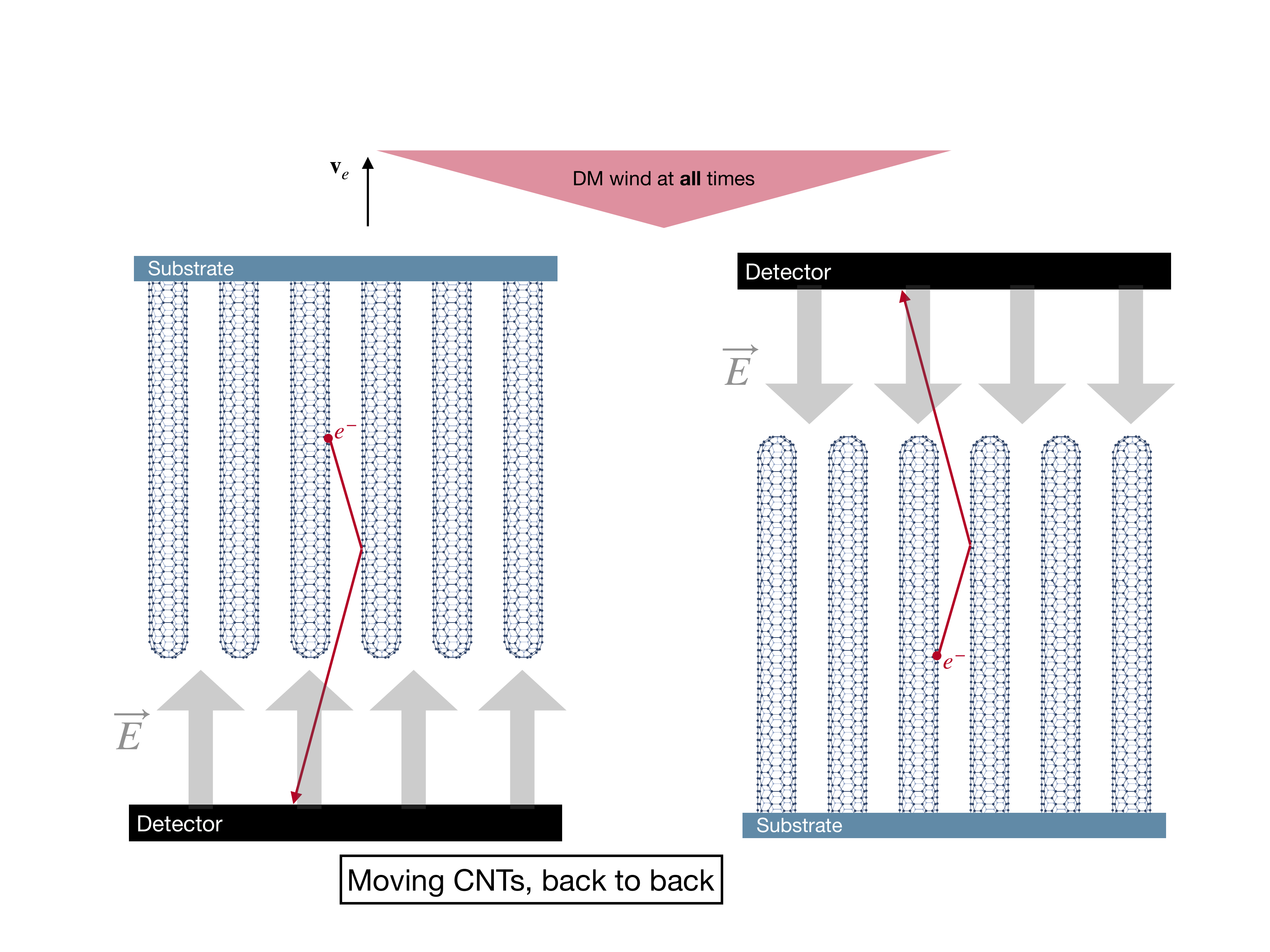}
    \includegraphics[scale=0.27]{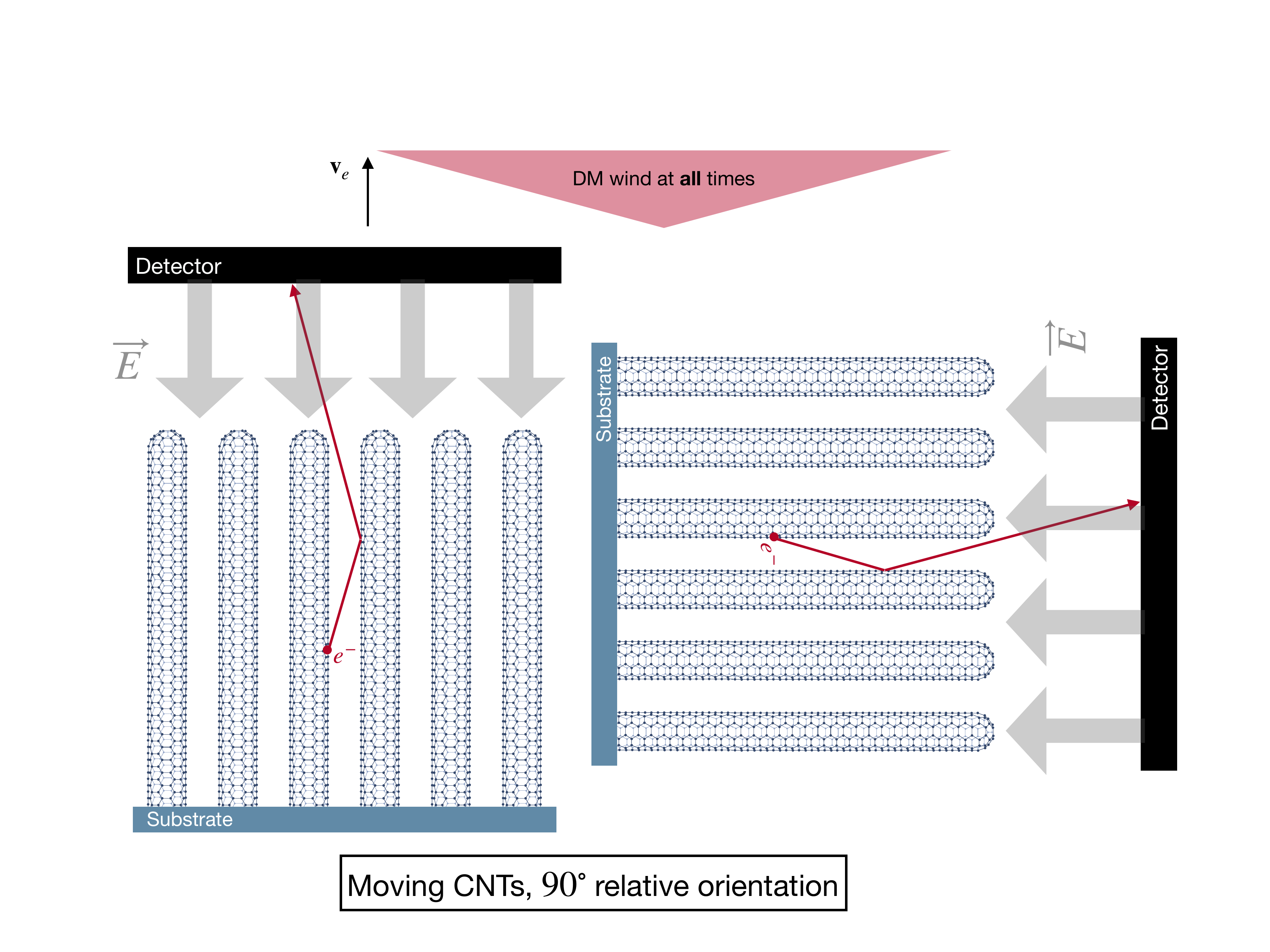}
    \includegraphics[scale=0.27]{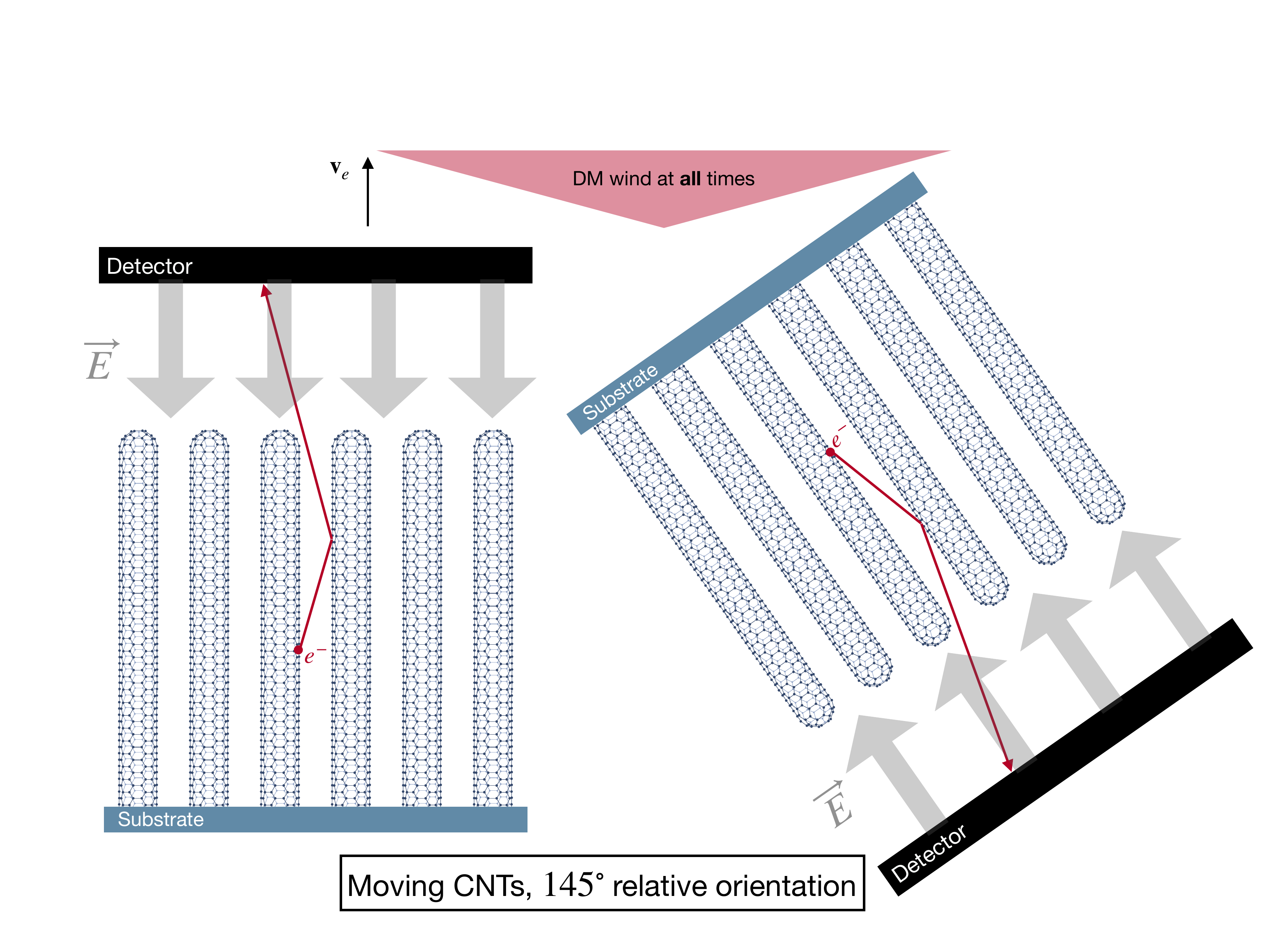}
    \caption{Illustrations of moving CNT setups considered in this appendix. The top panel was already shown in Fig.~\ref{fig: CNTs setups} (right) and is repeated here for comparison.}
    \label{fig: Moving CNT several setups}
\end{figure*}
\begin{figure*}
    \centering
    \includegraphics[width=0.48\textwidth]{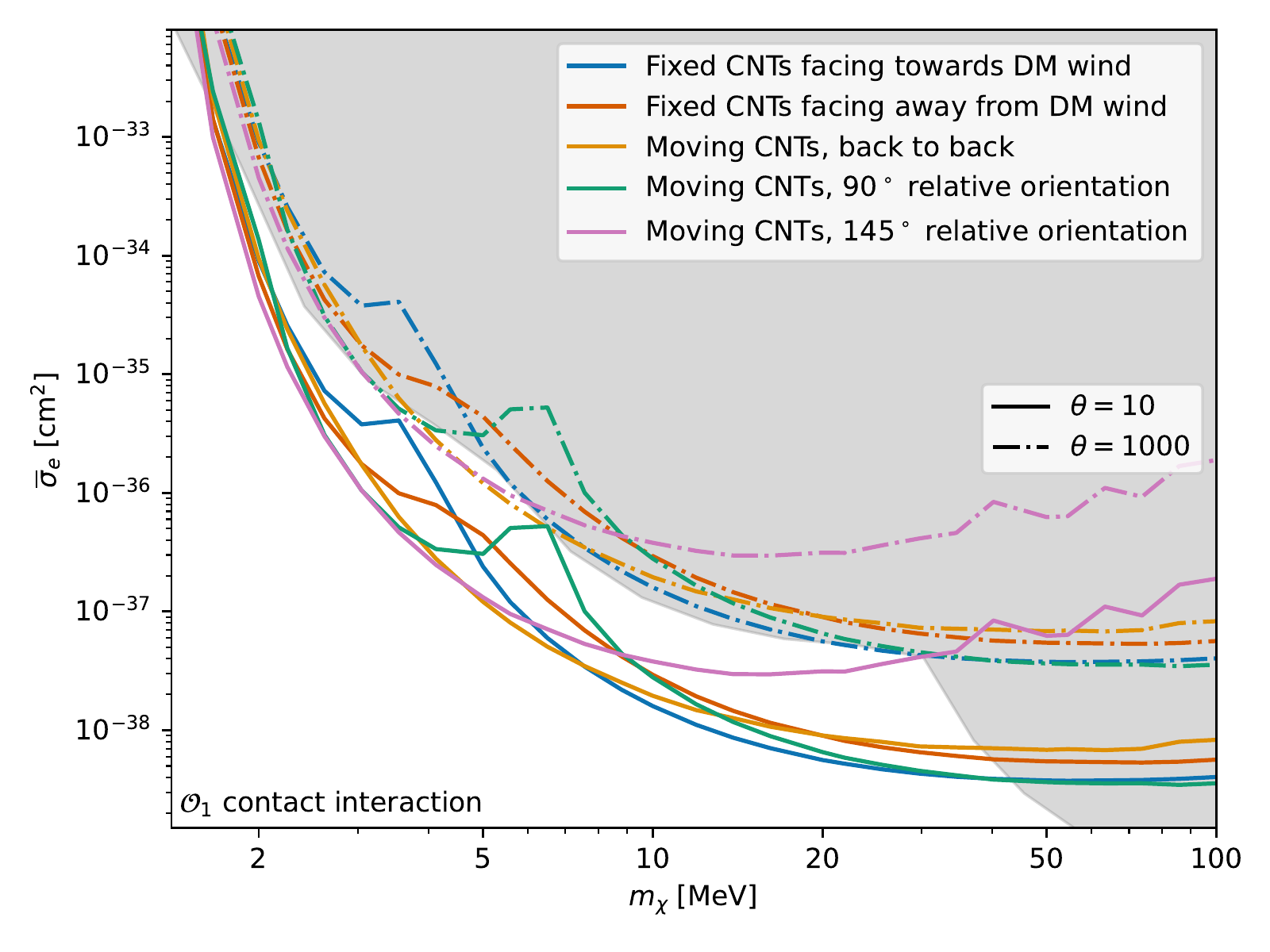}
    \includegraphics[width=0.48\textwidth]{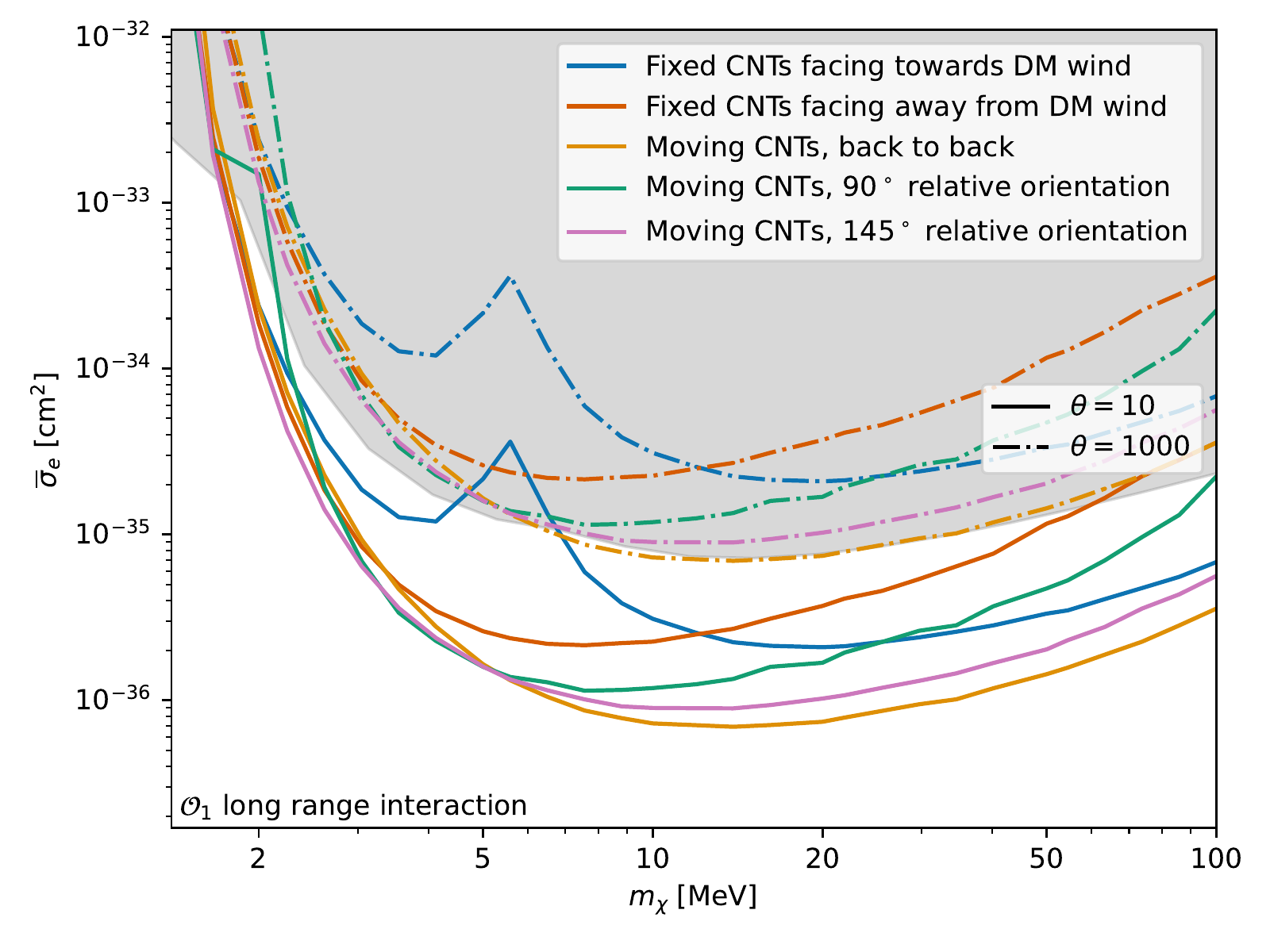}
    \includegraphics[width=0.48\textwidth]{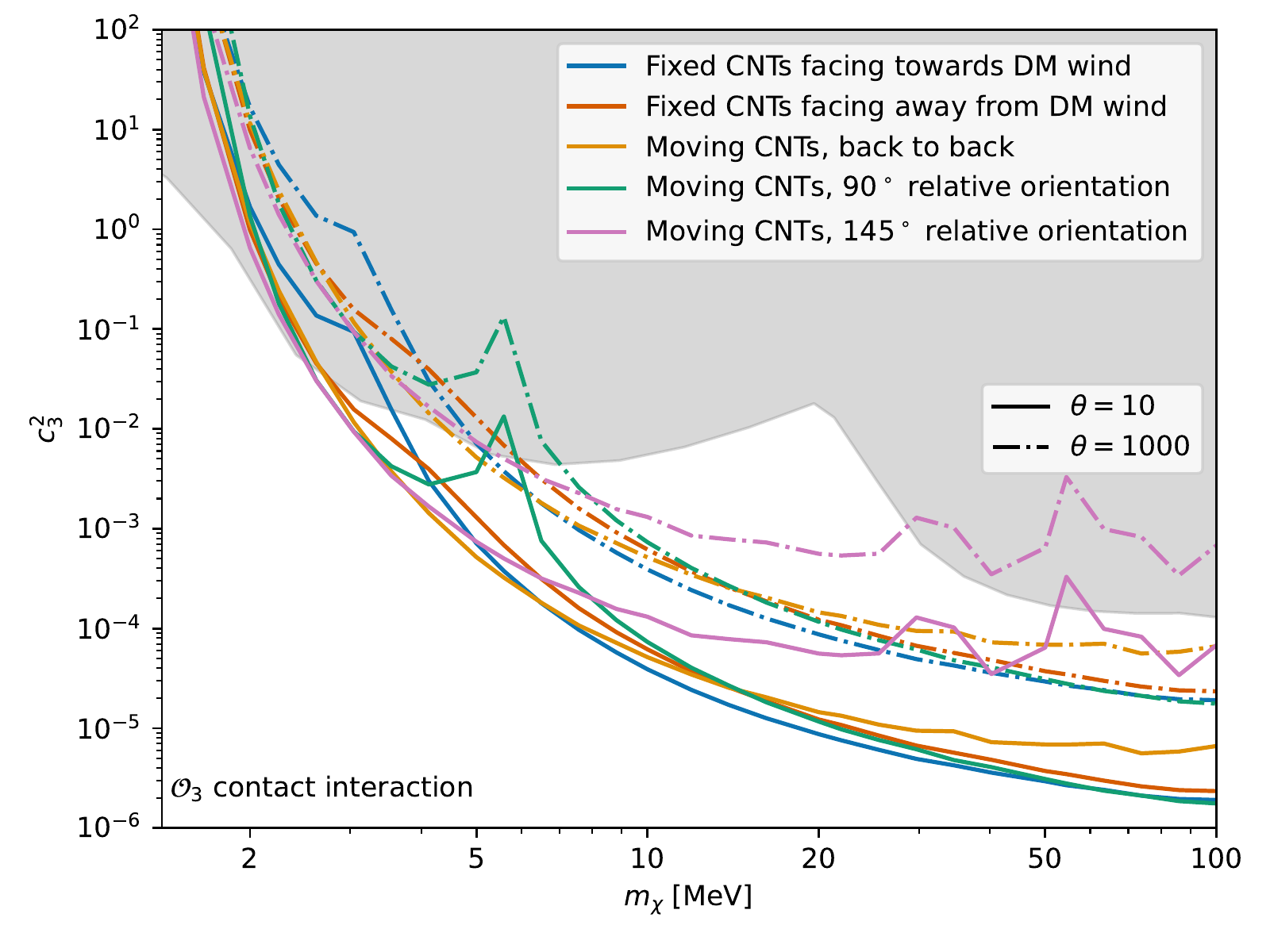}
    \includegraphics[width=0.48\textwidth]{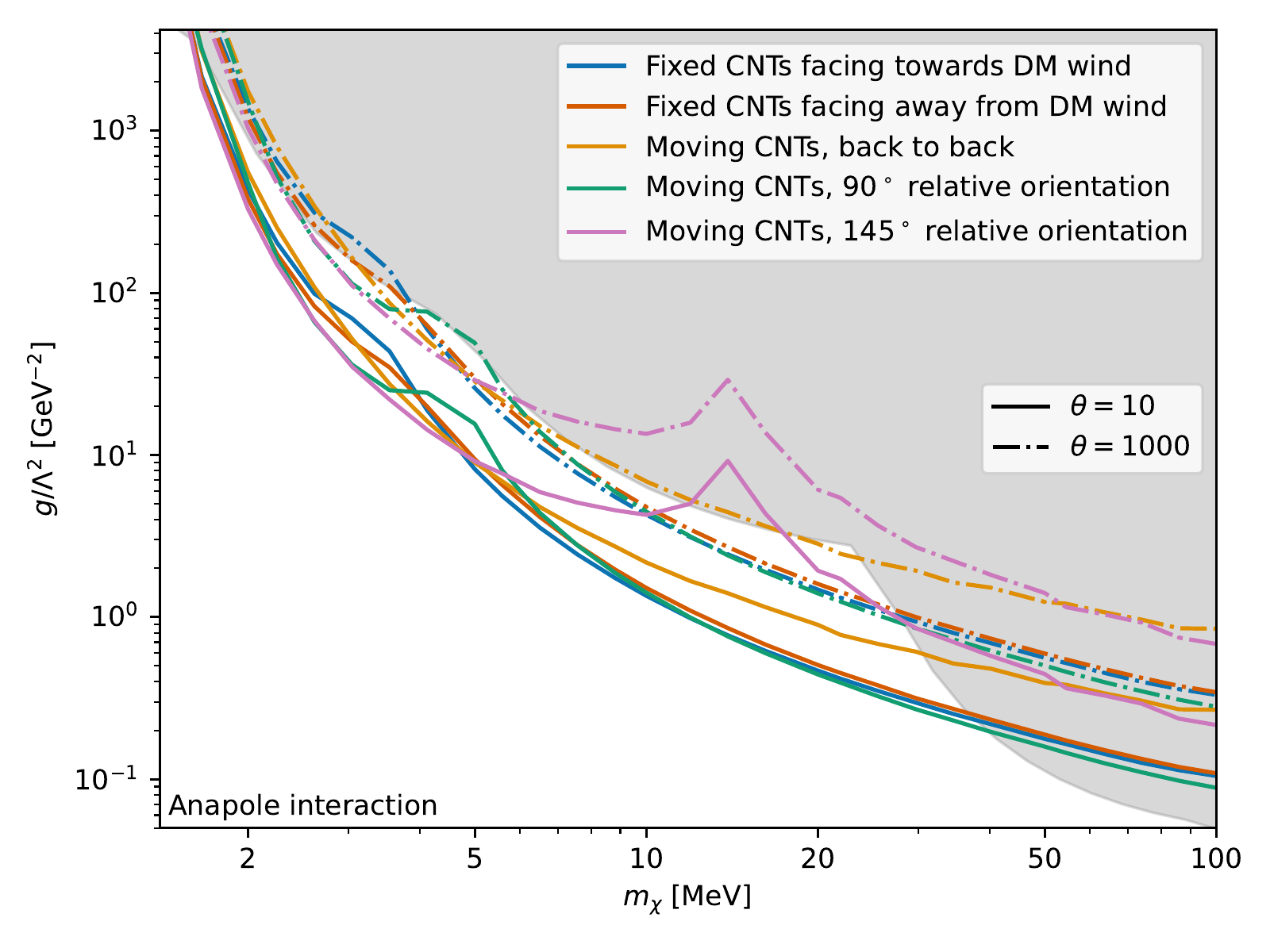}
    \includegraphics[width=0.48\textwidth]{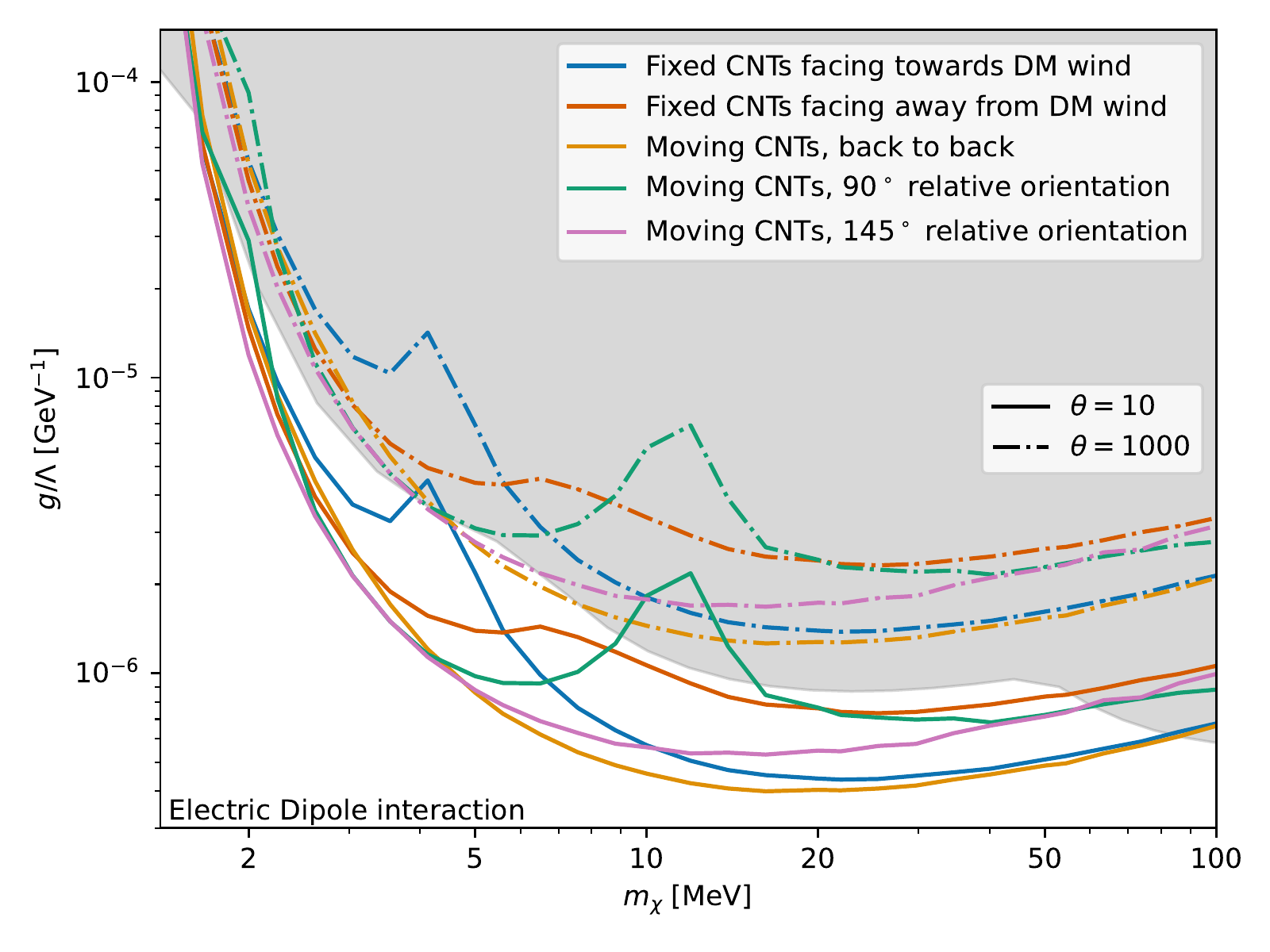}
    \includegraphics[width=0.48\textwidth]{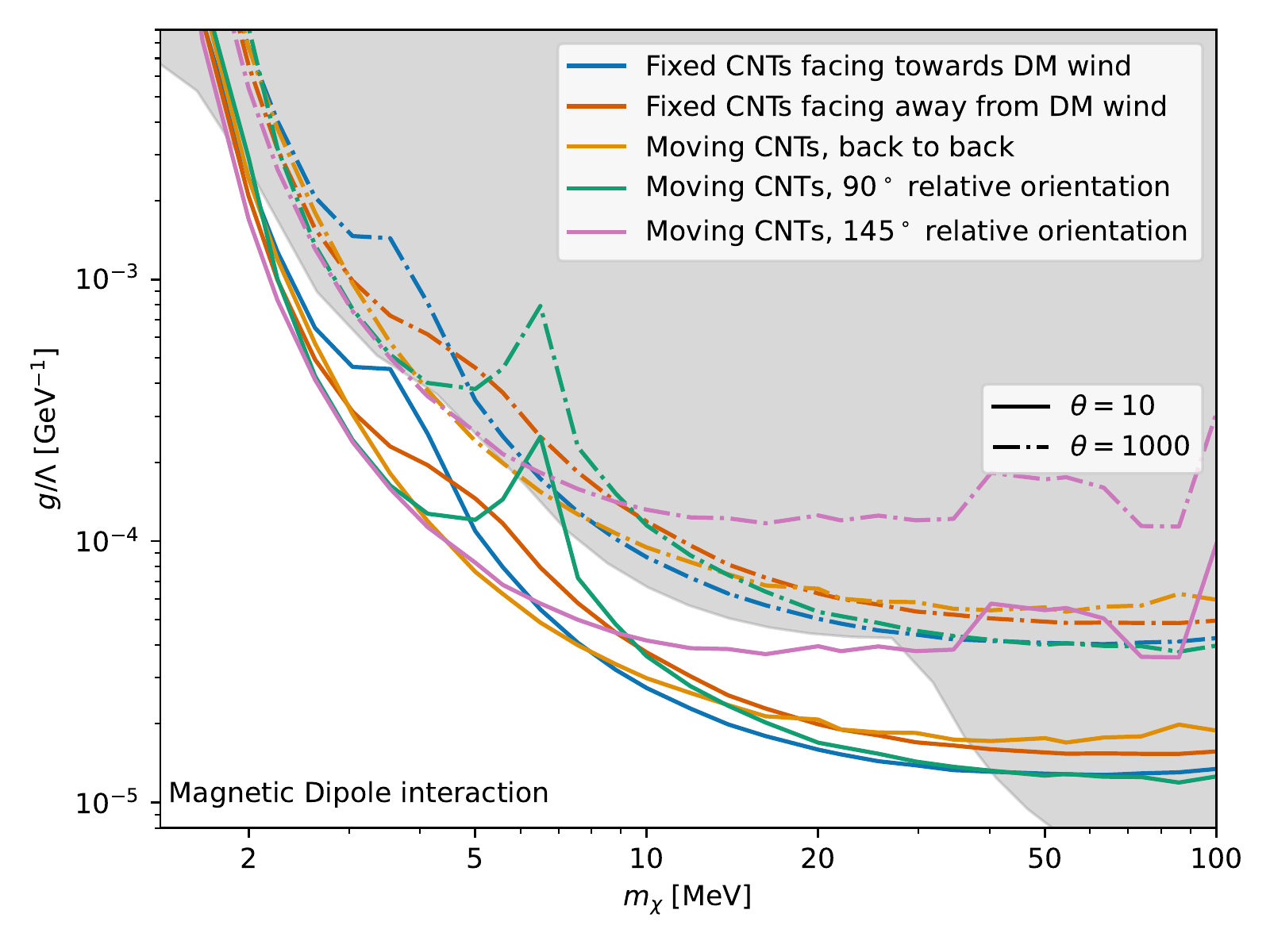}
    \caption{Same as Fig.~\ref{fig: sensitivity plot CNTs} but for the CNT based setups illustrated in Figs.~\ref{fig: Fixed  CNT several setups} and~\ref{fig: Moving CNT several setups}}
    \label{fig: CNT several setups sensitivities}
\end{figure*}
\begin{figure*}
    \centering
    \includegraphics[width=0.48\textwidth]{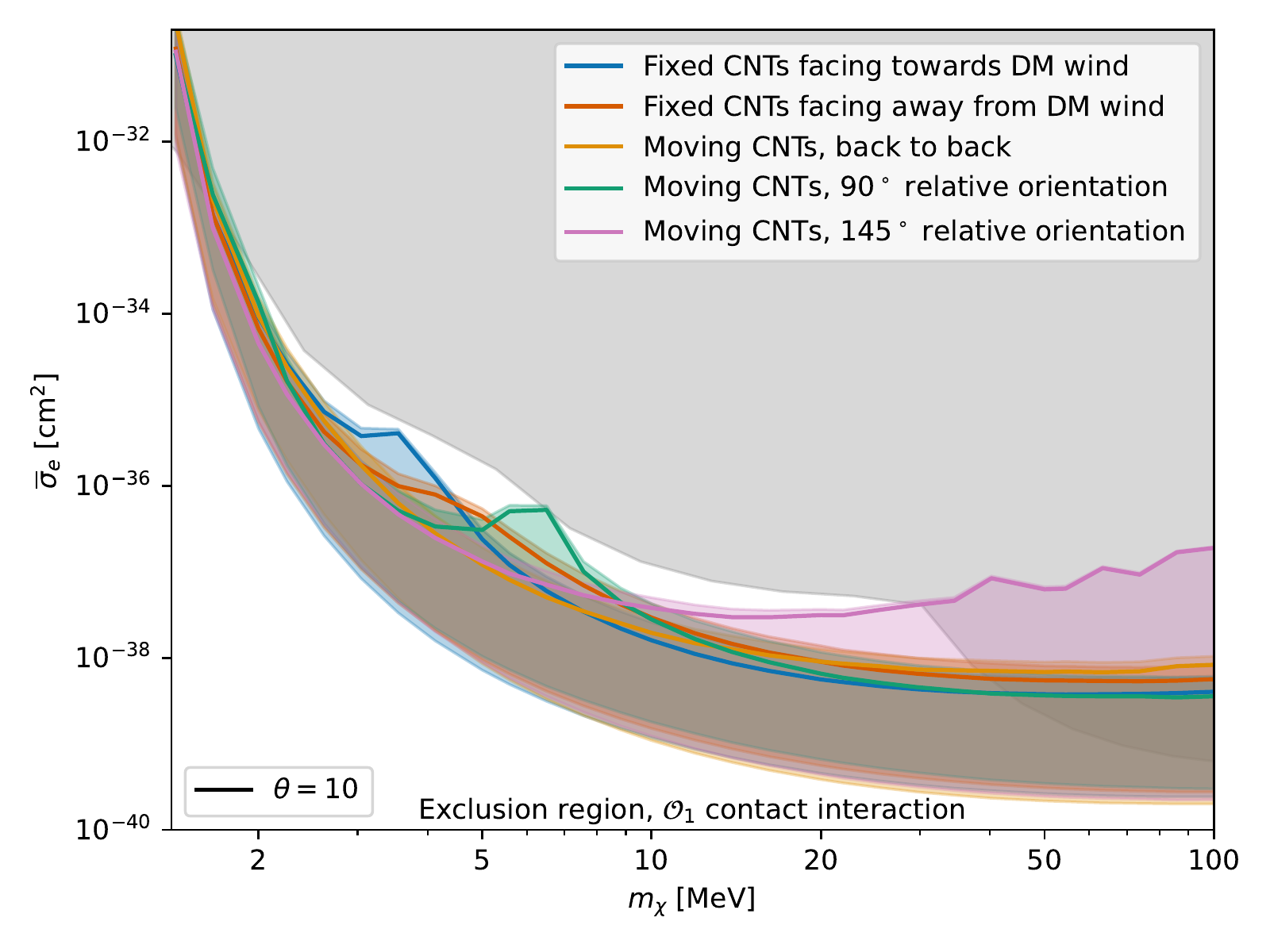}
    \includegraphics[width=0.48\textwidth]{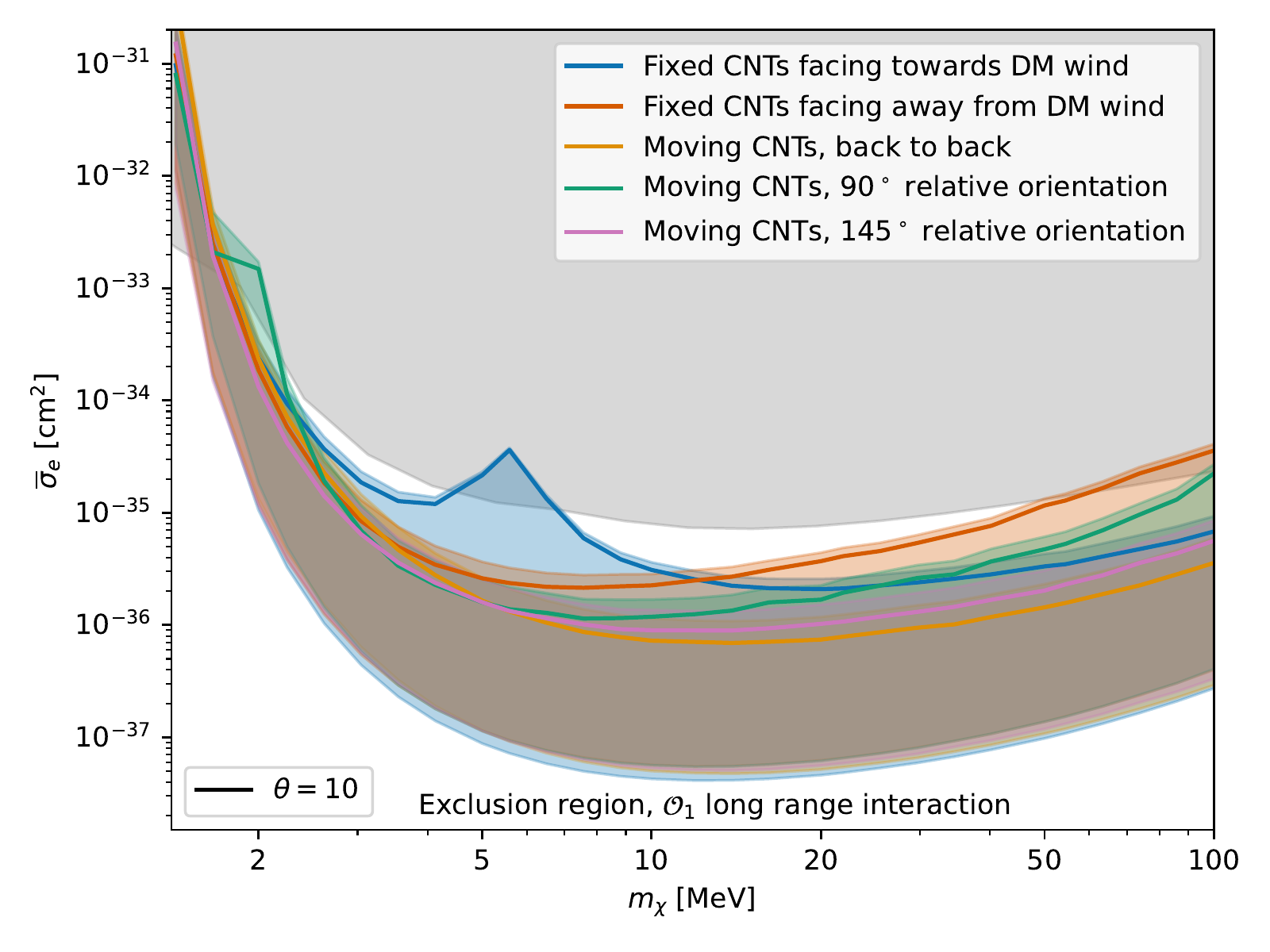}
    \includegraphics[width=0.48\textwidth]{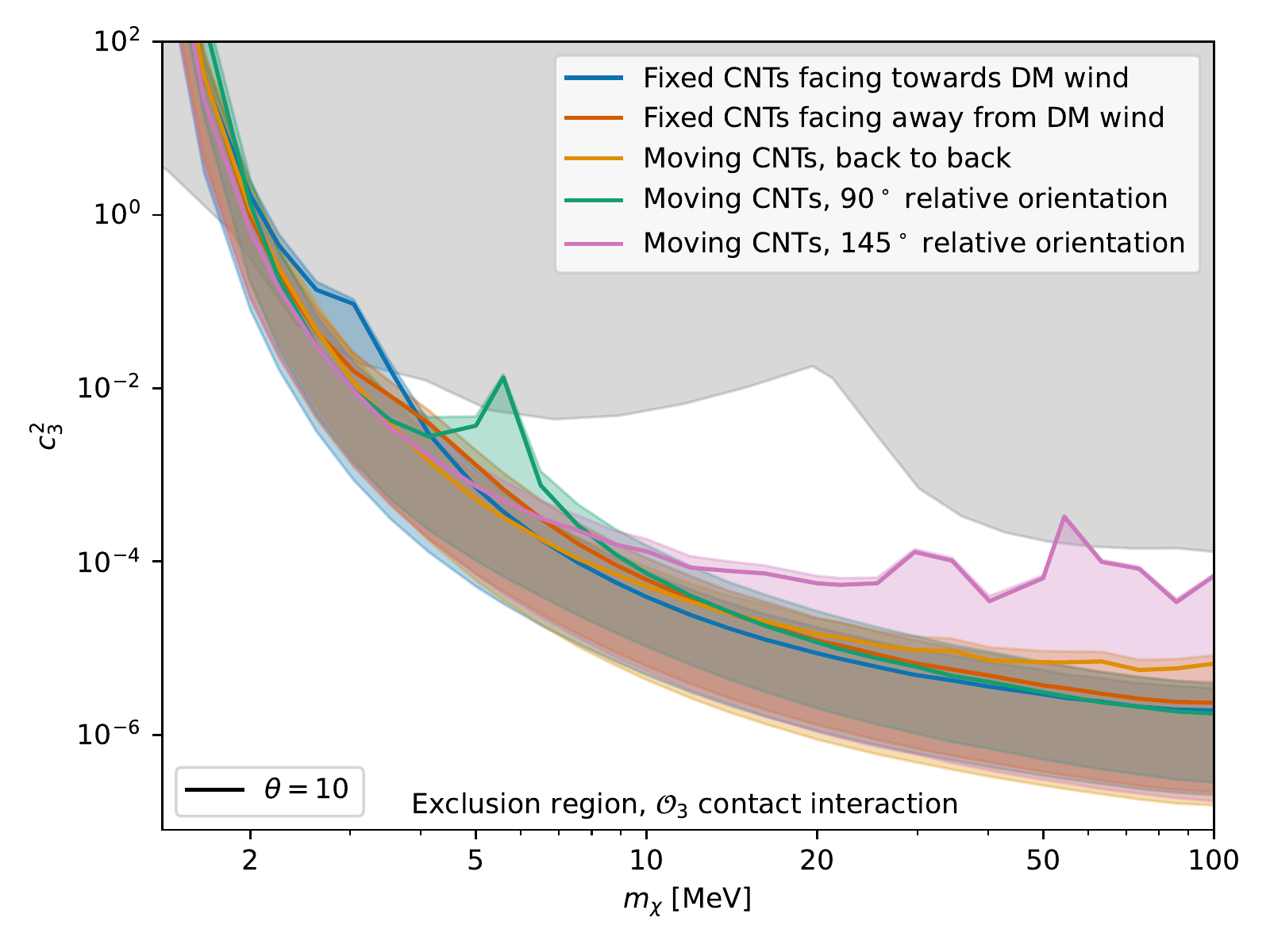}
    \includegraphics[width=0.48\textwidth]{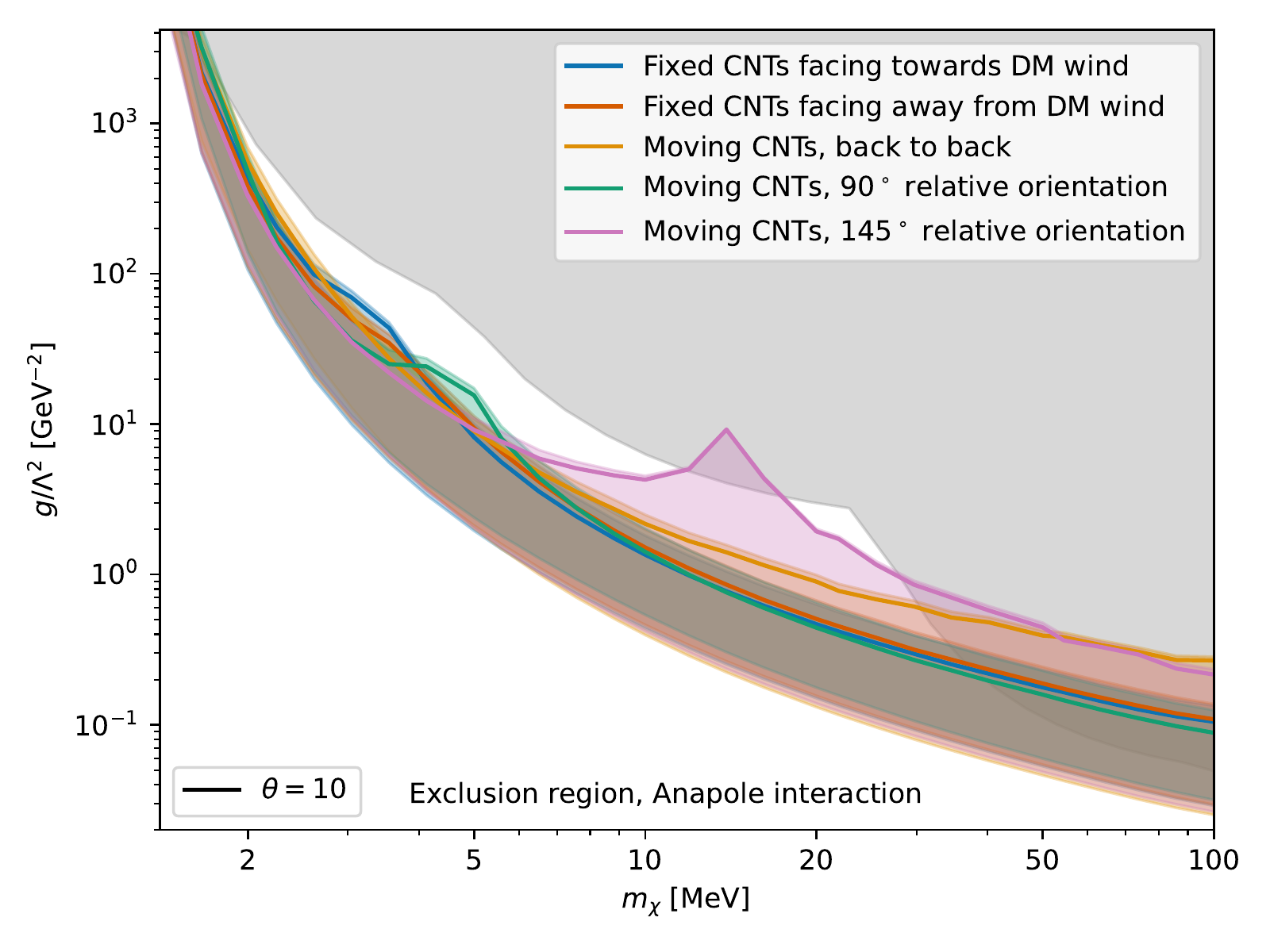}
    \includegraphics[width=0.48\textwidth]{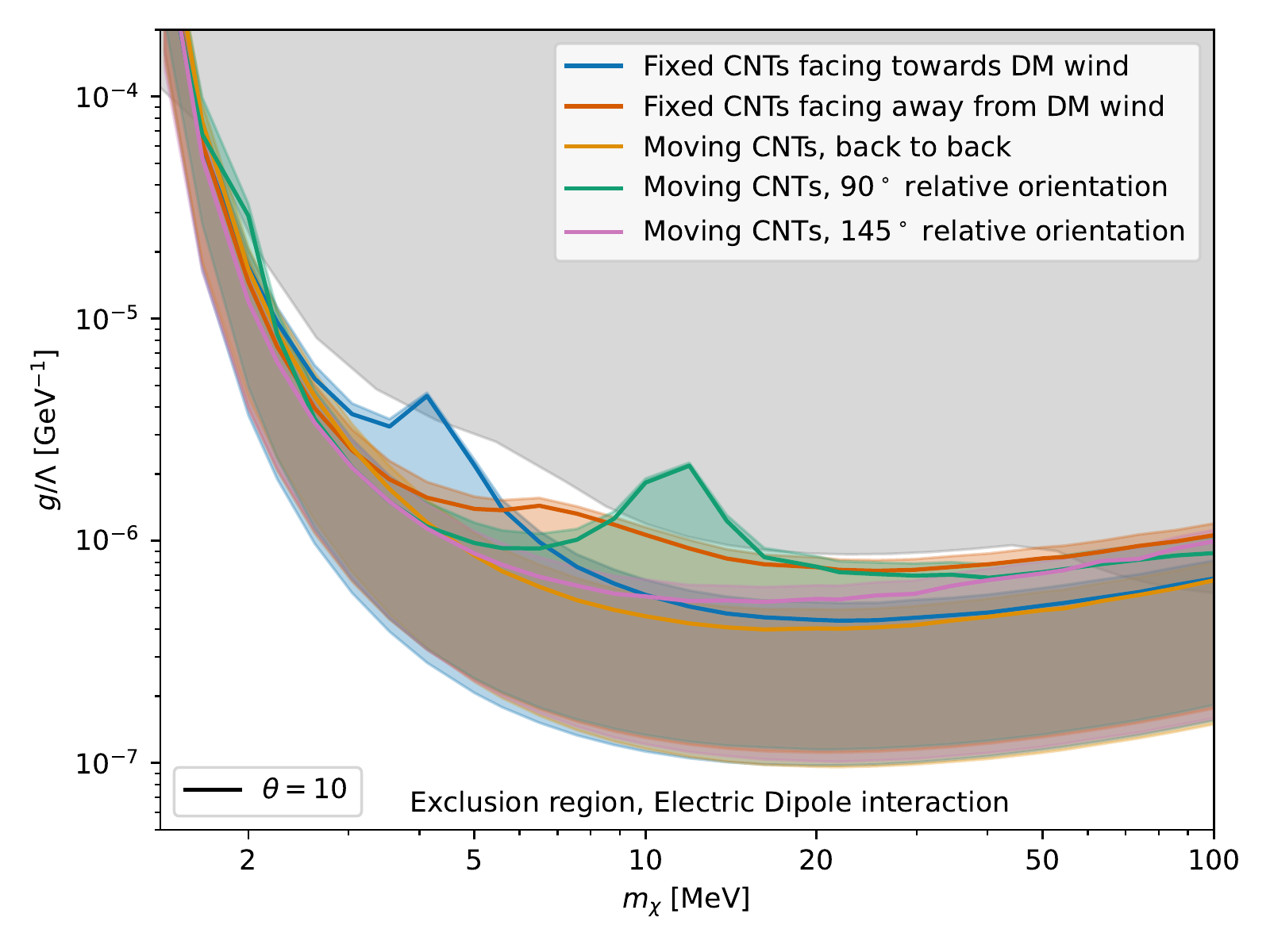}
    \includegraphics[width=0.48\textwidth]{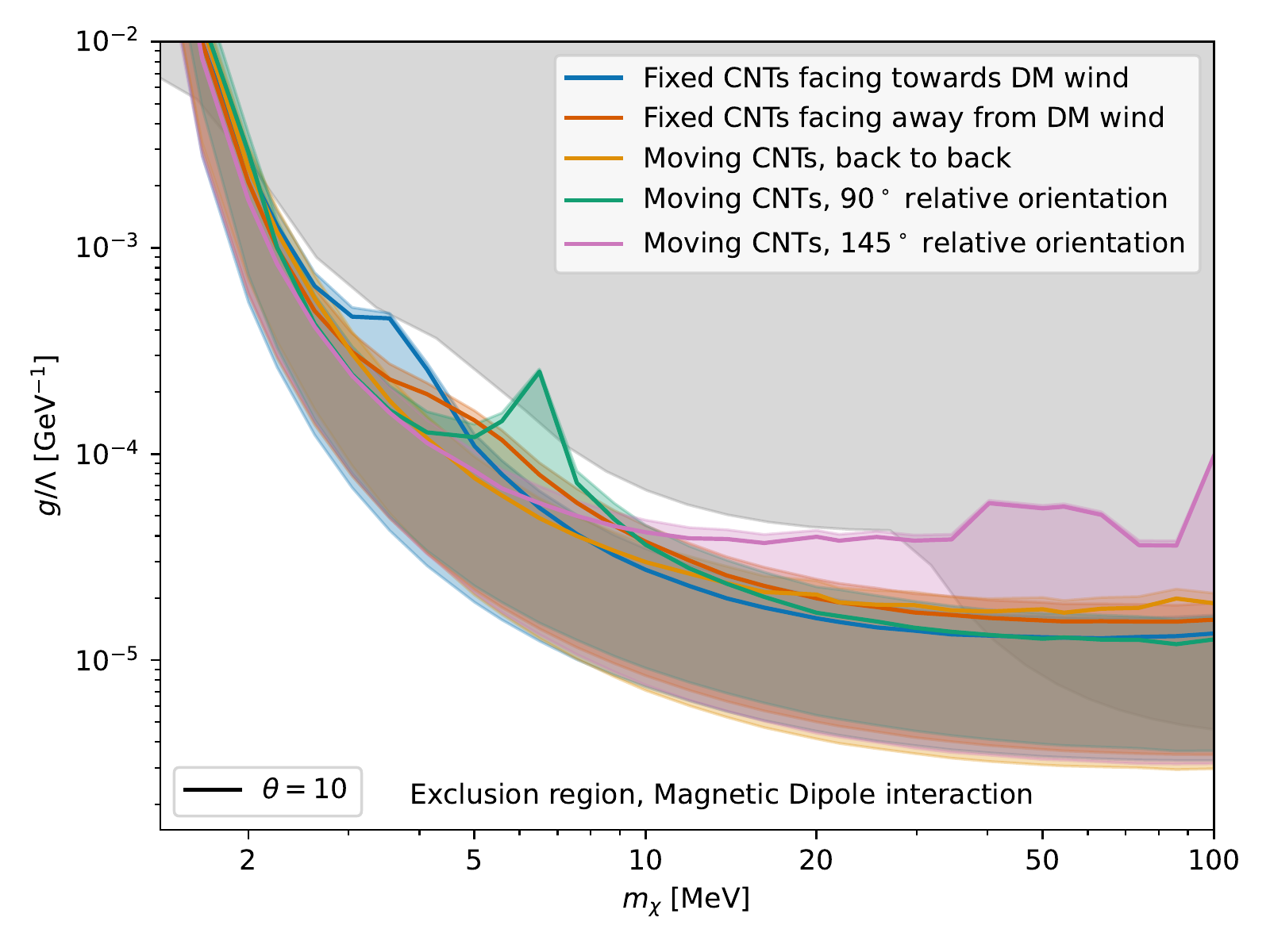}
    \caption{Same as Fig.~\ref{fig: Exclusion} but for the CNT based setups illustrated in Figs.~\ref{fig: Fixed  CNT several setups} and~\ref{fig: Moving CNT several setups}}
    \label{fig: CNT several setups exclusion}
\end{figure*}
\section{Summary and Conclusion}\label{sec:conclusions}
In summary, we performed a comprehensive theoretical study of graphene- and carbon nanotube-based targets as materials for the directional detection of DM, identifying the discovery potential and directionality for various types of interaction, as well as exploring different detector device setups. 

Based on our findings in Paper I, we chose DFT as our framework for obtaining the electronic structure of the graphene and CNT targets. We then modeled the DM-SM interaction within the generalized effective operator approach~\cite{Catena:2019gfa, Catena:2021qsr}, and described the ejected final-state electrons as free-electron plane waves. The use of plane-wave final states allowed us to reduce the formalism previously employed in~\cite{Catena:2021qsr} for 3D semiconductor targets, and describe all the operators listed in Tab.~\ref{tab:operators} with a single crystal response, Eq.~\eqref{eq: response function general}. 

Our main findings are as follows:

1) The directionality and characteristic spatial distribution of DM-induced electronic ejections from graphene and CNTs can be used to exclude many potential background sources, which in turn enables  the origin of the observed signal to be identified. In addition to either confirming or disproving the DM-origin hypothesis, the distribution yields important information about the nature of the DM candidate. In particular, different DM couplings and masses produce different angular distributions of the final-state electrons allowing for an interaction-type distinction once a signal has been established. 

2) The reach of the graphene- and CNT-based experiments currently under consideration is not expected to push the detection reach many orders of magnitude beyond its current experimental limits~\cite{SENSEI:2020dpa,Catena:2021qsr,XENON10:2011prx,Essig:2012yx,XENON:2019gfn,XENONCollaborationSS:2021sgk,Catena:2019gfa}. It will help, however, to resolve the observed signal excesses of other experiments in that mass range. 

3) In general, CNTs exhibit a slightly better detection reach than graphene sheets, with moving CNT-based detectors with the two detector arms oriented back-to-back and fixed CNTs facing toward the DM wind being particularly promising. Detector sensitivities depend heavily on the candidate mass and coupling type, and for a particular target material choice, one can switch between operation regimes in order to cover the largest area of the possible DM parameter space or to probe one specific DM candidate with more precision. 

4) The test statistics in the treatment of the measured data differ depending on whether a 3$\sigma$ modulation has been established or not. In the first case, the reach of the detector is given by the modulation of the signal and, in the latter, it is the absolute number of observed events and the ``null result'' of modulation that determines the achieved exclusion limits. We provide a quantification of the difference for the considered detector designs at an exposure of 10\,g-yr.

We hope that our analysis is useful in the optimization of graphene- and CNT-based detectors currently under consideration, and motivates consideration of other 2D materials for directional detection of light dark matter via electronic ejection.

An updated version of the \texttt{QEdark-EFT} tool that was developed for electron ejections from two-dimensional targets will be made available at the time of publication~\cite{QEdark-EFT}.

\acknowledgments
The authors thank Francesco Pandolfi and Christopher G. Tully for valuable discussions on the experimental setups.
R.C. and T.E. acknowledge support from the Knut and Alice Wallenberg project grant Light Dark Matter (Dnr.~KAW 2019.0080). Furthermore, R.C. acknowledges support from individual research grants from the Swedish Research Council, Dnr. 2018-05029 and Dnr. 2022-04299. TE was also supported by the Knut and Alice Wallenberg Foundation (PI, Jan Conrad). N.A.S. and M.M. were supported by the ETH Zürich and by the European Research Council (ERC) under the European Union’s Horizon 2020 research and innovation program Grant Agreement No. 810451.
T.E. thanks the Theoretical Subatomic Physics group at Chalmers University of Technology for its hospitality. 

\clearpage
\appendix

\section{Comparison of CNT based setups}
\label{app: Comparison of CNT based setups}

When constructing CNT-based experiments, the performance is influenced by the orientation of the detectors. In the main body of the text, we considered two CNT-based experiments shown in Fig.~\ref{fig: CNTs setups}, ``Fixed CNTs facing towards the DM wind'' and ``Moving CNTs, back to back''. We illustrate other possible setups, ``Fixed CNTs facing away from the DM wind'', ``Moving CNTs, $90^\circ$ relative orientation'' and ``Moving CNTs, $145^\circ$ relative orientation'' in Figs.~\ref{fig: Fixed CNT several setups} and ~\ref{fig: Moving CNT several setups}. The latter is tailored to have one detector hit the peak around $t=6\,\mathrm{h}$ present for low DM masses in Fig.~\ref{fig: daily modulation CNTs}, while the other detector faces away from the DM wind. The performance of these setups is shown in Figs.~\ref{fig: CNT several setups sensitivities} and~\ref{fig: CNT several setups exclusion}.

\section{Expanded matrix element}
\label{app: matrix element}

For convenience, here we include the explicit free particle response function $R_\mathrm{free}$ from Eq.~(\ref{eq:rate_general}) as taken from the Appendix of Paper I.

In order to avoid making the expressions too large, we divide $R_\mathrm{free}$ into three separate terms,
\begin{widetext}
\begin{align}
    R_\mathrm{free}=&\overline{\left|\mathcal{M}\right|^2}+ 2 m_e \overline{\Re\left[\mathcal{M}(\nabla_{\boldell}\mathcal{M}^*)_{\boldell=0}\cdot\frac{\mathbf{q}-\mathbf{k}^\prime}{m_e}\right]} + m_e^2 \overline{\left|(\nabla_{\boldell}\mathcal{M})_{\boldell=0}\cdot \frac{\mathbf{q}-\mathbf{k}^\prime}{m_e}\right|^2} \,,
\end{align}
where $\mathbf{q}$ is the momentum transfer, $m_e$ is the electron mass, $\mathbf{k}^\prime$ is the final state electron momentum, and $\boldell$ is the initial state electron momentum. The individual terms can then be expressed as
\begin{align}
    \overline{\left|\mathcal{M}\right|^2} =& c_1^2 + \frac{c_3^2}{4}\left( \frac{\mathbf{q}}{m_e}\times\vPerpEl \right)^2+ \frac{c_7^2}{4}\left(\vPerpEl\right)^2+ \frac{c_{10}^2}{4}\left(\frac{\mathbf{q}}{m_e}\right)^2 +\frac{j_\chi(j_\chi+1)}{12}\Bigg\{ 3 c_4^2+\left(4c_5^2-2c_{12}c_{15}\right)\left( \frac{\mathbf{q}}{m_e}\times\vPerpEl \right)^2 \nonumber\\
    & + c_6^2 \left(\frac{\mathbf{q}}{m_e}\right)^4 + \left(4 c_8^2+2c_{12}^2\right) \left(\vPerpEl\right)^2  + \left(2c_{9}^2+4c_{11}^2+2c_4c_6\right)\left(\frac{\mathbf{q}}{m_e}\right)^2+\left(c_{13}^2+c_{14}^2\right)\left(\frac{\mathbf{q}}{m_e}\right)^2\left(\vPerpEl\right)^2 \nonumber\\
    &+ c_{15}^2\left(\frac{\mathbf{q}}{m_e}\right)^2\left( \frac{\mathbf{q}}{m_e}\times\vPerpEl \right)^2 +2c_{13}c_{14}\left(\frac{\mathbf{q}}{m_e}\cdot \vPerpEl\right)\left(\frac{\mathbf{q}}{m_e}\cdot \vPerpEl\right)\Bigg\}\, ,
\end{align}

where $\vPerpEl=\mathbf{v}-\frac{\mathbf{q}}{2\mu_{\chi e}}-\frac{\boldell}{m_e}$ with $\mathbf{v}$ being the DM initial velocity in the detector rest frame and $\mu_{\chi e}$ the DM-electron reduced mass. $c_i$'s are the effective couplings, and $j_\chi$ is the DM spin which we set to 1/2.

\begin{align}
    2 m_e \overline{\Re\left[\mathcal{M}(\nabla_{\mathbf{p}_1}\mathcal{M}^*)_{\mathbf{p}_1=0}\cdot\frac{\mathbf{q}-\mathbf{k}^\prime}{m_e}\right]}=& \left[\frac{c_3^2}{2}\left(\left(\frac{\mathbf{q}}{m_e}\cdot \vPerpEl\right) \frac{\mathbf{q}}{m_e}-\left(\frac{\mathbf{q}}{m_e}\right)^2\vPerpEl\right) - \frac{c_7^2}{2}\vPerpEl\right]\cdot\frac{\mathbf{q}-\mathbf{k}^\prime}{m_e}\nonumber\\
 &+\frac{j_\chi(j_\chi+1)}{6}\Bigg\{ \bigg[\left(4c_5^2+c_{15}^2\left(\frac{\mathbf{q}}{m_e}\right)^2\right)\left(\left(\frac{\mathbf{q}}{m_e}\cdot \vPerpEl\right) \frac{\mathbf{q}}{m_e}-\left(\frac{\mathbf{q}}{m_e}\right)^2\vPerpEl\right)\nonumber\\
 &-\left(4c_8^2+2c_{12}^2+(c_{13}^2+c_{14}^2)\left(\frac{\mathbf{q}}{m_e}\right)^2\right)\vPerpEl\bigg]\cdot\frac{\mathbf{q}-\mathbf{k}^\prime}{m_e}\nonumber\\
 &-2c_{12}c_{15}\left(\left(\frac{\mathbf{q}}{m_e}\cdot \vPerpEl\right) \frac{\mathbf{q}}{m_e}-\left(\frac{\mathbf{q}}{m_e}\right)^2\vPerpEl\right)\cdot\frac{\mathbf{q}-\mathbf{k}^\prime}{m_e}\nonumber\\
 &-2c_{13}c_{14}\left(\frac{\mathbf{q}}{m_e}\cdot \vPerpEl\right)\frac{\mathbf{q}}{m_e}\cdot\frac{\mathbf{q}-\mathbf{k}^\prime}{m_e}\Bigg\}\,,
\end{align}

and 

\begin{align}
    m_e^2 \overline{\left|(\nabla_{\mathbf{p}_1}\mathcal{M})_{\mathbf{p}_1=0}\cdot \frac{\mathbf{q}-\mathbf{k}^\prime}{m_e}\right|^2} =&\left(\frac{c_3^2}{4}\left(\frac{\mathbf{q}}{m_e}\right)^2+\frac{c_7^2}{4}\right)\left(\frac{\mathbf{q}-\mathbf{k}^\prime}{m_e}\right)^2 -\frac{c_3^2}{4}\left(\frac{\mathbf{q}}{m_e}\cdot \frac{\mathbf{q}-\mathbf{k}^\prime}{m_e}\right)^2\nonumber\\
&+\frac{j_\chi(j_\chi+1)}{12}\Bigg\{ \left(\frac{\mathbf{q}-\mathbf{k}^\prime}{m_e}\right)^2 \Bigg[(4c_5^2+c_{13}^2+c_{14}^2-2c_{12}c_{15})\left(\frac{\mathbf{q}}{m_e}\right)^2+4c_8^2+2c_{12}^2 \nonumber\\
&+c_{15}^2\left(\frac{\mathbf{q}}{m_e}\right)^4\Bigg]+ \left(\frac{\mathbf{q}}{m_e}\cdot \frac{\mathbf{q}-\mathbf{k}^\prime}{m_e}\right)^2 \left[ -4c_5^2-c_{15}^2\left(\frac{\mathbf{q}}{m_e}\right)^2+2c_{12}c_{15}+2c_{13}c_{14})\right]  \Bigg\}\, .
\end{align}
\end{widetext}

Furthermore, to rewrite the above equations one can use the following relations
\begin{align}
    \left(\frac{\mathbf{q}}{m_e}\times \vPerpEl\right)^2=&\left(\frac{\mathbf{q}}{m_e}\right)^2\left(\vPerpEl\right)^2-\left(\frac{\mathbf{q}}{m_e}\cdot \vPerpEl\right)^2\,,\\
    \left(\vPerpEl\right)^2|_{\boldell=0}=&\mathbf{v}^2+\frac{\mathbf{q}^2}{4\mu_{\chi e}^2}\frac{m_\chi-m_e}{m_e+m_\chi}-\frac{\Delta E_e}{\mu_{\chi e}}\,,\\
    \left(\vPerpEl\cdot \mathbf{q}\right)|_{\boldell=0}=&\Delta E_e - \frac{\mathbf{q}^2}{2m_e}\,,\\
    \vPerpEl|_{\boldell=0}=&\mathbf{v}-\frac{\mathbf{q}}{2\mu_{\chi e}}\,,
\end{align}
where $\Delta E_e$ is the energy transferred to the target electron.

\bibliography{ref,ref2,bibliography,Nicola,references}

\end{document}